\pdfoutput=1 

\documentclass[%
 reprint,
nofootinbib,
 amsmath,amssymb,
 aps,
]{revtex4-1}

\usepackage{preprintcover}  
\PreprintCoverPaperTitle{
Measurements of $W\gamma$ and $Z\gamma$ production in $pp$ collisions at $\sqrt{s}=$ 7 TeV with the ATLAS detector at the LHC}  
\PreprintIdNumber{CERN-PH-EP-2012-345}  
\PreprintCoverAbstract{ 
The integrated and differential fiducial cross sections for the production of a $W$ or $Z$ boson in association
with a high-energy photon are measured using $pp$ collisions at $\sqrt{s}$ = 7 TeV.
The analyses use a data sample with an integrated luminosity of
4.6 fb$^{-1}$ collected by the ATLAS detector during the 2011 LHC data-taking period. Events are
selected using leptonic decays of the $W$ and $Z$ bosons [$W(e\nu,\mu\nu)$ and
$Z(e^{+}e^{-}, \mu^{+}\mu^{-}, \nu\bar{\nu})$] with the requirement of an
associated isolated photon. The data are used to test the electroweak
sector of the Standard Model and search for evidence for new phenomena.
The measurements are used to probe the anomalous $WW\gamma$, $ZZ\gamma$ and $Z\gamma\gamma$
triple-gauge-boson couplings and to search for the production of vector resonances decaying to $Z\gamma$ and $W\gamma$.
No deviations from Standard Model predictions are observed and
limits are placed on anomalous triple-gauge-boson couplings and on the production of new vector meson resonances.
}  

\PreprintJournalName{Phys. Rev. D.}  

\usepackage{amssymb}
\usepackage{color} 
\usepackage{atlasphysics}
\usepackage{subfigure}
\usepackage{graphicx}
\usepackage{longtable}
\usepackage{multirow}
\usepackage{textcomp}
\usepackage{units}
\usepackage{marvosym}
\usepackage{hyperref}

\usepackage{graphicx}
\usepackage{dcolumn}
\usepackage{bm}
\usepackage[mathlines]{lineno}

\usepackage{enumerate}

\newcommand{\tpion}{\ensuremath{\pi_{\mathsf{T}}}}

\newcommand{\trho}{\ensuremath{\rho_{\mathsf{T}}}}
\newcommand{\tomega}{\ensuremath{\omega_{\mathsf{T}}}}
\newcommand{\techa}{\ensuremath{a_{\mathsf{T}}}}
\newcommand{\Zgamma}{\ensuremath{\Zboson\gamma}}
\newcommand{\Wgamma}{\ensuremath{\Wboson\gamma}}

\begin{document}
\title{
Measurements of $W\gamma$ and $Z\gamma$ production in $pp$ collisions at $\sqrt{s}=$ 7 TeV with the ATLAS detector at the LHC
}

\collaboration{ATLAS Collaboration}
\date{August 21, 2012}

\begin{abstract}
The integrated and differential fiducial cross sections for the production of a $W$ or $Z$ boson in association
with a high-energy photon are measured using $pp$ collisions at $\sqrt{s}$ = 7 TeV.
The analyses use a data sample with an integrated luminosity of
4.6 fb$^{-1}$ collected by the ATLAS detector during the 2011 LHC data-taking period. Events are
selected using leptonic decays of the $W$ and $Z$ bosons [$W(e\nu,\mu\nu)$ and
$Z(e^{+}e^{-}, \mu^{+}\mu^{-}, \nu\bar{\nu})$] with the requirement of an
associated isolated photon. The data are used to test the electroweak
sector of the Standard Model and search for evidence for new phenomena.
The measurements are used to probe the anomalous $WW\gamma$, $ZZ\gamma$ and $Z\gamma\gamma$
triple-gauge-boson couplings and to search for the production of vector resonances decaying to $Z\gamma$ and $W\gamma$.
No deviations from Standard Model predictions are observed and
limits are placed on anomalous triple-gauge-boson couplings and on the production of new vector meson resonances.
\end{abstract}

\maketitle

\section{Introduction}
\label{sec:intro}
The Standard Model (SM) has proved to provide an accurate description
of the production of elementary particles observed in high energy
physics experiments.
The interactions of $W$ and $Z$ bosons with photons are particularly interesting as they
test the self-couplings of these bosons as 
predicted by the non-Abelian $SU(2)_L\times U(1)_Y$ gauge group of the electroweak sector.  
In particular, the high-energy proton--proton collisions provided by the LHC explore
the production of $W\gamma$ and $Z\gamma$ pairs in a new energy domain.
The high center-of-mass energy also allows searches for new particles,
for example, techni-mesons which are predicted in technicolor models~\cite{Weinberg:1979bn,Susskind:1978ms},
that decay to these final states. 

\begin{figure}[htbp]
  \centering
  \subfigure[u-channel]{\includegraphics[width=0.49\columnwidth]{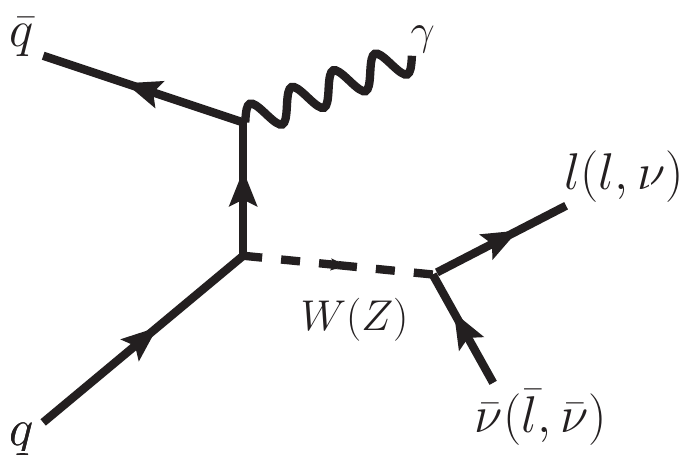}}
  \subfigure[t-channel]{\includegraphics[width=0.49\columnwidth]{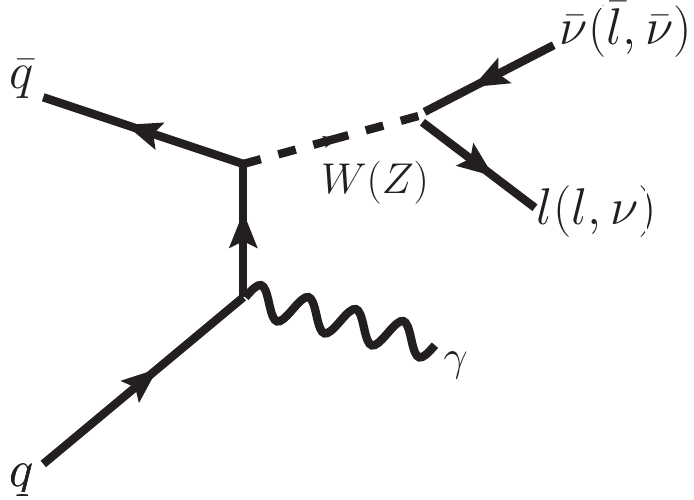}}
  \subfigure[FSR]{\includegraphics[width=0.40\columnwidth]{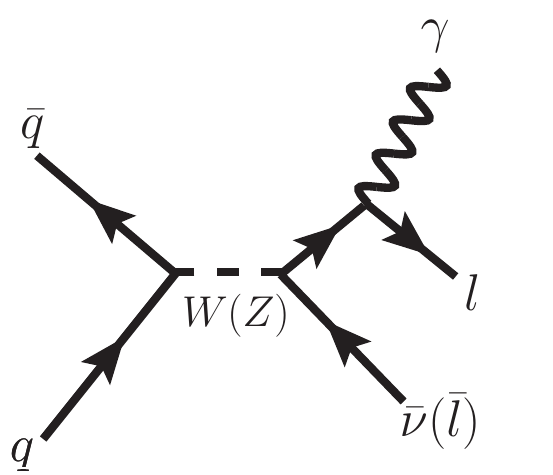}}
  \subfigure[s-channel]{\includegraphics[width=0.49\columnwidth]{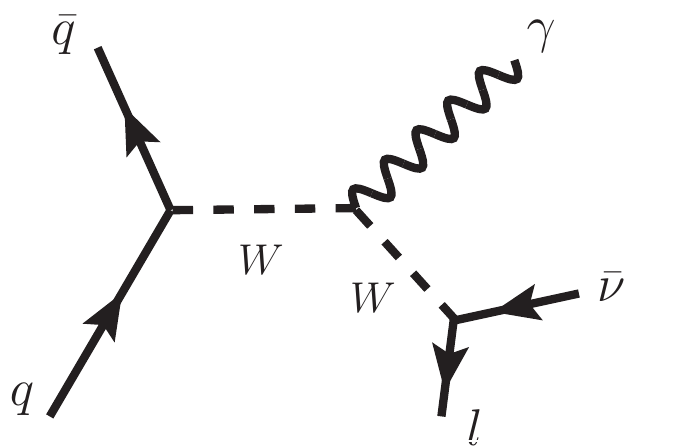}}
  \subfigure[fragmentation]{\includegraphics[width=0.49\columnwidth]{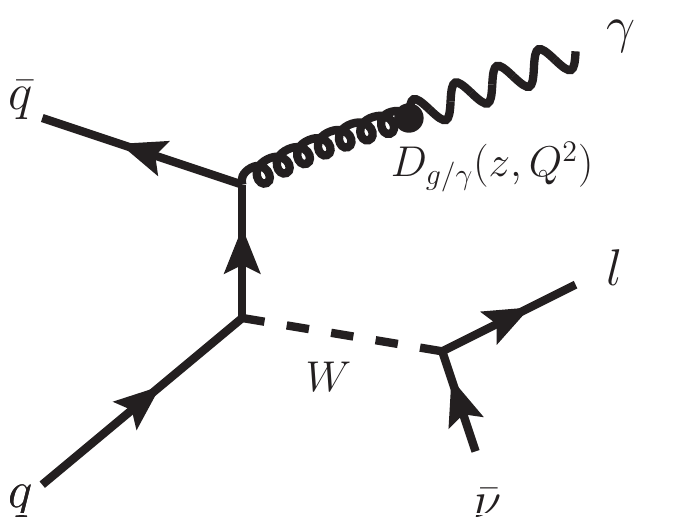}}
  \subfigure[fragmentation]{\includegraphics[width=0.49\columnwidth]{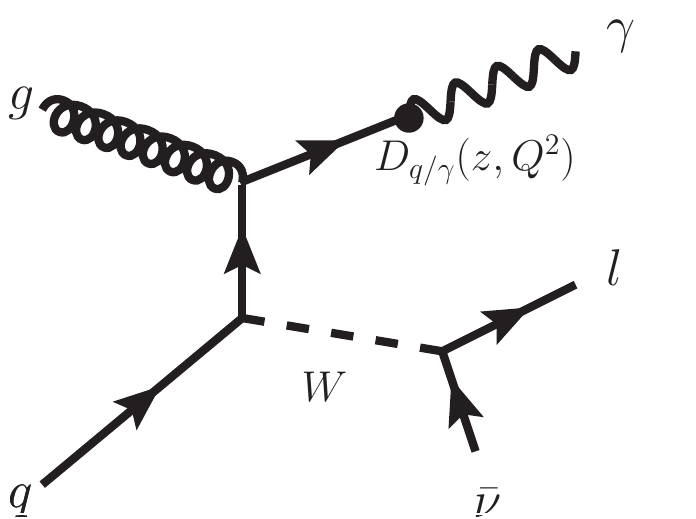}}
  \caption{Feynman diagrams of $W\gamma$ and $Z\gamma$ production in (a) u-channel
    (b) t-channel and (c) final state photon radiation (FSR) from the $W$ and $Z$ boson decay process.
    (d) Feynman diagram of $W\gamma$ production in the s-channel.
    Diagrams of the signal contributions from the
    $W+q(g)$~processes when a photon emerges from the fragmentation of (e) a gluon and (f) a quark
    in the final state.}
  \label{fig:fey_dia}
\end{figure}

The measurements presented here are improvements on previous studies of the hadroproduction of 
$W\gamma$ and $Z\gamma$ pairs, as more precise measurements are performed with a larger data sample.
The events used for the measurements were recorded in 2011 by the ATLAS detector~\cite{DetectorPaper:2008} from 4.6 fb$^{-1}$ of $pp$ collisions at a center-of-mass energy of 7 \TeV{}. 
The diboson candidate events are selected from the production processes 
$pp \to \ell \nu \gamma + X$ ($\ell=e$, $\mu$), $pp \to \ell^+  \ell^-  \gamma + X$
and $pp \to \nu\bar{\nu}  \gamma + X$. 
These final states include the production of $W$ and $Z$ bosons
with photon bremsstrahlung from the charged leptons from the $W/Z$ boson decays 
in addition to the $W\gamma$ and $Z\gamma$ 
diboson events of primary interest. In the SM, the latter originate from $W$ and $Z$ boson production with photons radiated from initial-state quarks (prompt photons), photons from the fragmentation of secondary quarks and gluons into isolated photons, and from photons radiated directly by $W$ bosons.
The diagrams of these production mechanisms are shown in Fig.~\ref{fig:fey_dia}.
Theories beyond the SM, such as technicolor, predict the decay of narrow resonances to $W\gamma$ or $Z\gamma$ pairs.
The data analyses presented here provide differential distributions of relevant kinematic variables,
corrected for detector effects, allowing the search for deviations from the SM predictions to be made with high sensitivity.

Previous measurements of $W\gamma$ and $Z\gamma$ final states from $p\bar{p}$ and $pp$ production have been made at the 
Tevatron, by the CDF~\cite{CDFpaper} and D$\O$~\cite{D0paper,D0paper2} collaborations, and at the LHC by the 
ATLAS~\cite{ATLASpaper,WZg1fbpaper} and CMS~\cite{CMSpaper}  collaborations. These experiments have set limits on anomalous triple gauge-boson couplings (aTGCs) that are improved on by the current analysis. 
The limits on new vector meson resonances that are presented in this paper improve on previous limits set at the Tevatron by the D$\O$~\cite{D0paperExoSearch} collaboration in the \Zgamma\ final state, and they are the first reported in the \Wgamma\ final state.

Throughout this paper the notations ``$\ell \nu \gamma$'', ``$\ell^+  \ell^-  \gamma$'' and ``$\nu\bar{\nu}  \gamma$''
specify the production channels ``$pp \to \ell \nu \gamma + X$'' , ``$pp \to \ell^+  \ell^-  \gamma + X$''
and ``$pp \to\nu\bar{\nu}  \gamma + X$'', respectively,
and the label ``$Z$'' refers to $Z/\gamma*$.
In addition, ``inclusive'' refers to production
with no restriction on the recoil system and ``exclusive'' refers to production restricted
to those events with no central jets with transverse energy greater than 30 GeV.
Measurements of integrated cross sections and differential kinematic distributions are performed within a fiducial region of the detector.
Events with high-transverse-energy photons are used to establish aTGC limits and to carry out
the searches for narrow $W\gamma$ and $Z\gamma$ resonances.

This paper is organized as follows: An overview of the ATLAS detector and the data samples used is given in Section~\ref{sec:atlasdet}. Section~\ref{sec:simulation} describes the signal and background Monte Carlo samples. Section~\ref{sec:eff} defines the selections of the physics objects such as photons, leptons, and jets. Section~\ref{sec:Event_Selection} describes the event selection criteria for $W\gamma$ and $Z\gamma$ candidates. Section~\ref{sec:background} presents the background estimations. Section~\ref{sec:cs} presents the measured $V\gamma$ ($V=W$ or $Z$) fiducial cross sections. Section~\ref{sec:theory} summarizes the comparisons between the measurements and SM predictions. The observed aTGC limits are presented in Section~\ref{sec:ATGC} and the limits on masses of new vector meson resonances are given in Section~\ref{sec:ExoticSearch}.

\section{The ATLAS Detector and the Data Sample}
\label{sec:atlasdet}
The ATLAS detector is composed of
an inner tracking system (ID) surrounded by a thin
superconducting solenoid providing a 2~T axial magnetic field,
electromagnetic (EM) and hadronic calorimeters, and a muon spectrometer (MS).  
The ID consists of three subsystems: the pixel and silicon
microstrip (SCT) detectors cover the pseudorapidity\footnote{ATLAS uses a right-handed coordinate system with its origin at the nominal interaction point (IP) in the center of the detector and the $z$-axis along the beam pipe. The $x$-axis points from the IP to the center of the LHC ring, and the $y$-axis points upward. Cylindrical coordinates $(r,\phi)$ are used in the transverse (x,y) plane, $\phi$ being the azimuthal angle around the beam pipe. The pseudorapidity is defined in terms of the polar angle $\theta$ as $\eta=-\ln\tan(\theta/2)$. The distance $\Delta R$ in the $\eta-\phi$ space is defined as $\Delta R = \sqrt{({\Delta \eta})^2 + ({\Delta \phi})^2}$.} range $|\eta|<2.5$, while the transition radiation tracker (TRT), which is made of straw tubes,
has an acceptance range of $|\eta|<2.0$.
The calorimeter system covers the range $|\eta|<4.9$.
The highly segmented electromagnetic calorimeter, which plays a crucial role in electron
and photon identification, comprises lead absorbers with liquid argon (LAr) as the active
material and covers the range $|\eta|<3.2$.
In the region $|\eta|<1.8$, a presampler detector using a thin layer of LAr is used to
correct for the energy lost by electrons and photons upstream of the calorimeter.
The hadronic tile calorimeter ($|\eta|<1.7$) is a steel/scintillating-tile detector and is located directly
outside the envelope of the barrel electromagnetic calorimeter.
The two end-cap hadronic calorimeters have LAr as the active material and copper absorbers.
The calorimeter coverage is extended to $|\eta|=4.9$ by a forward calorimeter with LAr as
active material and copper (EM) and tungsten (hadronic) as absorber material.
The MS is based on three large superconducting aircore toroid magnets,
a system of three stations of chambers for precise tracking
measurements in the range $|\eta|<2.7$, and a muon trigger system that
covers the range $|\eta|<$ 2.4.

The data used for the analyses presented in this paper were collected in 2011 from
$pp$ collisions at a center-of-mass energy of 7 \TeV{}.
The total integrated luminosity is 4.6 fb$^{-1}$ with an
uncertainty of 3.9$\%$~\cite{ATLASLumi1,ATLASLumi2}.
Events were selected by triggers requiring at least one identified
electron, muon, or photon.
The transverse energy ($E_{\mathrm{T}}$) threshold for the
single-electron trigger was initially 20 GeV, and was raised to 22 GeV
in the later part of 2011 to maintain a manageable trigger rate at higher
instantaneous luminosity.
The transverse momentum ($p_{\mathrm{T}}$) threshold for the single-muon trigger was 18 GeV.
Single-photon events were triggered with a transverse energy 
$E_{\mathrm{T}}>80$ GeV.

\section{Signal and Background modeling}
\label{sec:simulation}
 Monte Carlo (MC) event samples, including a full simulation~\cite{ATLASsim}
of the ATLAS detector with {\sc geant4}~\cite{geant4},
are used to compare the data to the SM signal and background expectations.
All MC samples are simulated with additional $pp$ interactions (pileup) in the same and neighboring bunch crossings.
The number of $pp$ interactions in the same bunch crossing averages 9 and extends up to about 20, as observed in the data.

The production of $pp \to \ell \nu  \gamma$ and $pp \to \tau \nu  \gamma$ is modeled with the
{\sc alpgen}~(2.14) generator~\cite{alpgen} interfaced to
{\sc herwig}~(6.520)~\cite{herwig}  for parton shower and fragmentation processes,
and to {\sc jimmy}~(4.30)~\cite{jimmy}  for underlying event simulation.
The modeling of $pp \to \ell^+ \ell^- \gamma$ and  $pp \to \nu \bar{\nu} \gamma$
processes is performed with the {\sc sherpa}~(1.4.0) generator~\cite{sherpa} 
since the simulation of these processes is not available in {\sc alpgen}.
An invariant mass cut of $m(\ell^{+}\ell^{-})>40$~GeV is applied at the generator level
when simulating the $pp \to \ell^+ \ell^- \gamma$ process.
The {\sc cteq6l1}~\cite{cteq6l1} and {\sc cteq6.6m}~\cite{cteq66m}
parton distribution functions (PDFs) are used for
samples generated with {\sc alpgen} and {\sc sherpa}, respectively.
The final-state radiation (FSR) photons from charged leptons are simulated by
{\sc photos}~(2.15)~\cite{photos} for the {\sc alpgen} sample,
and by the {\sc sherpa} generator~\cite{sherpafsr} for the {\sc sherpa} sample.
All the signal production processes, including the quark/gluon fragmentation into photons,
are simulated by these two generators. 
The {\sc alpgen} sample is generated with leading-order (LO) matrix elements
for final states with up to five additional partons, whereas the {\sc sherpa} sample is
generated with LO matrix elements for final states with up to three additional partons.  
In the search for technicolor, the signal processes are simulated using {\sc pythia} (6.425)~\cite{pythia} with a LO {\sc mrst2007}~\cite{pdfmrst2007} PDF set.

The $Z(\ell^{+}\ell^{-})$ and $Z(\tau^{+} \tau^{-})$ backgrounds are
modeled with {\sc pythia}.
The radiation of photons from charged leptons is treated in {\sc pythia}
using {\sc photos}.
{\sc tauola}~(1.20)~\cite{tau} is used to model $\tau$ lepton decays.
The {\sc powheg}~(1.0)~\cite{powheg} generator is used to simulate $t\bar{t}$ production and is
interfaced to {\sc pythia} for parton showering and fragmentation. 
The $WW$ and single top quark processes are modeled by
{\sc mc@nlo}~(4.02)~\cite{MCatNLO1,MCatNLO2}, interfaced to {\sc herwig}
for parton showering and fragmentation.
The LO {\sc mrst2007} PDF set is used to simulate the
$Z(\ell^{+}\ell^{-})$, $Z(\tau^{+} \tau^{-})$, and $W(\tau \nu)$
backgrounds, and the {\sc ct10}~\cite{CT10} PDF set is used in simulating $t\bar{t}$,
single top quark, and $WW$ production.
The next-to-leading-order (NLO) cross-section predictions~\cite{Hamberg:1990np,Anastasiou:2003ds,Bonciani:1998vc,Moch:2008qy}
are used to normalize the simulated background events.
Backgrounds where a jet or an electron is misidentified as a photon are derived from data as described in Sec.~\ref{sec:background}.

\section{Physics Object Reconstruction}   
\label{sec:eff}

The $W$ and $Z$ bosons are reconstructed from their leptonic decays.
The $\ell \nu\gamma$ final state consists of an isolated electron or muon,
large missing transverse momentum due to the undetected neutrino, and an isolated photon.
The $\ell^+ \ell^- \gamma$ final state contains one $e^{+}e^{-}$ or $\mu^{+}\mu^{-}$ pair
and an isolated photon. 
The $\nu \bar{\nu}\gamma$ final state contains at least one isolated photon and large
missing transverse momentum due to the undetected neutrinos.
Collision events are selected by requiring at least one reconstructed vertex with at 
least three charged particle tracks with $p_{\mathrm{T}}>0.4$~GeV.
If more than one vertex satisfies the vertex selection requirement,
the vertex with the highest sum of the $p_{\mathrm{T}}^{2}$ of the associated tracks is chosen
as the primary vertex.
Physics objects for the measurement are required to be associated with the primary vertex.

An electron candidate is obtained from an energy cluster in the EM calorimeter
associated with a reconstructed track in the ID.
The transverse energy of electrons is required to be greater than 25 GeV{}.
The electron cluster must lie outside the transition region between the
barrel and end-cap EM calorimeters and within the overall fiducial acceptance of the EM calorimeters and the ID,
so it must satisfy $|\eta|<1.37$ or $1.52<|\eta|<2.47$.
At the electron track's closest approach to the primary vertex, the ratio of the transverse impact parameter
$d_{0}$ to its uncertainty (the $d_{0}$ significance) must be smaller than 10,
and the longitudinal impact parameter $|z_{0}|$ must be less than 1 mm.
Tight\footnote{The definitions of tight and medium identification~\cite{atlas_electron} were reoptimized for 2011 data-taking conditions. They are based on information about calorimeter shower shapes, track quality, track--calorimeter-cluster matching, particle identification information from the TRT, and a photon conversion veto.} electron identification~\cite{atlas_electron} is used in the
$W(e \nu)\gamma$ analysis,
whereas medium identification~\cite{atlas_electron} is used to select electrons in the
$Z(e^+ e^-)\gamma$ analysis.
To reduce the background due to a jet misidentified as an electron,
a calorimeter-based isolation requirement $E_{\mathrm{T}}^{\mathrm{iso}}<6$ GeV{} is applied
to the electron candidate.
$E_{\mathrm{T}}^{\mathrm{iso}}$ is the total transverse energy recorded in the
calorimeters within a cone of radius $\Delta R=0.3$ around the electron position
excluding the energy of the electron itself.
$E_{\mathrm{T}}^{\mathrm{iso}}$ is corrected for leakage from the electron energy cluster's core into the isolation cone and for contributions from the underlying event and pileup~\cite{photonpaper,ATLASPhotonIDConf}.

Muon candidates are identified by associating complete tracks or track segments in the
MS to tracks in the ID~\cite{WZpaper}.
Each selected muon candidate is a combined track originating from the primary vertex
with transverse momentum $p_{\mathrm{T}}>25$ GeV and $|\eta|<2.4$.
It is required to be isolated by imposing $R^{\mathrm{iso}}<0.15$,
where $R^{\mathrm{iso}}$ is the sum of the $p_{\mathrm{T}}$ of the tracks
in a $\Delta R =0.3$ cone around the muon direction, excluding the track of the muon,
divided by the muon $p_{\mathrm{T}}$.
The $d_{0}$ significance must be smaller than 3, and $|z_{0}|$ must be less than 1 mm.

Photon candidates are based on clustered energy deposits in the EM calorimeter in the
range $|\eta|<2.37$ (excluding the calorimeter transition region $1.37<|\eta|<1.52$)
with $E_{\mathrm{T}}>$ 15 \GeV{}.
Clusters without matching tracks are directly classified as unconverted photon candidates.
Clusters that are matched to tracks that originate from reconstructed conversion vertices in the ID
or to tracks consistent with coming from a conversion are considered as converted photon candidates.
Tight requirements on the shower shapes~\cite{photonpaper} are
applied to suppress the background from multiple showers produced in
meson (e.g. $\pi^{0}$, $\eta$) decays. To further reduce this
background, a photon isolation requirement $E_{\mathrm{T}}^{\mathrm{iso}}<6$ GeV is applied.
The definition of photon isolation is the same as the electron isolation described above.

Jets are reconstructed from energy observed in the calorimeter cells using the anti-$k_t$ jet clustering
algorithm~\cite{antikt} with radius parameter $R=0.4$.
The selected jets are required to have $p_{\mathrm{T}}>30$ GeV with $|\eta| < 4.4$,
and to be well separated from the lepton and photon candidates
($\Delta R(e/\mu/\gamma,$ jet)$>0.3$).

The missing transverse momentum ($E_{\mathrm{T}}^{\mathrm{miss}}$)~\cite{METpaper}
magnitude and direction are measured from the vector sum of
the transverse momentum vectors associated with clusters of energy reconstructed
in the calorimeters with $|\eta|<4.9$.
A correction is applied to the energy of those clusters that are associated with
a reconstructed physical object (jet, electron, $\tau$ lepton, photon).
Reconstructed muons are also included in the sum, and any calorimeter energy deposits
associated with them are excluded to avoid double counting.

\section{$W\gamma$ and $Z\gamma$ event selection}
\label{sec:Event_Selection}

The $\ell \nu \gamma$ candidate events are selected by requiring exactly one lepton with $p_{\mathrm{T}}>25$ \GeV{}, at least one isolated photon with $E^{\gamma}_{\mathrm{T}}>15$ \GeV{} and $E_{\mathrm{T}}^{\mathrm{miss}}$ above 35 \GeV{}.  In addition, the transverse mass\footnote{ $m_{\mathrm{T}}= \sqrt{2p_{\mathrm{T}}(\ell) \times E_{\mathrm{T}}^{\mathrm{miss}} \times (1-\cos{\Delta{\phi}})}$, and $\Delta{\phi}$ is the azimuthal separation between the directions of the lepton
and the missing transverse momentum vector.} of the lepton--$E_{\mathrm{T}}^{\mathrm{miss}}$ system
is required to be greater than 40~\GeV{}.
A $Z$-veto requirement is applied in the electron channel of the $W\gamma$ analysis by requiring
that the electron--photon invariant mass ($m_{e\gamma}$) is not within 15~\GeV{} of the $Z$ boson mass.
This is to suppress the background where one of the electrons from the $Z$ boson decay
is misidentified as a photon.
The events selected by the criteria above are used for the inclusive $W\gamma$ cross-section measurements.

The $\ell^+ \ell^- \gamma$ candidates are selected by requiring exactly two oppositely charged same-flavor leptons with an invariant mass greater than 40~\GeV{} and one isolated photon with $E^{\gamma}_{\mathrm{T}}> 15$ \GeV{}.

The $\nu\bar{\nu} \gamma$ candidates are selected by requiring one isolated photon with $E^{\gamma}_{\mathrm{T}}> 100$ \GeV{} and $E_{\mathrm{T}}^{\mathrm{miss}}>90$ \GeV{}. The reconstructed photon, $E_{\mathrm{T}}^{\mathrm{miss}}$ and jets (if jets are found) are required to be well separated in the transverse plane with $\Delta\phi(E_{\mathrm{T}}^{\mathrm{miss}},\gamma)>2.6$ and $\Delta\phi(E_{\mathrm{T}}^{\mathrm{miss}}$, jet)$>0.4$, in order to reduce the $\gamma$+jets background. Events with identified electrons and muons are vetoed to reject $W+$jets and $W\gamma$ background. The selection criteria to identify the electrons and muons are the same as in the $Z(\ell^+ \ell^-)\gamma$ analysis.

In both the $W\gamma$ and $Z\gamma$ analyses, a selection requirement $\Delta R(\ell,\gamma)>0.7$ is applied to suppress the contributions from FSR photons in the $W$ and $Z$ boson decays. The events with no jets with $E_{\mathrm{T}}>30$ \GeV{} are used to measure the exclusive $V\gamma$ cross sections.
For $V\gamma$ production, events with a high-$E_{\mathrm{T}}$ photon tend to have more jet activity in the final state.
Contributions from aTGCs also enhance $V\gamma$ production with high-$E_{\mathrm{T}}$ photons.
Thus, the exclusive $V\gamma$ cross-section measurements are expected to be more sensitive to aTGC
than the inclusive measurements.
In the current analyses the sensitivity to aTGCs improves by $\sim 40\%$ when measurements are performed
using exclusive channels compared to inclusive channels.

\section{Background Estimation}
\label{sec:background}
In the measurements of $\ell \nu \gamma$, $\ell^{+} \ell^{-} \gamma$ and
$\nu \bar{\nu} \gamma$ production, the background contributions are estimated
either from simulation or from data.
The backgrounds estimated from data include $W+$jets and $\gamma+$jets
for the $\ell \nu \gamma$ final state, $Z+$jets for the $\ell^{+} \ell^{-} \gamma$ final state,
and $Z+$jets, multijets, $\gamma+$jets and events with an electron faking a photon
for the $\nu\bar{\nu} \gamma$ final state.
The remaining backgrounds are estimated from simulation.

For the differential fiducial cross sections, the contributions from each background
source are estimated in each bin used for the measurement.
The sources of backgrounds and the methods of estimating them are described in the following subsections.

\subsection{Background estimation for $pp \rightarrow \ell \nu \gamma$}
\label{sec:bglnugamma}
The primary backgrounds to the $\ell \nu \gamma$ signal come from the $W+$jets, $Z(\ell^+ \ell^-)$
and $\gamma+$jets processes.
\begin{enumerate}[i]
  \item Events from $W+$jets production can be misidentified as signal candidates when photons come
        from the decays of mesons produced in jet fragmentation (mainly $\pi^0 \to \gamma \gamma$);

  \item $Z(\ell^+ \ell^-)$ events mimic the $W\gamma$ signal when one of the leptons from the $Z$ boson
        decay is misidentified as a photon (in the case of the electron channel),
        or is not identified and the photon originates from initial-state radiation from a quark
        or from photon bremsstrahlung from a charged lepton;

  \item Events from $\gamma+$jets production can mimic the $W\gamma$ signal when there are leptons
        from heavy quark decays (or, in the electron channel, when charged hadrons or electrons from
        photon conversions are misidentified as prompt electrons), and large apparent
        $E_{\mathrm{T}}^{\mathrm{miss}}$  is created by a combination of real $E_{\mathrm{T}}^{\mathrm{miss}}$
        from neutrinos in heavy quark decays and of mis-measurement of jet energies;

  \item In addition, there are background contributions from $t\bar{t}$, single top quark,
        $WW$, $W(\tau \nu)$ and $Z(\tau \tau)$ processes. The $pp \to \tau \nu \gamma + X$ source of
        events is considered as a background since measurements of cross sections for
        $pp \to \ell \nu \gamma + X$ production are quoted for a single lepton flavor.
\end{enumerate}

The background contributions from $W+$jets and $\gamma+$jets events in the $W\gamma$ analysis are
estimated from data.

\begin{figure}
  \centering
  \includegraphics[width=0.9\columnwidth]{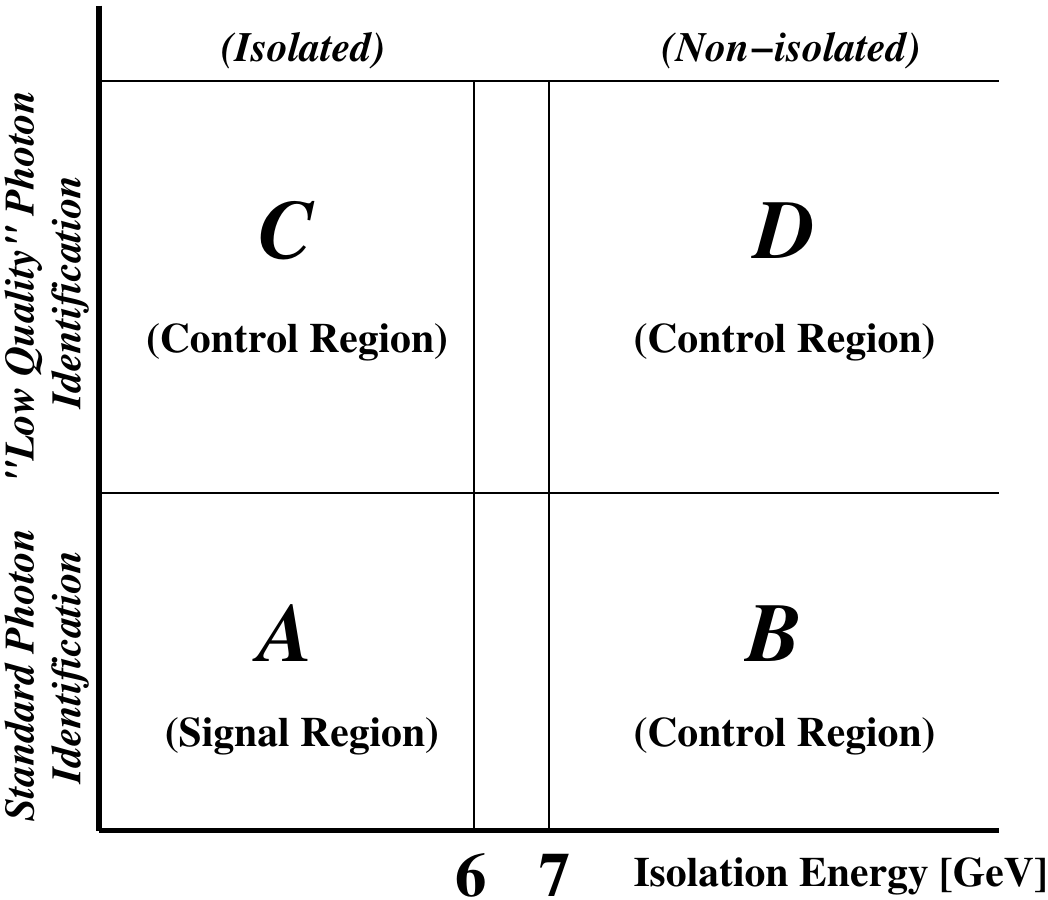}
  \caption{Sketch of the two-dimensional plane defining the four regions
    used in the sideband method.  Region A is the signal region. The
    nonisolated control regions (B and D) are defined for photons
    with $E^{\mathrm{iso}}_{\mathrm{T}}>7$ \GeV{}. The ``low quality photon
    identification'' control regions (C and D) include photon candidates
    that fail the full photon shower-shape selection criteria,
    but pass a subset of them. For the data driven $W+$jets background estimation to the inclusive $W\gamma$ measurement, about 1000 $W+$jets candidates are selected in the nonisolated control regions, and about 2000 $W+$jets candidates are selected in the ``low quality photon identification'' control regions.}
  \label{fig:twoD_phojet}
\end{figure}

{\em $W+$jets background:}
A two-dimensional sideband method is used for measuring the $W+$jets background
as in Refs.~\cite{photonpaper,WZg1fbpaper,35pbPhotonPaper,PhotonJetPaper} with the
two discriminating variables being the photon isolation and the photon identification
based on the shower shape (see Fig.~\ref{fig:twoD_phojet}).
The nonsignal regions are corrected for any contamination by signal events.
A quantity $f_\gamma$ is defined as the ratio of photon candidates passing the photon isolation
criteria to the number of candidates failing the isolation requirement. 
The ratio $f_\gamma$ is measured in $W(\ell\nu)$ events with one ``low quality'' photon candidate,
which is defined  as one that fails the full photon shower-shape selection criteria,
but passes a subset of them (C/D).
Monte Carlo simulation is used to correct $f_\gamma$ for signal contamination in the ``low quality'' photon sample.
The estimated contribution from $W+$jets in the signal region is obtained by multiplying the measured
$f_\gamma$ by the number of events passing all $W\gamma$ selections, except the photon isolation requirement (region B).

The main contribution to the uncertainty in the $W+$jets background estimate comes from the
potential bias in the $E_{\mathrm{T}}^{\mathrm{iso}}$ shape for the fake photons in background-enriched samples
due to effects from the detector (e.g. measurement of shower shapes) and physics (e.g. simulation of the underlying event).
This uncertainty is found to be less than 15\%  using a MC $W$+jets sample,
by comparing the $E_{\mathrm{T}}^{\mathrm{iso}}$ shape between the ``low quality'' photon sample
and the ``high quality'' photon sample.
The difference is used to modify the ratio $f_{\gamma}$ and a new $W+$jets background contribution
in the signal region is estimated.
The difference between the nominal estimate and the new estimate is taken to be the systematic uncertainty.

To estimate the uncertainty related to the selection of the background-enriched samples,
two alternative selections, with tighter and looser background selection requirements
based on the shower shapes are used. For the tighter selection, more shower-shape
variables are required to fail the selection cuts than for the looser background-enriched samples.
The $W+$jets background estimates from the alternative background-enriched samples are consistent
with those obtained from the nominal sample, and the differences  (10\%--15\%) are assigned
as a systematic uncertainty.
The changes in the background estimates from varying the photon isolation requirements
by $\pm 1$~GeV for the sideband (2\%--4\%) are also assigned as a systematic uncertainty.

{\em $\gamma+$jets background:}
Similarly, the $\gamma+$jets background is estimated from data using the two-dimensional sideband method,
with lepton isolation (using the measured ratio $f_l$) and $E_{\mathrm{T}}^{\mathrm{miss}}$ as the
independent variables.
The ratio $f_l$ is measured in a control sample, which requires
the events to pass all the $W\gamma$ selection criteria, except the $E_{\mathrm{T}}^{\mathrm{miss}}$ requirement, which is inverted.
The potential bias in the $E_{\mathrm{T}}^{\mathrm{iso}}$ shape for the fake lepton in the low-$E_{\mathrm{T}}^{\mathrm{miss}}$
background-enriched samples is found to be 10\%--15\% based on MC simulations.
By varying the $E_{\mathrm{T}}^{\mathrm{miss}}$ threshold, alternative control samples are
obtained to evaluate the systematic uncertainties on $f_l$.
In addition, the impact parameter requirements for the muon-candidate tracks
and the shower-shape selection criteria for electron candidates are also varied to obtain alternative control samples enriched in $\gamma+$jets events.
The differences between the $\gamma+$jets estimates (about 9\%) from those control samples
give one of the main systematic uncertainties. The change in the $\gamma+$jets estimates
from varying the lepton isolation requirements (about 4\%) is also assigned as a systematic uncertainty.

In the measurement of the differential fiducial cross section as a function of $E_{\mathrm{T}}^{\gamma}$,
the sideband method is used to estimate the $W/\gamma+$jets backgrounds in each
$E_{\mathrm{T}}^{\gamma}$ bin for the range $15<E_{\mathrm{T}}^{\gamma}<60$~GeV.
Extrapolation methods are used to estimate the $W/\gamma+$jets background
in the $E_{\mathrm{T}}^{\gamma}>60$ GeV region, where few events are available.
The statistical uncertainty on the background estimates become comparable to, or larger than, 
the systematic uncertainty at $E_{\mathrm{T}}^{\gamma}>40$ GeV.
The extrapolation from the low to the high $E_{\mathrm{T}}^{\gamma}$ regions is done
using the $E_{\mathrm{T}}^{\gamma}$ distribution shape
obtained from control samples []$W(\ell\nu)$ events with one ``low quality'' photon candidate
to estimate the $W+$jets background and $W(\ell\nu)$ events with a nonisolated lepton
to estimate the $\gamma+$jets background].
The difference between results (15\%--30\%) obtained from the sideband method and extrapolation methods
is treated as an additional uncertainty for the high-$E_{\mathrm{T}}^{\gamma}$ bins.

To measure the differential fiducial cross sections as a function of jet multiplicity
and the transverse mass of the $W\gamma$ system, the distributions of these kinematic variables for the $W/\gamma+$jets
backgrounds are taken from the control samples described in the previous paragraph.
The $W/\gamma+$jets distributions are then normalized to the predicted contributions
to the measurements.

{\em $Z(\ell^+ \ell^-)$ background:}
To understand background contributions from the $Z(\ell^+ \ell^-)$ process, MC simulation is needed to study the possibility of losing one lepton from $Z$ decay due to acceptance. Furthermore, two control regions are built to study the $E_{\mathrm{T}}^{\mathrm{miss}}$ modeling in $Z+\gamma$ and $Z+$jets events. The events in the $Z+\gamma$ control regions are selected by imposing the nominal $\ell^+ \ell^-\gamma$ event selection criteria, and the events in the $Z(e^+e^-)+$jets control regions are selected by imposing the nominal $e\nu\gamma$ selection criteria, except requiring that $m_{e\gamma}$ be within 15~\GeV{} of the $Z$ boson mass, assuming one of the electrons is misidentified as a photon.
A good agreement between the data and the MC simulation for the $E_{\mathrm{T}}^{\mathrm{miss}}$ distributions is found in these two $Z$ control regions, both in events with low pileup and in events with high pileup. 
Therefore their contributions are estimated from MC simulations. The uncertainties in $E_{\mathrm{T}}^{\mathrm{miss}}$ modeling in the $Z(\ell^+ \ell^-)$ process are studied by varying the energy scale and resolution of the leptons, photons, jets, and unassociated energy clusters\footnote{Unassociated energy clusters in the calorimeter are the energy deposits that are not matched to any reconstructed high-$p_{\mathrm{T}}$ object (jet, electron, muon, and photon).} in the calorimeter. 

{\em Other backgrounds:}
The background contributions from $t\bar{t}$, $WW$, single top quark, $Z(\tau^{+}\tau^{-})$, and $W(\tau \nu)$ processes are estimated from MC simulations. The systematic uncertainties arise mainly from theoretical uncertainties on the production cross sections of these background processes and uncertainties on the lepton, photon, jet, and $E_{\mathrm{T}}^{\mathrm{miss}}$ modeling in the simulation.

\begin{figure*}
  \centering
  \subfigure[]{\includegraphics[trim=30 20 60 5,clip=true,width=0.99\columnwidth]{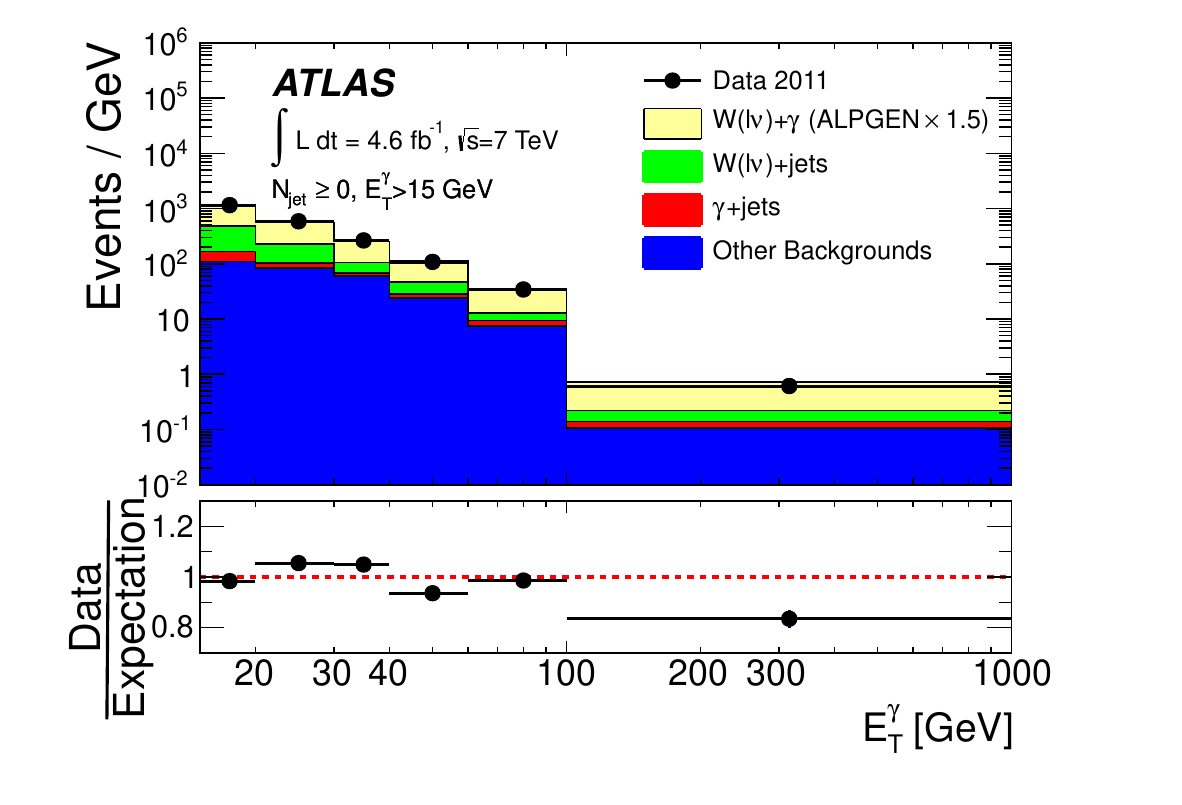}}
  \subfigure[]{\includegraphics[trim=30 20 60 5,clip=true,width=0.99\columnwidth]{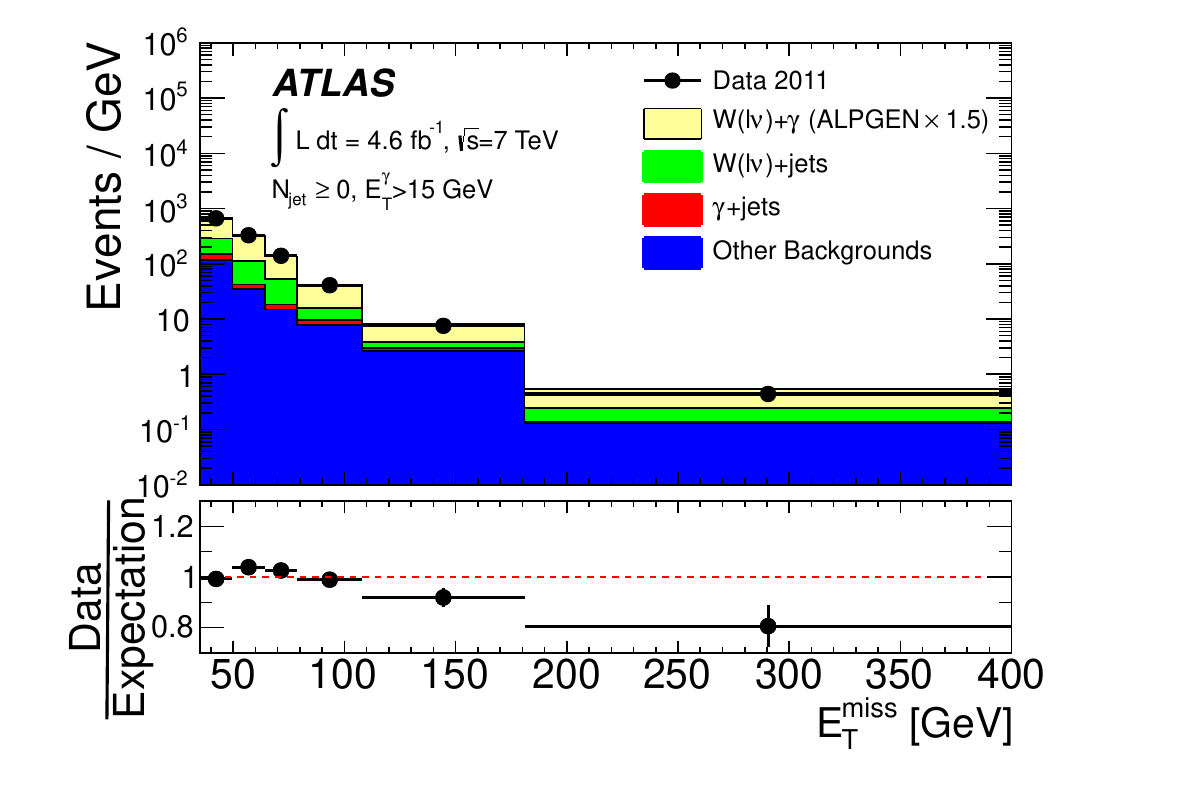}}
  \subfigure[]{\includegraphics[trim=30 20 60 5,clip=true,width=0.99\columnwidth]{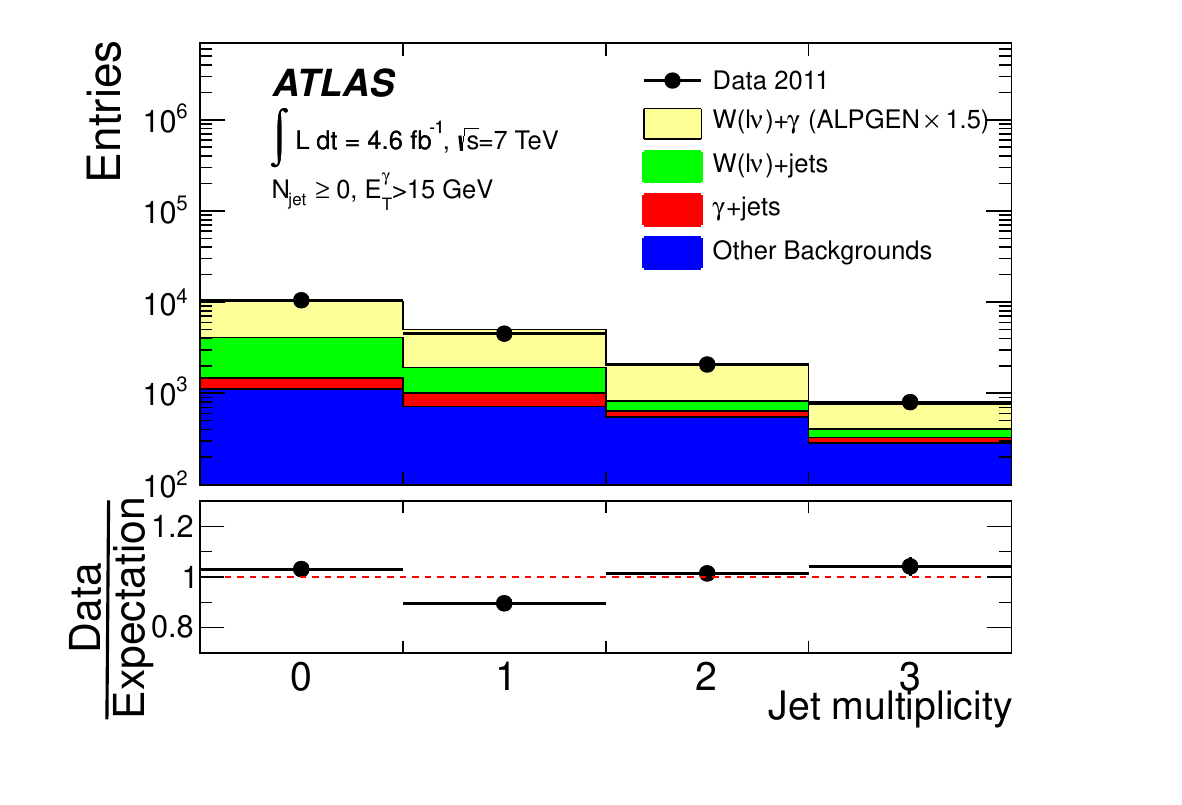}}
  \subfigure[]{\includegraphics[trim=30 20 60 5,clip=true,width=0.99\columnwidth]{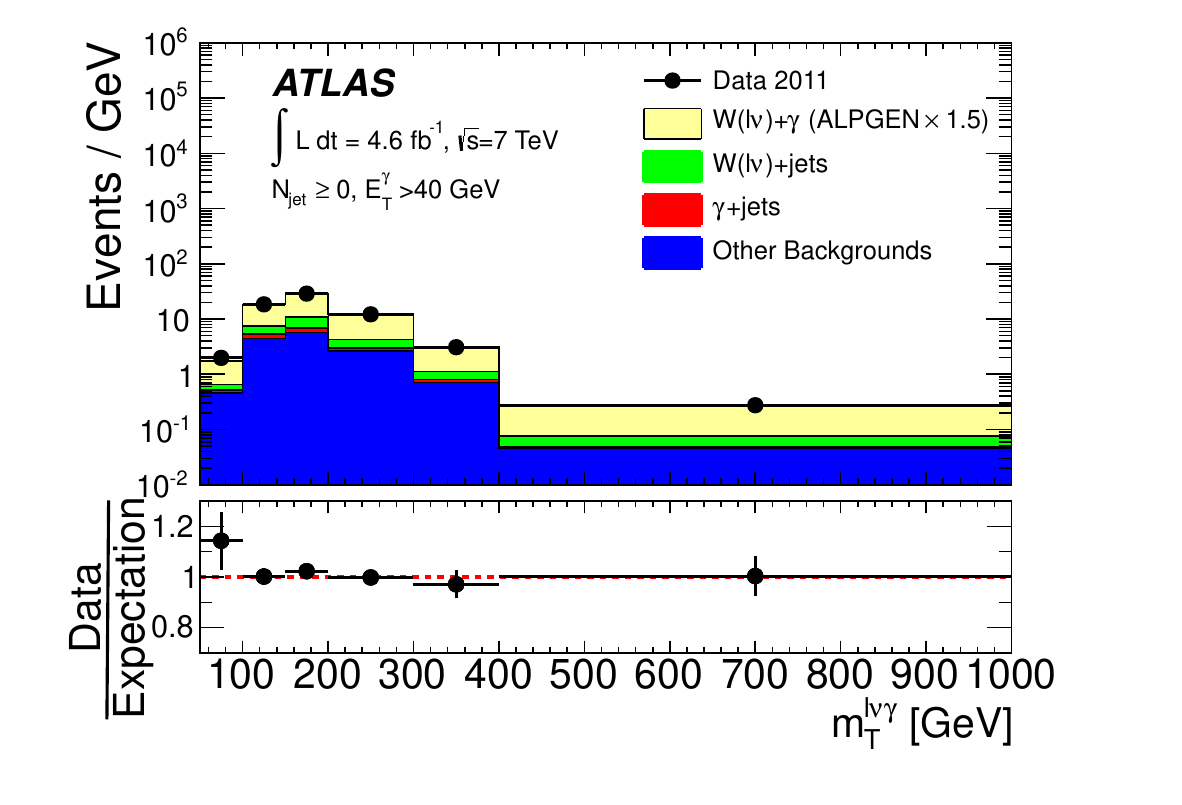}}

  \caption{Combined distributions for $\ell\nu\gamma$ candidate events in the electron and muon channels of (a) the photon transverse energy, (b) the missing transverse energy, (c) the jet multiplicity, and (d) the three-body transverse mass distribution as defined in Eq.~(\ref{equ:MT3}). The selection criteria are defined in Sec.~\ref{sec:Event_Selection}, in particular, the photon transverse energy is required to be $E_{\mathrm{T}}^{\gamma}>15$~\GeV{}, except for panel (d) where it is required to be $E_{\mathrm{T}}^{\gamma}>40$~\GeV{}. The distributions for the expected signals are taken from the ALPGEN MC simulation and scaled by a global factor ($\sim 1.5$) such that the total contribution from the predicted signal and background is precisely normalized to the data. The ratio of the number of candidates observed in the data to the number of expected candidates from signal and background processes is also shown. Only the statistical uncertainties on the data are shown for these ratios. As the expected signal is normalized to match the extracted number of signal events, the ratio provides a comparison only between the observed and predicted shapes of the distributions. The histograms are normalized by their bin width.}

  \label{fig:Wg_kin1}
\end{figure*}

A summary of background contributions and signal yields in the $W\gamma$ analysis is given in Table~\ref{tab:wgbg}. The estimated $W+$jets background is significantly smaller in the electron channel than in the
muon channel due to the $Z$--veto requirement in the electron channel, described in Sec.~\ref{sec:Event_Selection}. 
The distributions of the photon transverse energy, \met{}, jet multiplicity, and three-body transverse mass [see Eq.~(\ref{equ:MT3})] from the selected $W\gamma$ events are shown in Fig.~\ref{fig:Wg_kin1}. The data are compared to the sum of the backgrounds and the SM signal predictions. The distributions for the expected $W\gamma$ signal are taken from signal MC simulation and normalized to the total extracted number of signal events shown in Table~\ref{tab:wgbg} ($N^{\mathrm{sig}}_{W\gamma}$).

\begin{table*}[htbp]

 \begin{tabular}{ccccccccc}

       \hline
       & $ e\nu \gamma$  & $\mu\nu \gamma$ & $ e\nu \gamma$  & $ \mu\nu \gamma$\\
       &\multicolumn{2}{c}{$N_{\mathrm{jet}}\geq0$ }&\multicolumn{2}{c}{$N_{\mathrm{jet}}=0$} \\
       \hline

       $N^{\mathrm{obs}}_{W\gamma}$      &    7399                  &   10914                  & 4449                   &  6578    \\ \hline
        $W(\ell \nu)$+jets                         & $1240 \pm 160 \pm 210  $ & $2560 \pm 270 \pm 580  $ & $910 \pm 160 \pm 160 $  & $1690 \pm 210 \pm 270$   \\  
        $Z(\ell^+ \ell^-) + X$                  & $678 \pm 18 \pm 86$      & $779 \pm 19 \pm 93$      & $411 \pm 13 \pm 51$     & $577 \pm 16 \pm 73$    \\  
        $\gamma$+jets                    & $625 \pm 80 \pm 86  $    & $184 \pm 9 \pm 15 $      & $267 \pm 79 \pm 54 $    & $87 \pm 7 \pm 14 $    \\  
         $t\bar t$                       & $320 \pm 8 \pm 28$       & $653 \pm 11 \pm 57$      & $ 22 \pm 2\pm 4 $       & $44 \pm 3\pm 6$    \\  
        other background                 & $141 \pm 16 \pm 13$      & $291 \pm 29 \pm 26 $     & $ 52 \pm 5 \pm 6 $      & $140 \pm 22 \pm 18$    \\ 
\hline
      $N^{\mathrm{sig}}_{W\gamma}$       & $4390 \pm 200 \pm 250 $  & $6440 \pm 300 \pm 590 $  & $2780 \pm 190 \pm 180$  & $4040 \pm 230 \pm 280 $\\  

\hline

\end{tabular}

\caption{
Total number of events passing the selection requirements in the data ($N^{\mathrm{obs}}_{W\gamma}$), expected number of background events and observed number of signal events ($N^{\mathrm{sig}}_{W\gamma}$) in the $e\nu \gamma$ and the $\mu\nu \gamma$ channels for inclusive ($N_{\mathrm{jet}} \geq 0$) and exclusive ($N_{\mathrm{jet}} = 0$) events. 
$N^{\mathrm{sig}}_{W\gamma}$ is defined as the difference between $N^{\mathrm{obs}}_{W\gamma}$ and the total number of expected background events.
The first uncertainty is statistical and the second uncertainty represents an estimate of the systematic effects. 
The ``other background'' includes contributions from $WW$, single top quark, $W(\tau \nu)$, and $Z(\tau^{+}\tau^{-})$ production.}
  \label{tab:wgbg} 
\end{table*}

\begin{table*}[htbp]
 \begin{tabular}{ccccccccc}
       \hline

      \textbf{} & $e^{+}e^{-}\gamma$  & $ \mu^{+}\mu^{-}\gamma$ &
                                    $ e^{+}e^{-}\gamma$  & $\mu^{+}\mu^{-}\gamma$ \\
           &\multicolumn{2}{c}{$N_{\mathrm{jet}}\geq0$ }&\multicolumn{2}{c}{$N_{\mathrm{jet}}=0$} \\
       \hline
       $N^{\mathrm{obs}}_{Z\gamma}$  & 1908    &  2756  & 1417  &  2032   \\
       \hline
        $N_{Z\gamma}^{\mathrm{BG}}$    & $311 \pm 57 \pm 68$ &  $ 366 \pm 83 \pm 73  $  & $ 156\pm 43 \pm 32 $  &  $ 244 \pm 41 \pm 49$    \\
       \hline
       $N^{\mathrm{sig}}_{Z\gamma}$     &  $1600 \pm 71\pm 68$ &  $ 2390 \pm 97\pm 73$  &  $ 1260 \pm 56 \pm 32$    &  $  1790 \pm 59 \pm 49$   \\
       \hline

\end{tabular} 
\caption{Total number of events passing the selection requirements in the data ($N^{\mathrm{obs}}_{Z\gamma}$), expected number of background events ($N^{\mathrm{BG}}_{Z\gamma}$), and observed number of signal events ($N^{\mathrm{sig}}_{Z\gamma}$) in the $e^{+}e^{-} \gamma$ channel and the $\mu^+\mu^- \gamma$ channel with inclusive ($N_{\mathrm{jet}} \geq 0$) and exclusive ($N_{\mathrm{jet}} = 0$) selections.
$N^{\mathrm{sig}}_{Z\gamma}$ is defined as the difference between $N^{\mathrm{obs}}_{Z\gamma}$ and the total number of expected background events.
The first uncertainty is statistical and the
second uncertainty represents an estimate of the systematic effects.
}
  \label{tab:zgbg}
\end{table*}

\begin{table}[htbp]
\centering
\small
 \begin{tabular}{ccc}
       \hline
 & $ \nu\bar{\nu} \gamma$ & $ \nu\bar{\nu} \gamma$  \\
          &$N_{\mathrm{jet}}\geq0 $ & $ N_{\mathrm{jet}}=0 $  \\
            \hline

             $N^{\mathrm{obs}}_{Z\gamma}$            & 1094   & 662 \\ 
             \hline
            $W(e \nu)$ & $171 \pm 2 \pm 17$   & $132 \pm 2\pm 13$ \\ 
	    $Z(\nu\bar{\nu})+$jets, multi-jet   & $70\pm 13\pm 14 $ & $29\pm 5 \pm 3$ \\
	    $W\gamma $   & $238 \pm 12 \pm 37$ & $104 \pm 9 \pm 24$ \\
            $\gamma+$jets & $168 \pm 20\pm 42$   & $26 \pm 7 \pm 11$ \\ 
            $Z(\tau^+\tau^-)\gamma$ &  $11.7 \pm 0.7\pm 0.9$  & $6.5 \pm 0.6\pm 0.6$ \\
            $t \bar t$ & $11\pm 1.2 \pm 1.0$ & $0.9\pm 0.6\pm 0.1$  \\
       \hline
       $N^{\mathrm{sig}}_{Z\gamma}$     &  $420 \pm 42\pm 60 $ &  $360 \pm 29\pm 30 $   \\
            \hline
\end{tabular}

\caption{Total number of events in the data ($N^{\mathrm{obs}}_{Z\gamma}$), expected number of background events from various SM processes, and
observed signal yields ($N^{\mathrm{sig}}_{Z\gamma}$) after all $\nu\bar{\nu}\gamma$ selection criteria are applied for inclusive ($N_{\mathrm{jet}} \geq 0$) and exclusive ($N_{\mathrm{jet}} = 0$) events.
$N^{\mathrm{sig}}_{Z\gamma}$ is defined as the difference between $N^{\mathrm{obs}}_{Z\gamma}$ and the total number of expected background events. The first uncertainty is statistical and the second uncertainty represents an estimate of the systematic effects. }
  \label{tab:zvvbg} 
\end{table}

\begin{figure*}
  \centering
  \subfigure[]{\includegraphics[trim=30 20 60 5,clip=true,width=0.99\columnwidth]{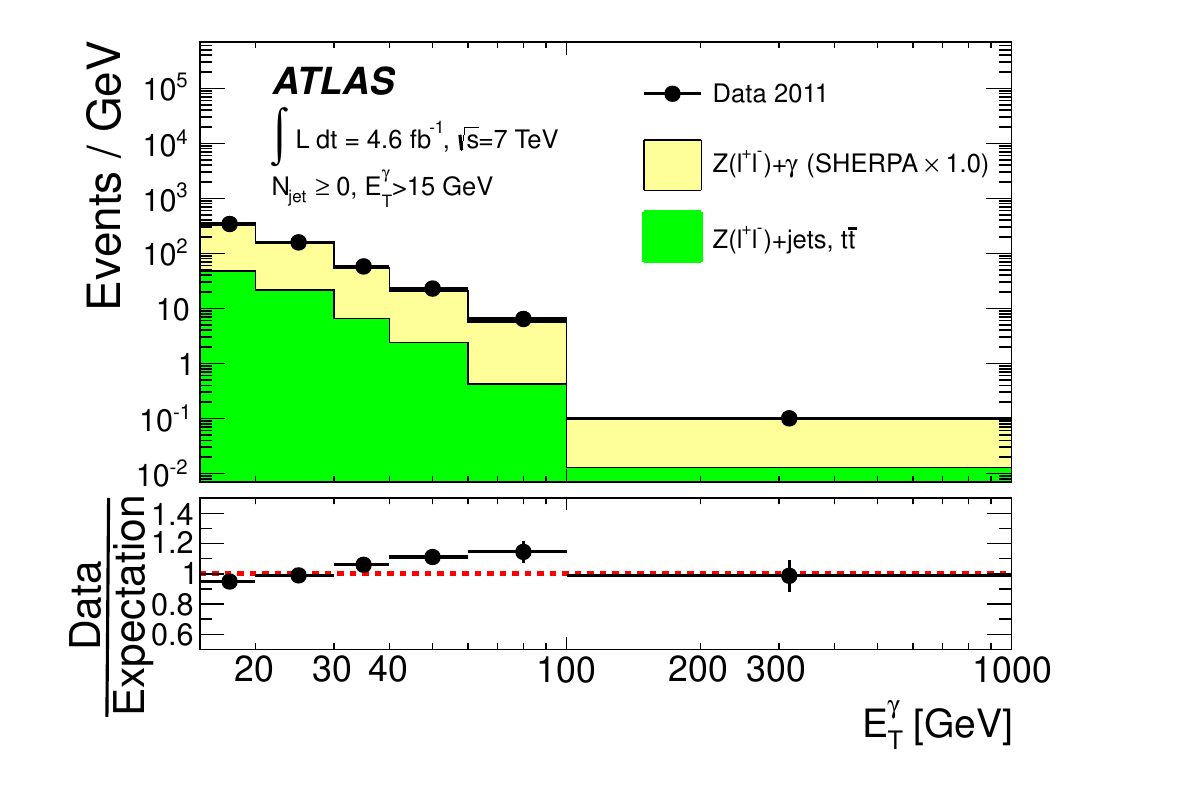}}
  \subfigure[]{\includegraphics[trim=30 20 60 5,clip=true,width=0.99\columnwidth]{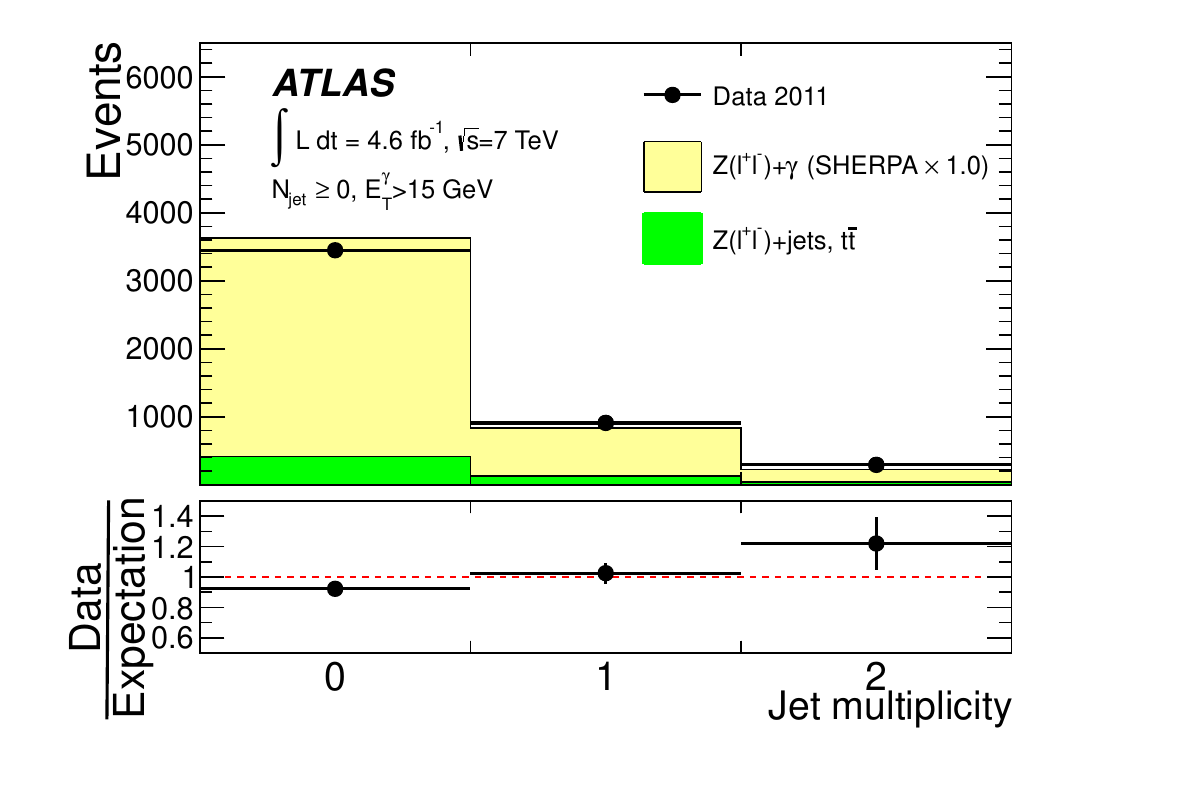}}
  \subfigure[]{\includegraphics[trim=30 20 60 5,clip=true,width=0.99\columnwidth]{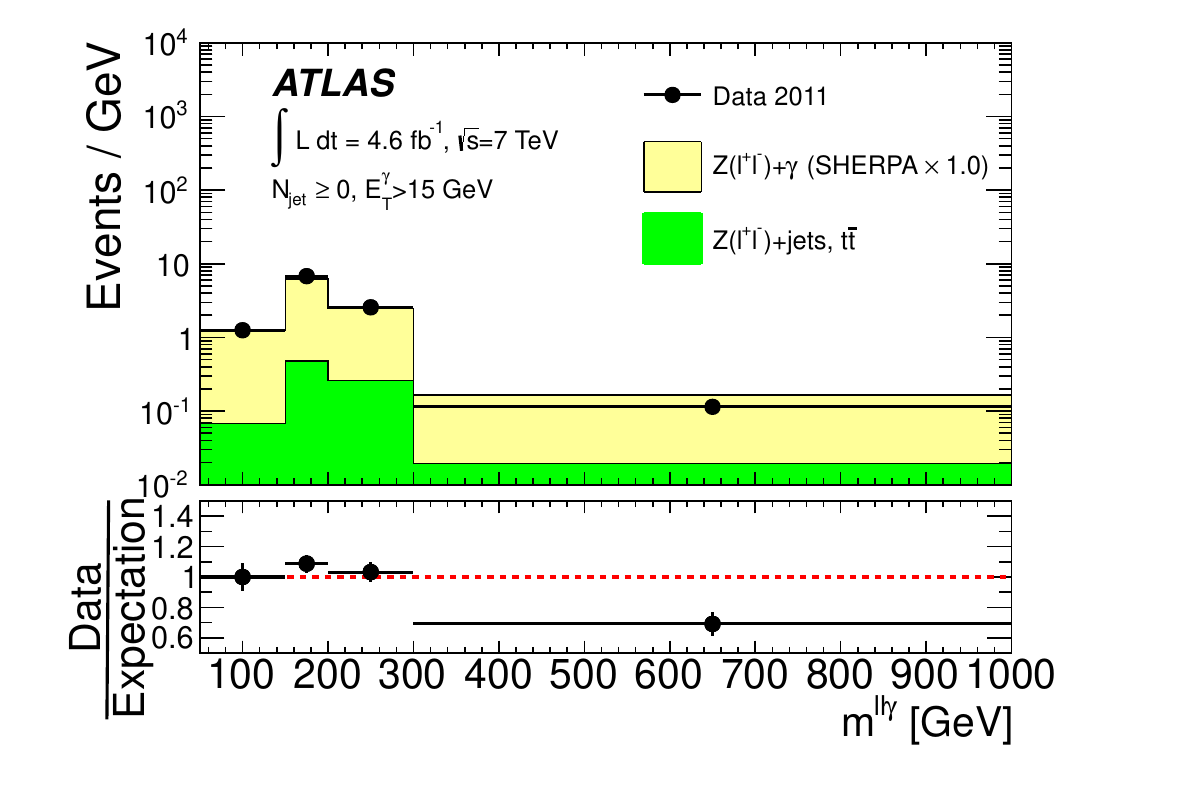}}

  \caption{Distribution for $\ell^{+}\ell^{-}\gamma$ candidate events combining the electron and muon channels of (a) the photon transverse energy, (b) the jet multiplicity, and (c) the three-body mass distribution. The selection criteria are defined in Sec.~\ref{sec:Event_Selection}, in particular, the photon transverse energy is required to be $E_{\mathrm{T}}^{\gamma}>15$~\GeV{}, except for panel (c) where it is required to be $E_{\mathrm{T}}^{\gamma}>40$~\GeV{}. The distributions for the expected signals are taken from the SHERPA MC simulation and scaled by a global factor ($\sim 1.0$) such that the total contribution from the predicted signal and background is precisely normalized to the data. The ratio of the number of candidates observed in the data to the number of expected candidates from signal and background processes is also shown. Only the statistical uncertainties on the data are shown for these ratios. The histograms are normalized by their bin width.}
  \label{fig:Zg_kin1}
\end{figure*}

\begin{figure*}
  \centering
  \subfigure[]{\includegraphics[trim=30 30 60 5,clip=true,width=0.99\columnwidth]{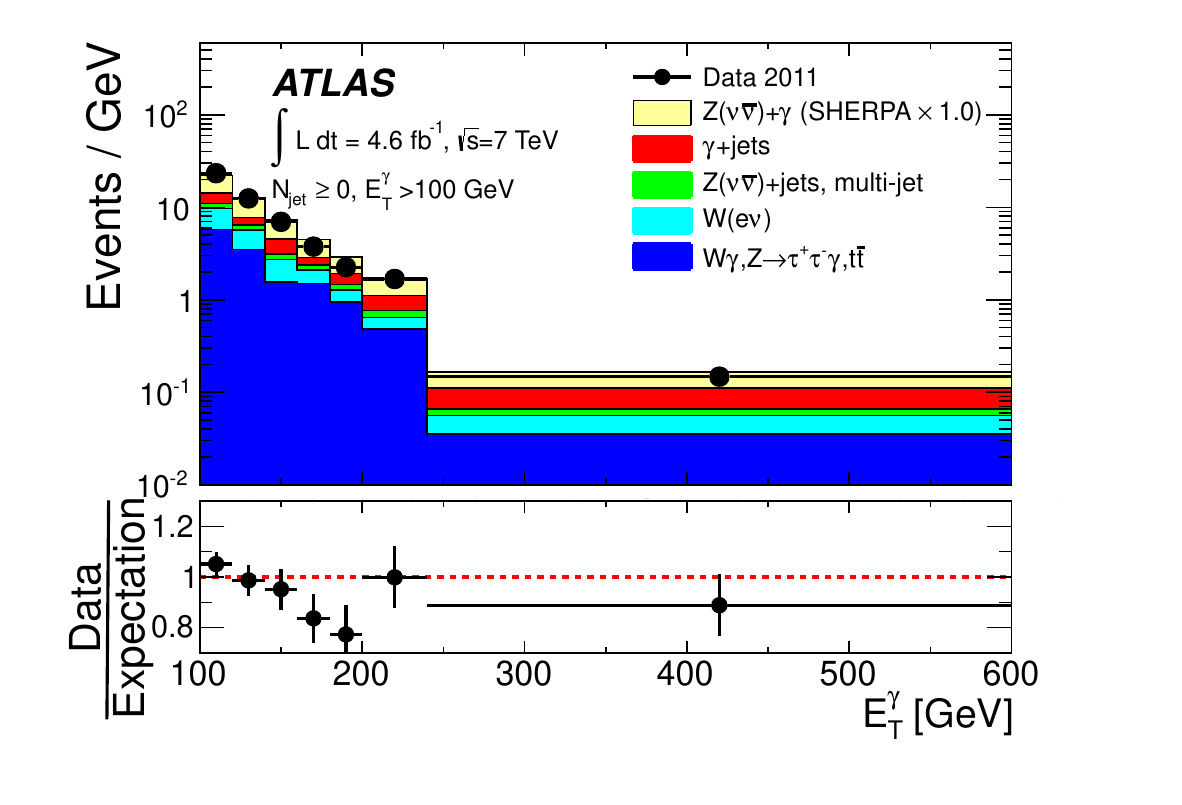}}
   \subfigure[]{\includegraphics[trim=30 30 60 5,clip=true,width=0.99\columnwidth]{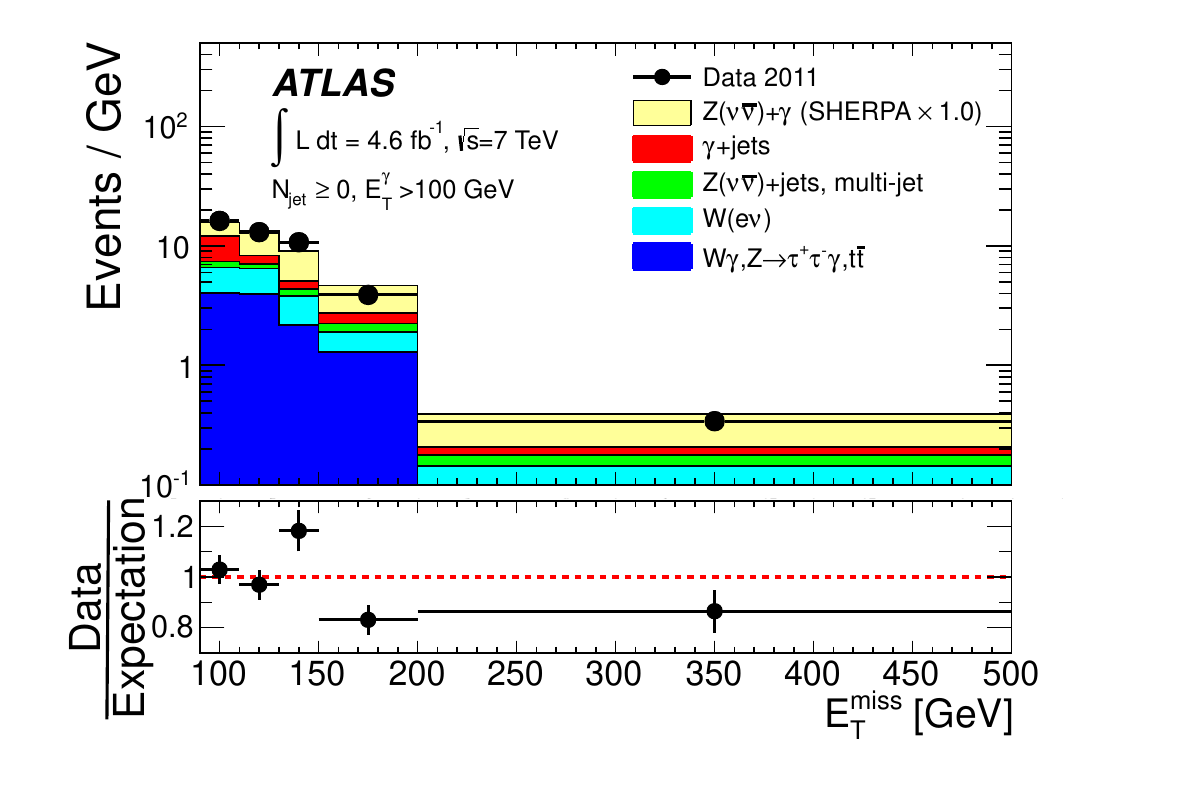}}
  \subfigure[]{\includegraphics[trim=30 30 60 5,clip=true,width=0.99\columnwidth]{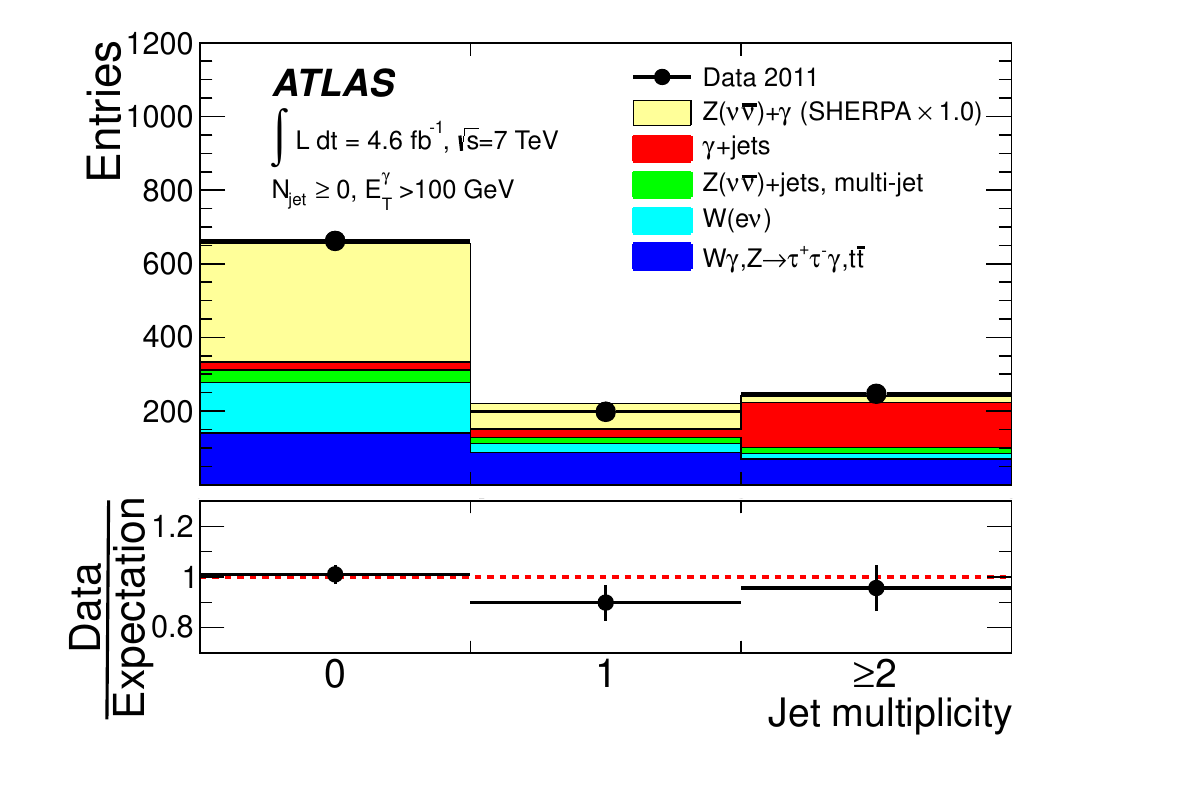}}

  \caption{Distributions of inclusive $\nu\bar{\nu}\gamma$ candidate events of (a) the photon transverse energy, (b) the missing transverse energy $E_{\mathrm{T}}^{\mathrm{miss}}$, and (c) the jet multiplicity. The selection criteria are defined in Sec.~\ref{sec:Event_Selection}. The distributions for the expected signals are taken from the SHERPA MC simulation and scaled by a global factor ($\sim 1.0$) such that the total contribution from the predicted signal and background is precisely normalized to the data. The ratio of the number of candidates observed in the data to the number of expected candidates from signal and background processes is also shown. Only the statistical uncertainties on the data are shown for these ratios. The histograms are normalized by their bin width.}

  \label{fig:Wg_kin2}
\end{figure*}

\subsection{Background estimation for $pp \rightarrow \ell^+ \ell^- \gamma$} 
The main background to the $\ell^+ \ell^- \gamma$ signal (amounting to $98\%$--$99\%$ of the total background)
originates from events with $Z+$jets where jets are misidentified as photons.
The $Z+$jets contamination is estimated from data using a sideband method similar to the one described
in Sec.~\ref{sec:bglnugamma}.
The main uncertainty (20\%) is due to the bias in the $E_{\mathrm{T}}^{\mathrm{iso}}$ shape for the fake photons in
background-enriched control samples defined by the ``low quality'' selection criteria.
The small contribution from $t\bar{t}+X$ production (mainly from $t\bar{t}+\gamma$) is estimated from MC simulation.
A summary of background contributions and signal yields in the $\ell^+ \ell^- \gamma$ analyses is given in Table~\ref{tab:zgbg}. The distributions of the photon transverse energy, jet multiplicity, and three-body mass from the selected $Z\gamma$ events are shown in Fig.~\ref{fig:Zg_kin1}.
The data and simulation agree within the uncertainty of the background estimate.

\subsection{Background estimation for $pp \rightarrow \nu \bar{\nu} \gamma$}
Background to the $\nu \bar{\nu} \gamma$ signal originates mainly from the following processes: 
\begin{enumerate}[i]
  \item $W(e \nu)$ events, when the electron is misidentified as a photon; 

  \item $Z(\nu\bar{\nu})+$jets and multijet events, when one of the jets in the event is misidentified as a photon; 

  \item $\tau \nu \gamma$ and $\ell \nu \gamma$ events from $W\gamma$ production,
        when the $\tau$ decays into hadrons or when the electron or muon from
        $\tau$ or $W$ decay is not reconstructed; 

  \item $\gamma+$jets events, when large apparent $E_{\mathrm{T}}^{\mathrm{miss}}$ is created by a combination of
        real $E_{\mathrm{T}}^{\mathrm{miss}}$ from neutrinos in heavy quark decays and mismeasured jet energy. 
\end{enumerate}

{\em $W(e \nu)$ background:}
To estimate the background contribution from $W(e \nu)$, the following dedicated studies are performed to determine the probability for an electron to be identified as a photon in the final state. A sample of $Z \to e^+e^-$ event candidates, with one of the $e$ replaced by a photon, taken from data is used to estimate the fraction of electrons from the $Z$ boson decay that are reconstructed as photons. The events are selected if the reconstructed invariant mass of the photon and the electron is close to the $Z$ mass. This fraction ($f_{e \to \gamma}$) increases from 2\% to 6\% as $|\eta|$ increases. These fake rates are used to determine the $W(e \nu)$ background in the signal region, by weighting the electron candidates in the control region with the misidentification rate corresponding to their $|\eta|$. The events in the $W(e \nu)$ control region are selected by nominal $\nu \bar{\nu} \gamma$ selection criteria, except an electron is used instead of a photon in the final state. The data-driven estimates of the $W(e \nu)$ background are limited mainly by the accuracy of the measurement of the misidentification rate. The combined statistical and systematic uncertainties of the determination of $f_{e \to \gamma}$ are used to evaluate the systematic uncertainties of the $W(e \nu)$ background estimate. 

{\em $Z(\nu\bar{\nu})+$jets and multijets backgrounds:}
A data-driven method similar to the one described in Sec.~\ref{sec:bglnugamma} is used to
determine the background contribution from $Z(\nu\bar{\nu})+$jets and multijets events.
The main systematic uncertainty (20\%) comes from the differences between $f_\gamma$ values measured
in various control samples obtained by varying the selection criteria for ``low quality'' photons. 

{\em $W\gamma$ background:}
Misidentified events from the $W\gamma$ process are one of the dominant background contributions to the $\nu \bar{\nu} \gamma$ signal.
A large fraction (about 65\%) of the $W\gamma$ contamination comes from $\tau \nu \gamma$ events.
The branching fractions of the $\tau$ decay modes are well known and modeled by MC simulation.
The main uncertainty on the $\tau \nu \gamma$ contamination is due to the uncertainty on the MC normalization factor.
By assuming lepton universality for the $W$ boson decays, the MC scale factor for $\tau \nu \gamma$
events and its uncertainty are taken from the measurement of $\ell \nu \gamma$ events.
The scale factor is defined to correct the yield of $\ell \nu \gamma$ events estimated by MC simulation
to match the $\ell \nu \gamma$ event yield measured in data as shown in Table~\ref{tab:wgbg}.
About 35\% of $W\gamma$ contamination comes from $\ell \nu \gamma$ events.
Most of the $\ell \nu \gamma$ contamination consists of events with a low-$E_\mathrm{T}$ lepton below 25~\GeV{} (70\%) or with a high-$E_\mathrm{T}$ central lepton that failed to pass the identification or isolation criteria (20\%). Less than 5\% of $\ell \nu \gamma$ contamination comes from events with a forward lepton outside the detector's fiducial volume.

{\em $\gamma+$jets background:}
Due to the high-$E_{\mathrm{T}}^{\mathrm{miss}}$ requirement in $\nu \bar{\nu} \gamma$ event selection, $\gamma+$jets contamination is suppressed, especially in the exclusive measurement with a jet veto cut. 
In order to measure this background from data, a sample is selected by applying all signal-region selection criteria except for requiring $\Delta\phi(E_{\mathrm{T}}^{\mathrm{miss}}$, jet)$<0.4$. By requiring the $E_{\mathrm{T}}^{\mathrm{miss}}$ direction to be close to the jet direction, the selected events in the control region are dominated by $\gamma+$jets background. The yield of $\gamma+$jets obtained in control regions is then scaled by an extrapolation factor to predict the $\gamma+$jets background yield in the signal region, where the extrapolation factor is taken from a $\gamma+$jets MC sample.
By varying the $E_{\mathrm{T}}^{\mathrm{miss}}$ threshold from 60 \GeV{} to 100 \GeV{} and varying
the jet multiplicity requirement for the events from
$N_{jet}\geq0$ to $N_{jet}\geq1$,
alternative control samples are obtained to evaluate the systematic uncertainties.
The main systematic uncertainty in the $\gamma+$jets estimate comes from the different background yields
in different control regions.
The systematic uncertainty on the extrapolation factor is obtained by comparing the predictions
from {\sc sherpa} and {\sc pythia} $\gamma+$jets MC samples and varying the energy scale and resolution for jets and $E_{\mathrm{T}}^{\mathrm{miss}}$ in MC samples.

{\em Other backgrounds:}
Background contributions from other processes are determined from MC samples. The contributions from $Z (\tau^+\tau^-)\gamma$, and $t\bar{t}$ are found to be small (about $1\%$ of the total background). The contributions from the other processes such as $Z (\ell^+ \ell^-)\gamma$, $\gamma\gamma$, and diboson production, are found to be negligible due to the strict cuts applied to the $\met$ and the photon transverse energy.

To investigate the possibility of non-collision backgrounds,
the distributions of the direction of flight
as well as quality criteria (e.g. shower shapes) of the photon candidates in data
are compared to those expected from the signal simulation to search for discrepancies.
The direction of flight, which is determined by using the depth segmentation
of the EM calorimeter, can show if the photon appears to 
be coming from a vertex other than the primary vertex.
The spectra of the direction of flight as well as the quality criteria are found to be
completely consistent with those photons produced in events with real photons
[e.g. $W(\ell \nu)+\gamma$ and $Z(\ell^{+} \ell^{-})+\gamma$] leading to the conclusion that
if there are noncollision background events, they are negligible.

A summary of background contributions and signal yields in the $\nu \bar{\nu} \gamma$ analysis is given in Table~\ref{tab:zvvbg}. The photon transverse energy, the jet multiplicity and the missing transverse energy distributions from the selected $ \nu\bar{\nu}\gamma$ events are shown in Fig.~\ref{fig:Wg_kin2}.

\section{Cross-Section Measurements}
\label{sec:cs}

The cross-section measurements for the $W\gamma$ and $Z\gamma$ processes are performed in the fiducial region, defined at particle level using the object and event kinematic selection criteria
described in Sec.~\ref{sec:Event_Selection}. They are then extrapolated to an extended
fiducial region (defined in Table~\ref{tab:fiducialcut}) common to the electron and muon final states.
In this analysis, particle level refers to stable particles, defined as having lifetimes exceeding 10 ps, that are produced from the hard scattering or after the hadronization but before their interaction with the detector.
The extrapolation corrects for the signal acceptance losses in the calorimeter
transition region ($1.37<|\eta|<1.52$) for electrons and photons, and in the high-$\eta$ region ($2.4<|\eta|<2.47$) for muons. It also corrects for the $Z$-veto requirement in the $W\gamma$ electron channel, for the transverse mass selection criteria in both channels in the $W\gamma$ analysis, and for the acceptance loss due to the selection requirements on $\Delta\phi(E_{\mathrm{T}}^{\mathrm{miss}},\gamma)$ and $\Delta\phi(E_{\mathrm{T}}^{\mathrm{miss}}$, jet) in the $\nu \nu \gamma$ analysis.
Jets at particle level are reconstructed in MC-generated events by applying
the anti-$k_t$ jet reconstruction algorithm with a radius parameter $R$ = 0.4
to all final-state stable particles. To account for the effect of final-state QED radiation,
the energy of the generated lepton at particle level is defined as the energy of the
lepton after radiation plus the energy of all radiated photons within a $\Delta R<0.1$ cone around the lepton direction.
Isolated photons with $\epsilon_h^p<0.5$~\cite{BaurLO,BaurNLO} are considered as signal, where $\epsilon_h^p$ is defined
at particle level as the sum of the energy carried by final state particles
in a $\Delta R<0.4$ cone around the photon direction (not including the photon) divided by the energy carried by the photon.

\begin{table}[!htbp]
  \centering 
  \begin{tabular}{lcccccc}
    \hline 
Cuts   &  $pp \rightarrow \ell \nu\gamma$  & $pp \rightarrow \ell^+ \ell^- \gamma$ & $pp \rightarrow \nu\bar{\nu} \gamma$\\
\hline
Lepton  & $p_{\mathrm{T}}^{\ell} >25\GeV{}$ & $p_{\mathrm{T}}^{\ell} >25\GeV{}$ & ---\\
        & $|\eta_{\ell}|<2.47$  & $|\eta_{\ell}|<2.47$ & ---\\
        & $N_\ell =1$& $N_{\ell^+} =1 , N_{\ell^-}=1$ & $N_\ell =0$ \\
\hline
Neutrino   & $p_{\mathrm{T}}^\nu>35\ \GeV{}$    & ---  & --- \\
\hline
Boson &       ---     &   $m_{\ell^+ \ell^-}>40$ \GeV{}  &  $p_{\mathrm{T}}^{\nu\bar{\nu}}>90\ \GeV{}$\\
\hline
Photon  & $E^{\gamma}_{\mathrm{T}} > 15$ \GeV{}& $E^{\gamma}_{\mathrm{T}} > 15$ \GeV{}& $E^{\gamma}_{\mathrm{T}} > 100$ \GeV{}\\
        & \multicolumn{3}{c}{$|\eta^\gamma|<2.37$, $\Delta R(\ell,\gamma)>0.7$} \\
        & \multicolumn{3}{c}{  $\epsilon_h^p < 0.5$}\\
\hline

Jet & \multicolumn{3}{c}{  $E_{\mathrm{T}}^{\mathrm{jet}}>30$ \GeV{}, $|\eta^{\mathrm{jet}}|<4.4$ }\\
 & \multicolumn{3}{c}{$\Delta R(e/\mu/\gamma,\mathrm{jet})>0.3$ }\\
 & \multicolumn{3}{c}{ Inclusive : $N_{\mathrm{jet}}\geq 0$, Exclusive : $N_{\mathrm{jet}}=0$}\\

\hline

  \end{tabular} 
  \caption{Definition of the extended fiducial region where the cross sections are
  evaluated; $p_{\mathrm{T}}^\nu$ is the transverse momentum of the neutrino from $W$ decays; $p_{\mathrm{T}}^{\nu\bar{\nu}}$ is the transverse momentum of the $Z$ boson that decays into two neutrinos; $N_\ell$ is the number of leptons in one event; $\epsilon_h^p$ is the photon isolation fraction.}
\label{tab:fiducialcut}
\end{table}
\subsection{Integrated fiducial cross section}

The cross-section measurements for the processes $pp \to
\ell \nu\gamma+X$ and $pp \to (\ell^+ \ell^-\gamma/\nu\bar{\nu}\gamma)+X$ are calculated as

\begin{equation}
\sigma_{pp \rightarrow \ell \nu\gamma(\ell^{+} \ell^{-}\gamma/\nu\bar{\nu}\gamma)}^{\mathrm{ext-fid}} = \frac{N^{\mathrm{\mathrm{sig}}}_{V\gamma}} {A_{V \gamma}\cdot  C_{V \gamma} \cdot \int {\cal L} dt  .}
\label{Equ:cs_fid}
\end{equation}
where 
\begin{enumerate}[i]
\item $N_{W\gamma}^{\mathrm{\mathrm{sig}}}$ and $N_{Z\gamma}^{\mathrm{sig}}$ denote the number
  of background-subtracted signal events passing the selection
  criteria of the $W \gamma$ and $Z \gamma$ analyses.
  These numbers are listed in Tables~\ref{tab:wgbg}, ~\ref{tab:zgbg} and~\ref{tab:zvvbg}.
\item $\int {\cal L} dt$ is the integrated
  luminosity for the channels of interest (4.6 fb$^{-1}$).
\item $C_{V\gamma}$ is defined as the number of reconstructed MC events passing all selection requirements divided by the number of generated events at particle level within the fiducial region. These ratios, which are corrected with scale factors to account for small discrepancies between data and simulation, are shown in Table~\ref{tab:inputCA}.

\item $A_{V\gamma}$ are the acceptances, defined at particle level as the number of generated events found within the fiducial region divided by the number of generated events within the extended fiducial region. These acceptances are listed in Table~\ref{tab:inputCA}.
\end{enumerate}

The correction factors $C_{V\gamma}$ are determined by using $W/Z+\gamma$ signal MC events, corrected with scale factors to account for small discrepancies between data and simulation.
These discrepancies include the differences in the lepton and photon reconstruction,
identification, and isolation efficiencies, as well as trigger efficiencies.

Table~\ref{tab:Wg_xsecSys_excl} summarizes the systematic uncertainties on $C_{V\gamma}$ from different sources,
on the signal acceptance $A_{V\gamma}$, and on the background estimates.
The dominant uncertainties on $C_{V\gamma}$ come from photon identification and isolation efficiency.  The photon identification efficiency is determined from the signal MC samples where the shower-shape distributions of the photon are corrected to account for the observed small discrepancies between data and simulation. The systematic uncertainty is determined by comparing the corrected nominal value from MC simulation with the efficiency measurement using a pure photon sample from radiative $Z$ decays in data~\cite{ATLASPhotonIDConf}. The uncertainty on the photon identification efficiency is found to be about 6\% for all $V\gamma$ measurements. By doing a similar study, the uncertainty on the photon isolation efficiency is found to be less than 3\%.

The uncertainties coming from the jet energy scale (JES) and resolution (JER) are important for all exclusive $V\gamma$ measurements. Uncertainties associated with the JES and JER affect the efficiency of the jet veto criteria and have an impact on $E_{\mathrm{T}}^{\mathrm{miss}}$. By separately varying the JES and JER within one standard deviation and propagating them to the $E_{\mathrm{T}}^{\mathrm{miss}}$, the uncertainties on $C_{V\gamma}$ due to these effects are found to be less than 4\% for exclusive $\ell \nu \gamma$, and 3\% for exclusive $\ell^+ \ell^- \gamma$ and $\nu\bar{\nu} \gamma$ measurements.

The uncertainties on energy scale and resolution for unassociated energy clusters in the calorimeter and for additional $pp$ collisions are propagated to $E_{\mathrm{T}}^{\mathrm{miss}}$, with an impact on $C_{V\gamma}$ of less than 2\% for the $\ell \nu \gamma$ and $\nu\bar{\nu} \gamma$ measurements.

The muon momentum scale and resolution are studied by comparing the invariant mass distribution of $Z\to \mu^+\mu^-$ events in data and MC simulation~\cite{WZpaper}. The impact on $\ell \nu \gamma$ and $\ell^+ \ell^- \gamma$ signal events due to the muon momentum scale and resolution uncertainty is smaller than $1\%$. The uncertainties due to the EM energy scale and resolution, which affect both the electron and photon, are found to be 2\%--3\%.

The efficiencies of the lepton selections, and the lepton triggers, are first estimated from the signal MC events and then corrected with scale factors derived using high-purity lepton data samples from $W$ and $Z$ boson decays to account for small discrepancies between the data and the MC simulation~\cite{atlas_electron,WZpaper,ATLAS_Wjet,photonpaper}.
In the $\ell \nu \gamma$ and $\ell^+ \ell^- \gamma$ measurement, the uncertainty due to lepton identification and reconstruction is found to be about 2\% in the electron channel, and less than 1\% in the muon channel, and the uncertainty due to lepton isolation is found to be less than 2\% in the electron channel and less than 0.5\% in the muon channel. 

The uncertainty due to single-muon trigger efficiencies is 2\% for $\ell \nu \gamma$ and 0.6\% for $\ell^+ \ell^- \gamma$, while the uncertainty from single-electron trigger efficiencies is 0.7\% for $\ell \nu \gamma$ and $0.1$\% for $\ell^+ \ell^- \gamma$~\cite{ATLAS_leptrigger,ATLASMuonTrigger,ATLASElectronTrigger}.
The uncertainty from photon trigger efficiencies for $\nu\bar{\nu} \gamma$ is 1\%.

The systematic uncertainties for $A_{V\gamma}$ are dominated by PDF uncertainties ($<$0.8\%), by the renormalization and factorization scale uncertainties ($<$0.5\%) and by the uncertainties on the size of the contributions from fragmentation photons ($<$0.3\%). The PDF uncertainty is estimated using the CT10 error eigenvectors at their 90$\%$ confidence-level (C.L.) limits and rescaled appropriately to 68$\%$ C.L., with variations of $\alpha_s$ in the range 0.116--0.120. The renormalisation and factorisation scales are varied by factors of 2 around the nominal scales to evaluate the scale-related uncertainties.

\begin{table*}
  \centering 
  \begin{tabular}{cccccccccc}
    \hline 
    & $pp \rightarrow  e \nu\gamma$ $\;\;\;\;\;$& $pp \rightarrow  \mu \nu\gamma $ $\;\;\;\;\;$& $pp \rightarrow  e^+ e^-\gamma $ $\;\;\;\;\;$& $pp \rightarrow  \mu^+ \mu^-\gamma $ $\;\;\;\;\;$& $pp \rightarrow  \nu \bar{\nu} \gamma $   \\
\hline

   & \multicolumn{5}{c}{ $N_{\mathrm{jet}}\geq 0$}  \\
    
$C_{V\gamma}$ $\;\;\;\;\;$& $0.51 \pm 0.04$ $\;\;\;\;\;$& $0.58 \pm 0.04$ $\;\;\;\;\;$& $0.33\pm 0.02$ $\;\;\;\;\;$& $0.43\pm 0.03$ $\;\;\;\;\;$& $0.71 \pm 0.05$ \\
$A_{V\gamma}$ $\;\;\;\;\;$& $0.68 \pm 0.01$ $\;\;\;\;\;$& $0.86 \pm 0.01$ $\;\;\;\;\;$& $0.83 \pm 0.01 $ $\;\;\;\;\;$& $0.91 \pm 0.01$ $\;\;\;\;\;$& $0.97 \pm 0.01$ \\

\hline
   & \multicolumn{5}{c}{ $N_{\mathrm{jet}}=0$}  \\
$C_{V\gamma}$ $\;\;\;\;\;$& $0.46 \pm 0.04$ $\;\;\;\;\;$& $0.55 \pm 0.04$ $\;\;\;\;\;$& $0.31 \pm 0.02$ $\;\;\;\;\;$& $0.40 \pm 0.03$ $\;\;\;\;\;$& $0.69 \pm 0.05 $\\
$A_{V\gamma}$ $\;\;\;\;\;$& $0.73 \pm 0.01$ $\;\;\;\;\;$& $0.91 \pm 0.01$ $\;\;\;\;\;$& $0.83 \pm 0.01$ $\;\;\;\;\;$& $0.91 \pm 0.01$ $\;\;\;\;\;$& $0.98\pm 0.01$ \\

\hline
  \end{tabular} 
  \caption{Summary of correction factors $C_{W\gamma}$ ($C_{Z\gamma}$) and acceptance $A_{W\gamma}$ ($A_{Z\gamma}$) for the calculation of the $W\gamma$ ($Z\gamma$) production cross sections. The combined statistical and systematic uncertainties are also shown.}
  \label{tab:inputCA} 
\end{table*}

The cross-section measurements of each leptonic decay channel and the combined (electron, muon) channels
are extracted using a likelihood method.
A negative log-likelihood function is defined as
\begin{widetext}

\begin{equation}
\label{eq:loglike}
-{\rm ln}~L(\sigma, \bold{x}) = \sum_{i=1}^{n}-{\rm ln}\Bigg( \frac{e^{-(N_{s}^i(\sigma, \bold{x}) + N_{b}^i(\bold{x}))} \times (N_{s}^i(\sigma, \bold{x}) + N_{b}^i(\bold{x}))^{N_{\mathrm{obs}}^i}}{(N_{\mathrm{obs}}^i)!} \Bigg) + \frac{\bold{x} \cdot \bold{x}}{2}.
\end{equation}
\end{widetext}
The expression inside the natural logarithm in Eq.~(\ref{eq:loglike}) is the Poisson probability
of observing $N^{i}_{\mathrm{obs}}$ events in channel $i$ when $N^{i}_{s}$ signal and $N^{i}_{b}$
background events are expected.
The nuisance parameters $\bold{x}$, whose distribution is assumed to be Gaussian, affect $N^{i}_{s}$ and $N^{i}_{b}$ as
\begin{equation}
\label{eq:Nsx}
  N^{i}_{s}(\sigma, \bold{x}) = N^{i}_{s}(\sigma, 0)(1 + \sum_{k}x_{k}S^{i}_{k}),
\end{equation}
\begin{equation}
\label{eq:Nbx}
  N^{i}_{b}(\bold{x}) = N^{i}_{b}(0)(1 + \sum_{k}x_{k}B^{i}_{k}),
\end{equation}
where $S^{i}_{k}$ and $B^{i}_{k}$ are, respectively, the relative systematic uncertainties
on the signal and background due to the $k$th source of systematic uncertainty.
The quantity $n$ in Eq.~(\ref{eq:loglike}) is the number of channels to combine.
By varying the nuisance parameters $\bold{x}$, the negative log-likelihood in Eq.~(\ref{eq:loglike}) is minimized to obtain the most
probable value of the measured cross section.

For the combination, it is assumed that the uncertainties on the
lepton trigger and identification efficiencies are uncorrelated between
different leptonic decay channels.
All other uncertainties, such as the ones on the photon efficiency,
background estimation, and jet energy scale, are assumed to be fully correlated.
The measured production cross sections in the extended fiducial region defined in Table~\ref{tab:fiducialcut}
for the $\ell \nu \gamma$, $\ell^+ \ell^- \gamma$
and $\nu \bar{\nu} \gamma$ processes are summarized in Table~\ref{tab:cs}. 
These cross-section measurements are the most extensive made to date for the study of
V$+\gamma$ production at the LHC.

\begin{table*}[htbp]
\renewcommand{\arraystretch}{1.4}
\centering
\begin{tabular}{lccccc}

\hline
Source & $pp \rightarrow e\nu\gamma$  & $pp \rightarrow \mu\nu\gamma$  & $pp \rightarrow e^+ e^-\gamma$ 
& $pp \rightarrow \mu^+ \mu^-\gamma$  & $pp \rightarrow \nu\bar{\nu}\gamma$  \\

\hline

\multicolumn{6}{c}{Relative systematic uncertainties on the signal correction factor $C_{V\gamma}$ [\%]} \\

\hline 
$\gamma$ identification efficiency &  6.0 (6.0)  &  6.0 (6.0) &  6.0 (6.0) &  6.0 (6.0) &  5.3 (5.3) \\       

$\gamma$ isolation efficiency      &  1.9 (1.8) &  1.9 (1.7) &  1.4 (1.4) &  1.4 (1.4) &  2.8 (2.8) \\   

Jet energy scale                   &  0.4 (2.9) &  0.4 (3.2)  &   - (2.2) &   - (2.4) &  0.6 (2.0) \\

Jet energy resolution            &  0.4 (1.5) &  0.6 (1.7) &   - (1.7) &   - (1.8) &  0.1 (0.5) \\  

Unassociated energy cluster in $E_{\mathrm{T}}^{\mathrm{miss}}$  &  1.5 (1.6)  &  0.5 (1.0) &   - (-)   &   - (-)  &  0.3 (0.2) \\

$\mu$ momentum scale and resolution            &   - (-)  &  0.5 (0.4) &   - (-)  &  1.0 (0.8) &   - (-)  \\

EM scale and resolution                        &  2.3 (3.0)  &  1.3 (1.6) &  2.8 (2.8) &  1.5 (1.5)  &  2.6 (2.7)  \\

Lepton identification efficiency              &  1.5 (1.6) &  0.4 (0.4) &  2.9 (2.5) &  0.8 (0.8) &   - (-)  \\

Lepton isolation efficiency                &  0.8 (0.8) &  0.3 (0.2) &  2.0 (1.6) &  0.5 (0.4) &   - (-)  \\

Trigger efficiency               &  0.8 (0.1)  &  2.2 (2.1) &  0.1 (0.1) &  0.6 (0.6) &  1.0 (1.0) \\

\hline

Total                            &  7.1 (8.0) &  6.8 (7.8) &  7.6 (7.9) &  6.5 (7.1) &  6.6 (7.0) \\

\hline

\end{tabular}
\caption{Relative systematic uncertainties in $\%$ on the signal correction factor $C_{V\gamma}$
         for each channel in the inclusive $N_{\mathrm{jet}}>=0$ (exclusive $N_{\mathrm{jet}}=0$) $V\gamma$ measurement.}
\label{tab:Wg_xsecSys_excl}
\end{table*}

\begin{table*}
  \centering 
  \begin{tabular}{cccccccccc}
    \hline
   & \multicolumn{1}{c}{$\sigma^{\mathrm{ext-fid}}$[pb]} & \multicolumn{1}{c}{  $\sigma^{\mathrm{ext-fid}}$[pb]}  \\

   & \multicolumn{1}{c}{ Measurement } & \multicolumn{1}{c}{ {\sc mcfm} Prediction}  \\
    \hline
   & \multicolumn{2}{c}{ $N_{\mathrm{jet}} \geq 0$ }   \\
    \hline
$e\nu\gamma$    &  2.74  $\pm$ 0.05 (stat) $\pm$ 0.32 (syst) $\pm$ 0.14 (lumi)    & 1.96 $\pm$ 0.17  \\ 
$\mu\nu\gamma$    &  2.80  $\pm$ 0.05 (stat) $\pm$ 0.37 (syst) $\pm$ 0.14 (lumi)    & 1.96 $\pm$ 0.17  \\ 
$\ell \nu\gamma$    &  2.77  $\pm$ 0.03 (stat) $\pm$ 0.33 (syst) $\pm$ 0.14 (lumi)   & 1.96 $\pm$ 0.17  \\ 
    
$e^+e^-\gamma$    &  1.30  $\pm$ 0.03 (stat) $\pm$ 0.13 (syst) $\pm$ 0.05 (lumi)   &  1.18 $\pm$ 0.05\\ 
$\mu^+\mu^-\gamma$    &  1.32  $\pm$ 0.03 (stat) $\pm$ 0.11 (syst) $\pm$ 0.05 (lumi)   &  1.18 $\pm$ 0.05\\ 
$\ell^+ \ell^-\gamma$    &  1.31  $\pm$ 0.02 (stat) $\pm$ 0.11 (syst) $\pm$ 0.05 (lumi)   &  1.18 $\pm$ 0.05\\ 
    
$\nu \bar{\nu} \gamma$     &  0.133  $\pm$ 0.013 (stat) $\pm$ 0.020 (syst) $\pm$ 0.005 (lumi)  & 0.156 $\pm$ 0.012  \\ 
    \hline
   & \multicolumn{2}{c}{ $N_{\mathrm{jet}} = 0$ }   \\
    \hline
$e\nu\gamma$    &  1.77  $\pm$ 0.04 (stat) $\pm$ 0.24 (syst) $\pm$ 0.08 (lumi)    & 1.39 $\pm$ 0.13  \\ 
$\mu\nu\gamma$    &  1.74  $\pm$ 0.04 (stat) $\pm$ 0.22 (syst) $\pm$ 0.08 (lumi)    & 1.39 $\pm$ 0.13  \\ 
$\ell \nu\gamma$    &  1.76  $\pm$ 0.03 (stat) $\pm$ 0.21 (syst) $\pm$ 0.08 (lumi)    & 1.39 $\pm$ 0.13  \\ 
    
$e^+e^-\gamma$    &  1.07  $\pm$ 0.03 (stat) $\pm$ 0.12 (syst) $\pm$ 0.04 (lumi)    & 1.06 $\pm$ 0.05 \\ 
$\mu^+\mu^-\gamma$    &  1.04  $\pm$ 0.03 (stat) $\pm$ 0.10 (syst) $\pm$ 0.04 (lumi)    & 1.06 $\pm$ 0.05 \\ 
$\ell^+ \ell^-\gamma$    &  1.05  $\pm$ 0.02 (stat) $\pm$ 0.10 (syst) $\pm$ 0.04 (lumi)    & 1.06 $\pm$ 0.05 \\ 
    
$\nu \bar{\nu} \gamma$    &  0.116  $\pm$ 0.010 (stat) $\pm$ 0.013 (syst) $\pm$ 0.004 (lumi)    &   0.115 $\pm$ 0.009\\ 

    \hline
    
  \end{tabular} 
  \caption{Measured cross sections for the $\ell \nu\gamma$, $\ell^{+} \ell^{-}\gamma$ and $\nu \bar{\nu}\gamma$
           processes at $\sqrt{s}=7 \TeV{}$ in the extended fiducial region defined in Table ~\ref{tab:fiducialcut}.
           The statistical uncertainty of each measurement corresponds to the statistical uncertainty of the
           data sample used by the measurement.
           The SM predictions from {\sc mcfm}~\cite{MCFM}, calculated at NLO, are also shown in the table with systematic
           uncertainties. All {\sc mcfm} predictions are corrected to particle level using parton-to-particle
           scale factors as described in Sec.~\ref{sec:theory}.
}
  \label{tab:cs} 
\end{table*}

\subsection{Differential fiducial cross section}
Differential cross sections provide a more detailed comparison of the theoretical predictions to measurements, allowing a generic comparison of the kinematic distributions both in shape and normalization of the spectrum. For this purpose, the measured distributions are corrected to the underlying particle-level distributions by unfolding the effects of the experimental acceptance and resolution. A Bayesian iterative unfolding technique~\cite{unfolding} is used. 
  In the unfolding of binned data, effects of the experimental resolution are expressed by a response matrix, each element of which is the probability of an event in the $i$th bin at the particle level being reconstructed in the $j$th measured bin. In the iterative Bayesian unfolding, the initial prior for the underlying particle-level distribution is chosen to be the particle-level spectrum from the signal Monte Carlo sample. The posterior probability is obtained by Bayesian theory given the prior distribution, the measured distribution, and the response matrix. The posterior is then used by the unfolding algorithm as a prior for the next iteration. Two iterations are used in the unfolding procedure because tests have shown that the unfolded spectrum becomes insensitive to the initial prior probability after two iterations. 

The Bayesian unfolding is not sensitive to the MC simulation modeling of the spectrum shape. To estimate a potential bias due to MC modeling, the unfolding method was tested using a data-driven closure test. In this test the particle-level spectrum in the MC simulation is reweighted and convolved through the folding matrix such that significantly improved agreement between the data and the reconstructed spectrum from the MC simulation is attained. The reweighted, reconstructed spectrum in the MC simulation is then unfolded using the same procedure as for the data. The comparison of the result with the reweighted particle-level spectrum from the MC simulation provides the estimate of the bias due to the MC modeling. The typical size of the bias is less than 0.5\%.

The $E_{\mathrm{T}}^\gamma$ bins are chosen to be large compared to the the detector resolution to minimize migration effects and to maintain a sufficient number of events in each bin.

The differential fiducial cross section is then defined in Eq.~(\ref{eqn:absdiff}), where $x$ is the variable of the measurement,
$dx$ is the width of the $i$th bin of $x$,
and $N_i^{\rm unfold}$ is the unfolded number of events in the $i$th bin. 
\begin{equation}
\frac{d\sigma_i}{dx}= \frac{N_i^{\rm unfold}}{\int {\cal L}dt \cdot{} d x}.
\label{eqn:absdiff}
\end{equation}

\begin{figure*}[htbp]
  \centering
\subfigure[]{\includegraphics[width=1.4\columnwidth]{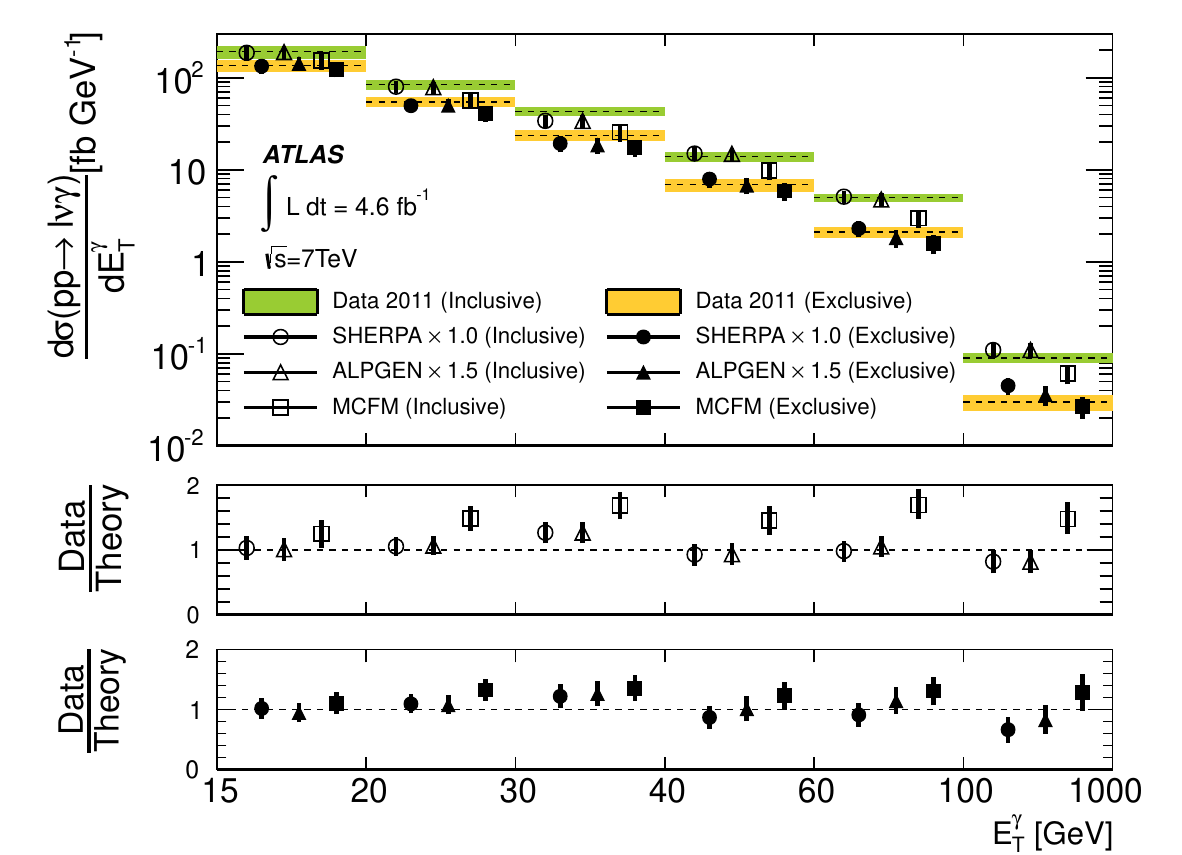}}
\subfigure[]{\includegraphics[width=1.4\columnwidth]{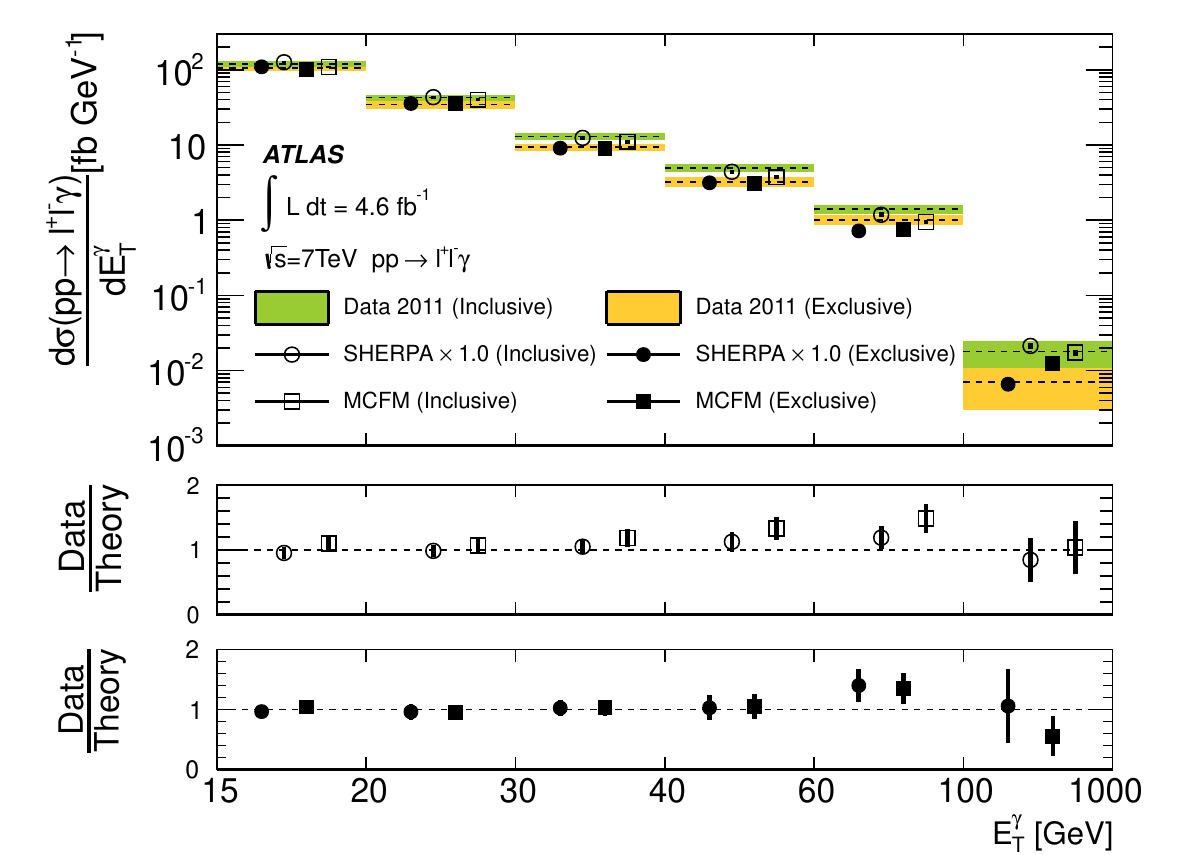}}
  \caption{Measured $E^{\gamma}_{\mathrm{T}}$ differential cross sections of (a) the $pp \rightarrow \ell \nu \gamma$ process and of (b) the $pp \rightarrow \ell^+ \ell^- \gamma$ process, using combined electron and muon measurements in the inclusive ($N_{\mathrm{jet}} \geq 0$) and exclusive ($N_{\mathrm{jet}} = 0$) extended fiducial regions. The lower plots show the ratio of the data to the predictions by different generators. The Monte Carlo uncertainties are shown only in the ratio plots. The cross-section predictions of the {\sc sherpa} and {\sc alpgen} generators have been scaled by a global factor to match the total number of events observed in data. The global factor is 1.5 for the {\sc alpgen} $\ell \nu\gamma$ signal sample and 1.0 for the {\sc sherpa} $\ell^+ \ell^-\gamma$ signal sample. No global factor is applied for {\sc mcfm} predictions.}
  \label{fig:diff_pt_wgamma_xsection_combined}
\end{figure*}

 Fig.~\ref{fig:diff_pt_wgamma_xsection_combined} shows the  differential fiducial cross sections as a function of $E_{\mathrm{T}}^\gamma$ in $V\gamma$ processes with the inclusive selection and with the exclusive zero-jet selection, as well as a comparison to the SM prediction. The corresponding numerical values ($\frac{d \sigma_i}{d E_{\mathrm{T}}^\gamma}$) are summarized in Table ~\ref{tab:nor_Vgamma}.
The systematic uncertainties on the differential fiducial cross sections are dominated by
the uncertainties on the $W+$jet, $\gamma+$jet, $Z(\ell^{+} \ell^{-})$ background normalization,
on the photon identification, and on the EM and jet energy scales.
The statistical uncertainties on the spectrum are propagated through the unfolding procedure by performing pseudoexperiments. Pseudoexperiments are generated by fluctuating the content of each bin in the data spectrum according to a Poisson distribution with a mean that is equal to the bin content.  The content of the response matrix is also fluctuated in pseudoexperiments according to their statistical uncertainties.  The unfolding procedure is then applied to each pseudoexperiment, and the standard deviation of the unfolded results is taken as the statistical uncertainty. 
The systematic uncertainties on the spectrum are evaluated by varying the response matrix for each source of uncertainty and by combining the resulting changes in the unfolded spectrum.

The normalized differential fiducial cross section
($\frac{1}{\sigma} \times \frac{ d \sigma_i}{d x}$ and $\frac{1}{\sigma} \times { d \sigma_i (x)}$,
where $\sigma= \sum \sigma_i(x)=\int \frac{d \sigma_i}{ dx} dx$ and $x$ is the variable under consideration such as $E_{\mathrm{T}}^\gamma$) is also provided for shape comparisons.
Some generators ({\sc sherpa} and {\sc alpgen}) can provide precise predictions for the kinematic variable shapes but are less accurate for the normalization.
Table~\ref{tab:nor_Vgamma} shows the normalized differential fiducial cross sections as a function of
$E_{\mathrm{T}}^\gamma$ for the $\ell \nu \gamma$ and $\ell^+ \ell^- \gamma$ processes.

\begin{figure*}[htbp]
  \centering
\subfigure[]{\includegraphics[trim=25 0 30 5,clip=true,width=0.99\columnwidth]{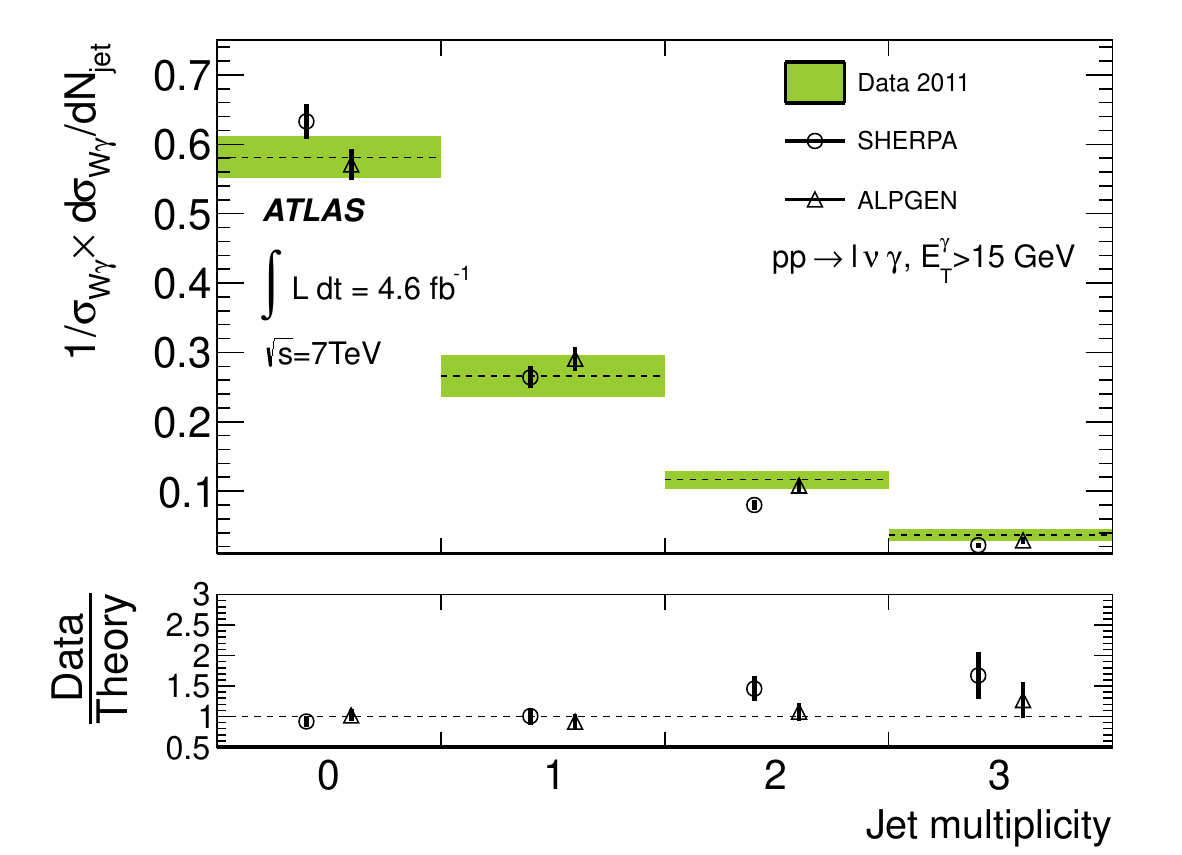}}
\subfigure[]{\includegraphics[trim=25 0 30 5,clip=true,width=0.99\columnwidth]{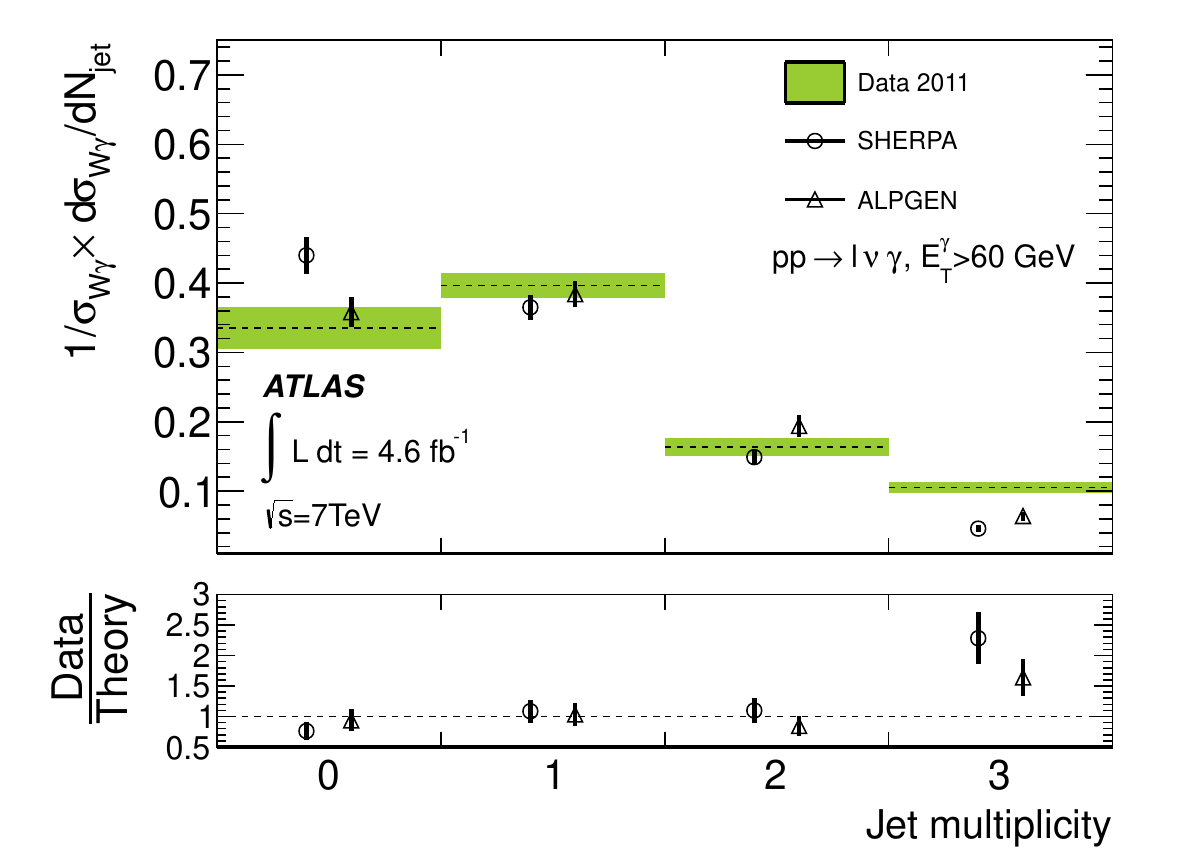}}
\subfigure[]{\includegraphics[trim=25 0 30 5,clip=true,width=0.99\columnwidth]{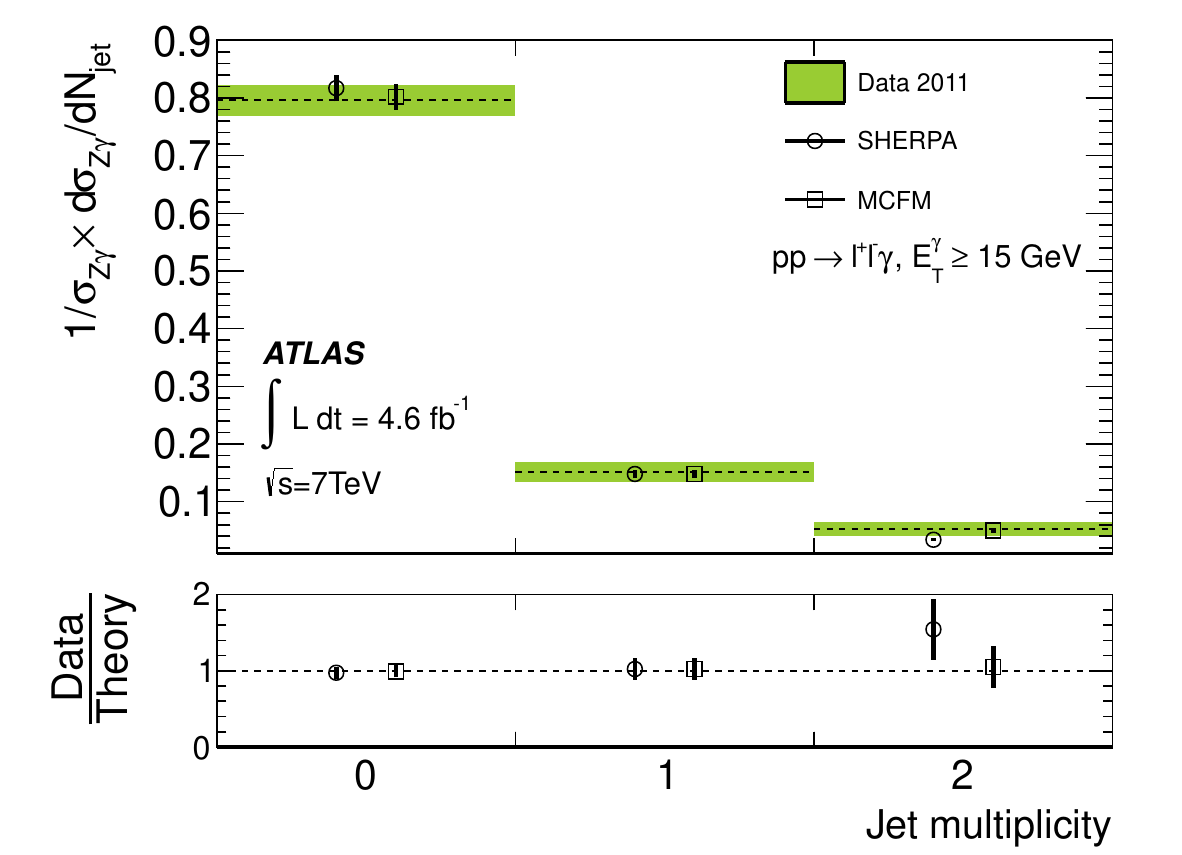}}
\subfigure[]{\includegraphics[trim=25 0 30 5,clip=true,width=0.99\columnwidth]{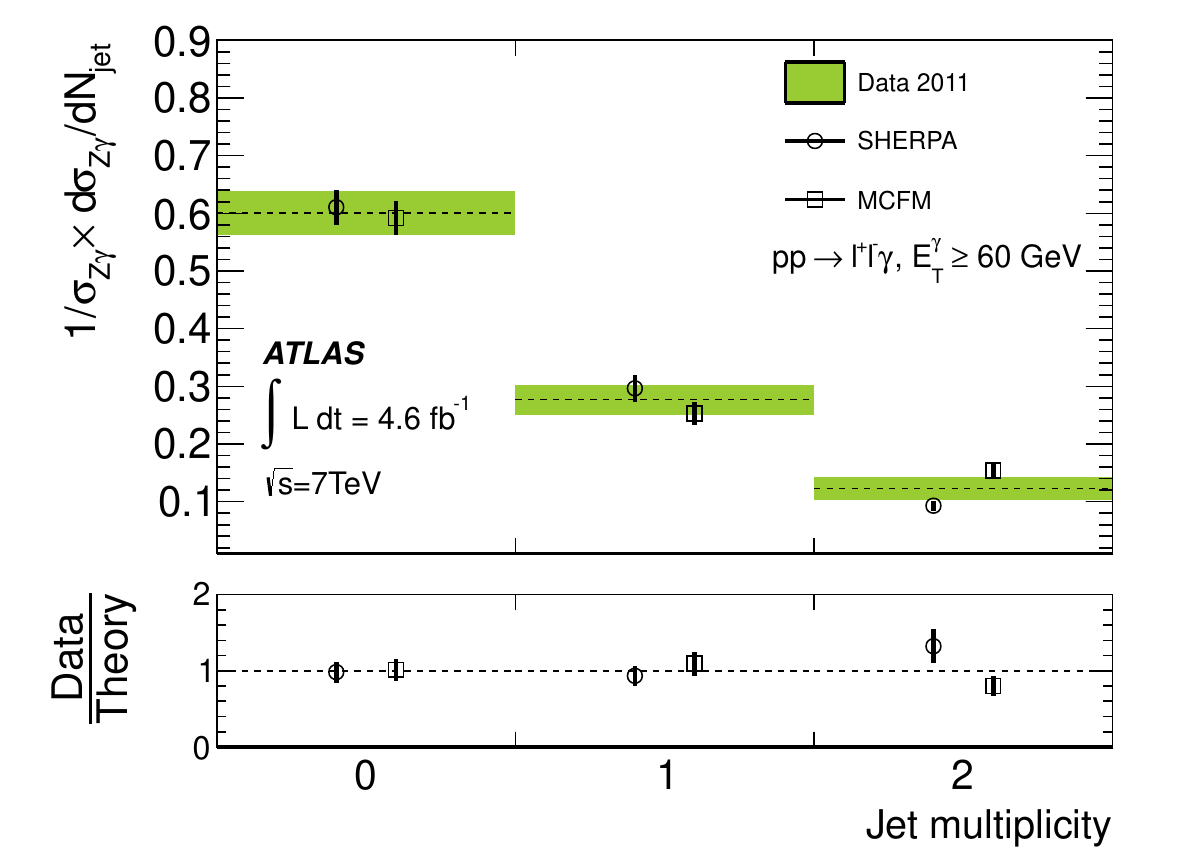}}
  \caption{ The differential cross section measurements as a function of the jet multiplicity for  the $pp \rightarrow \ell \nu \gamma$ and $pp \rightarrow \ell^+ \ell^- \gamma$ processes, for (a) $E^{\gamma}_{\mathrm{T}}>15$~\GeV{}, $pp \rightarrow \ell \nu \gamma$, (b) $E^{\gamma}_{\mathrm{T}}>60$~\GeV{}, $pp \rightarrow \ell \nu \gamma$, (c) $E^{\gamma}_{\mathrm{T}}>15$~\GeV{}, $pp \rightarrow \ell^+ \ell^- \gamma$, and (d) $E^{\gamma}_{\mathrm{T}}>60$~\GeV{}, $pp \rightarrow \ell^+ \ell^- \gamma$. The lower plots show the ratio of the data to the predictions by different generators. The {\sc mcfm} prediction for inclusive (exclusive) $\ell \nu \gamma$ cross section with ${p}_{\mathrm{T}}^\gamma >60$~\GeV{} is $171 \pm 23$~fb ($80 \pm 22$~fb). The corresponding predictions for ${p}_{\mathrm{T}}^\gamma >15$~\GeV{} are given in Table ~\ref{tab:cs}. {\sc mcfm} does not provide the predictions for two and three jet bins for the $pp \rightarrow \ell \nu\gamma$ process, therefore only {\sc alpgen} and {\sc sherpa} predictions are shown in (a) and (b). }
  \label{fig:diff_njet}
\end{figure*}

The normalized cross sections measured in bins of jet multiplicity in $V\gamma$ events are presented in Fig.~\ref{fig:diff_njet} and Table~\ref{tab:nor_njet}. The measurements are performed in the extended fiducial phase spaces defined in Table~\ref{tab:fiducialcut}, with $E_{\mathrm{T}}^\gamma > 15$~\GeV{} for the low-$E_{\mathrm{T}}^\gamma$ region, and with $E_{\mathrm{T}}^\gamma > 60$~\GeV{} for the high-$E_{\mathrm{T}}^\gamma$ region. The systematic uncertainties on the jet multiplicity measurement are dominated by the uncertainties on the jet energy scale, the jet energy resolution and the background shape.

The transverse mass $m_{\mathrm{T}}^{W\gamma}$ spectrum and the invariant mass $m^{Z\gamma}$ spectrum are also measured in the $\ell \nu \gamma$ and in the $\ell^+ \ell^- \gamma$ processes, respectively. The transverse mass is defined in Eq.~(\ref{equ:MT3}), where $m_{\ell \gamma}$ is the invariant mass of the lepton--photon system:

\begin{eqnarray}
 (m_{\mathrm{T}}^{W\gamma})^{2} &=& ( \sqrt{m^2_{\ell \gamma}+|\vec{p}_{\mathrm{T}}(\gamma)+\vec{p}_{\mathrm{T}}(\ell)|^2}+E_{\mathrm{T}}^{\mathrm{miss}})^2      \nonumber \\
   && -| \vec{p}_{\mathrm{T}}(\gamma)+\vec{p}_{\mathrm{T}}(\ell)+\vec{E}_{\mathrm{T}}^{\mathrm{miss}}|^2\nonumber  .\\ 
\label{equ:MT3}
\end{eqnarray}

\begin{figure*}[htbp]
  \centering
\subfigure[]{\includegraphics[trim=25 0 5 5,clip=true,width=0.99\columnwidth]{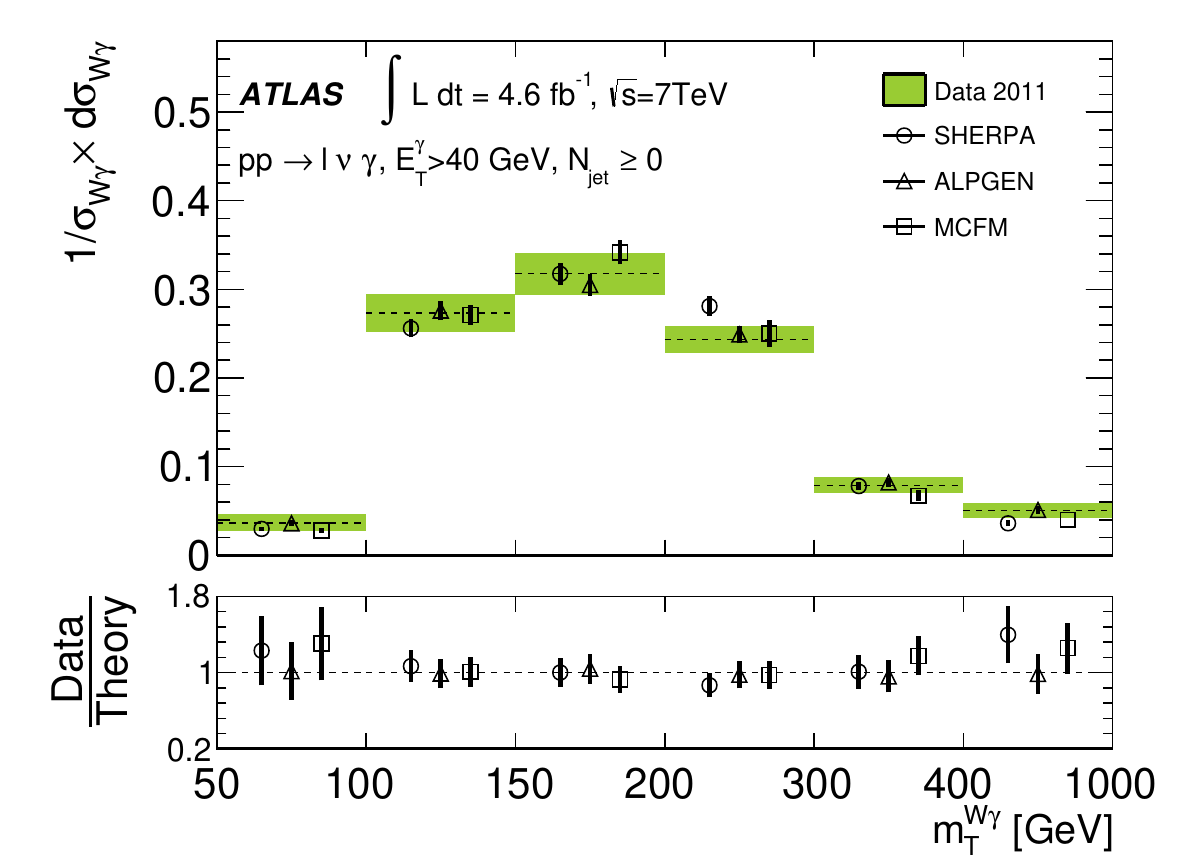}}
\subfigure[]{\includegraphics[trim=25 0 5 5,clip=true,width=0.99\columnwidth]{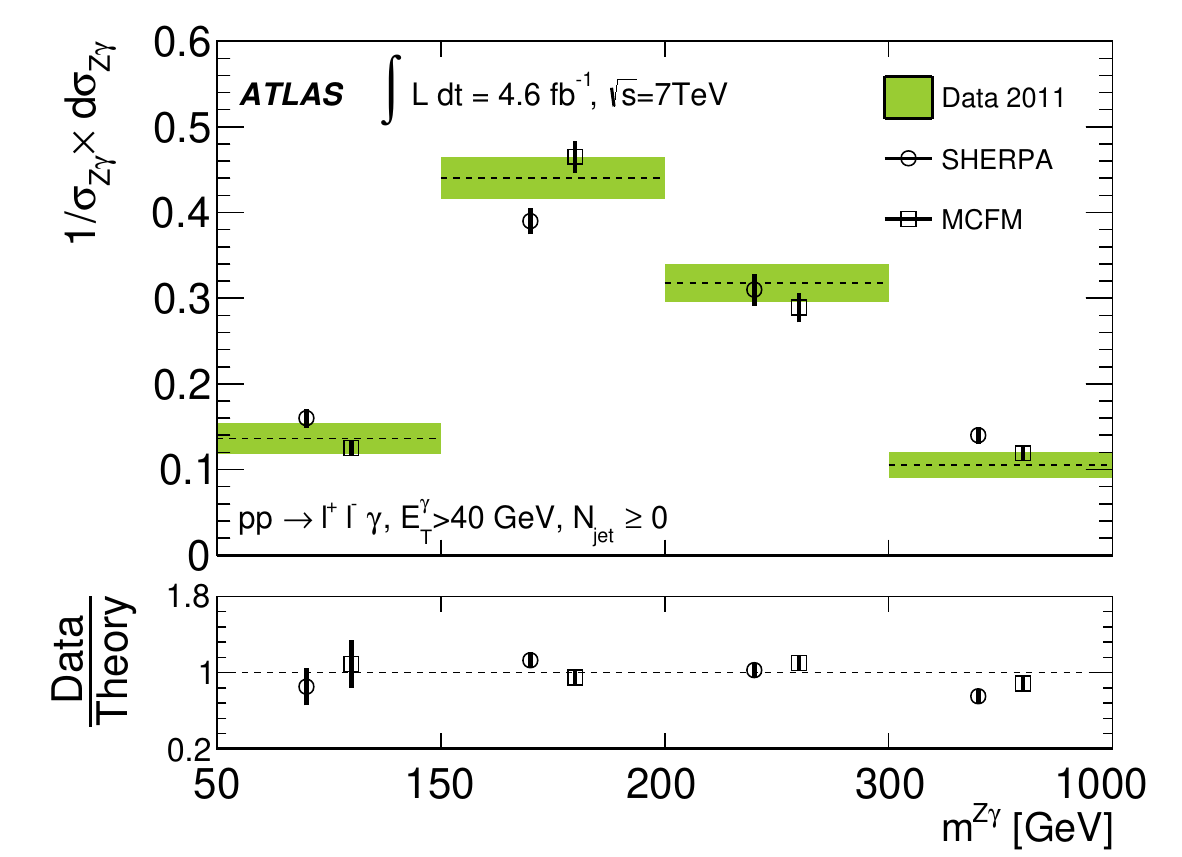}}
  \caption{The inclusive normalized differential cross section of (a) the $pp \rightarrow \ell \nu \gamma$ process as a function of $m_{\mathrm{T}}^{W\gamma}$ and (b) the $pp \rightarrow \ell^+ \ell^- \gamma$ process as a function of $m^{Z\gamma}$. The lower plots show the ratio of the data to the prediction by different generators.}
  \label{fig:diff_mass}
\end{figure*}

These measurements are performed in the extended fiducial phase space defined in Table~\ref{tab:fiducialcut}, with $E^{\gamma}_{\mathrm{T}} > 40$~\GeV{}. 
The distribution of $m_{\mathrm{T}}^{W\gamma}$ for the $\ell \nu\gamma$ candidates is shown in Fig.~\ref{fig:Wg_kin1}(d); the expected numbers of signal and background events are also shown. The unfolded $m_{\mathrm{T}}^{W\gamma}$ spectrum is presented in Fig.~\ref{fig:diff_mass}(a) and Table~\ref{tab:mass_Wgamma}. The systematic uncertainties of $m_T^{W\gamma}$ spectrum measurements are dominated by the uncertainties on the EM energy scale, the jet energy scale, the $E_{\mathrm{T}}^{\mathrm{miss}}$ energy scale and the background shape.

The distribution of $m^{Z\gamma}$ for the $\ell^{+} \ell^{-}\gamma$ candidates is presented in Fig.~\ref{fig:Zg_kin1}(c), together with the expected $m^{Z\gamma}$ distributions of the signal and background events. The unfolded $m^{Z\gamma}$ spectrum is presented in Fig.~\ref{fig:diff_mass}(b) and Table~\ref{tab:mass_zgamma}. The uncertainties in the $m^{Z\gamma}$ spectrum measurement arise predominantly from the uncertainties on the EM energy scale.

\begin{table*}[htbp]
  \begin{tabular}{ccccccc}
       \hline
  $E_{\mathrm{T}}^{\gamma}$ [GeV]        & [15,20]& [20,30]  & [30,40] & [40,60] & [60,100] & [100,1000]   \\
       \hline
    &  \multicolumn{6}{c}{$pp \rightarrow \ell \nu \gamma$, $N_{\mathrm{jet}} \geq 0$} \\

$d \sigma_{W\gamma}/d E_{\mathrm{T}}^{W\gamma}$ [fb \GeV{}$^{-1}$] & $192 \pm 32$ & $84 \pm 11$ & $43.0 \pm 5.0$ & $13.9 \pm 1.8$  & $5.0 \pm 0.5$  & $0.090 \pm 0.012$  \\ 

1/$\sigma_{W\gamma} \times d \sigma_{W\gamma}$     &  0.34      & 0.30   & 0.15 & 0.10 &  0.072  & 0.029   \\     
  
Rel. Uncertainty      &  7.4\%      &  5.4\%   &  10\%    & 6.6\%         & 9.1\%    &  10\% \\  
       
\hline
    &  \multicolumn{6}{c}{$pp \rightarrow \ell \nu \gamma$, $N_{\mathrm{jet}} = 0$} \\
       
$d \sigma_{W\gamma}/d E_{\mathrm{T}}^{W\gamma}$ [fb \GeV{}$^{-1}$] & $136 \pm 22$ & $54.5 \pm 7.1$ & $23.6 \pm 3.3$ & $6.9 \pm 1.2$  & $2.1 \pm 0.3$  & $0.030 \pm 0.006$  \\

1/$\sigma_{W\gamma} \times d \sigma_{W\gamma}$  &   0.40     & 0.32  & 0.14  & 0.081  & 0.050    &  0.016  \\

Rel. Uncertainty      &  8.8\%      &  8.5\%   &  11\%    & 9.1\%     &  11\%    &  18\% \\   
       \hline
    &  \multicolumn{6}{c}{$pp \rightarrow \ell^+ \ell^- \gamma$, $N_{\mathrm{jet}} \geq 0$} \\

$d \sigma_{Z\gamma}/d E_{\mathrm{T}}^{Z\gamma}$ [fb \GeV{}$^{-1}$] & $120 \pm 12$ & $ 42.5 \pm 4.2   $ & $ 13.0 \pm 1.4  $ & $  4.94 \pm 0.61 $  & $ 1.40 \pm 0.19   $  & $ 0.018 \pm  0.007  $   \\

1/$\sigma_{Z\gamma} \times d \sigma_{Z\gamma}$           &   0.45     & 0.32          & 0.098          &  0.075        &  0.042        & 0.012   \\   

Rel. Uncertainty      &   5.9\%     & 6.2\%          & 10\%         &  12\%    & 12\%    & 36\%  \\   
       \hline
    &  \multicolumn{6}{c}{$pp \rightarrow \ell^+ \ell^- \gamma$, $N_{\mathrm{jet}} = 0$} \\

$d \sigma_{Z\gamma}/d E_{\mathrm{T}}^{Z\gamma}$ [fb \GeV{}$^{-1}$] & $   106 \pm 11  $ & $    34.3 \pm 4.1  $ & $    9.3 \pm 1.1 $ & $    3.24 \pm 0.47  $   & $   1.01 \pm 0.16   $   & $   0.007 \pm 0.004 $   \\

1/$\sigma_{Z\gamma} \times d \sigma_{Z\gamma}$      &   0.49     & 0.32         & 0.087          &  0.060        &   0.038       & 0.0059   \\        

Rel. Uncertainty      &   6.5\%     & 11\%      & 12\%      & 14\%       &   15\%       & 54\%   \\   
  
       \hline
    
\end{tabular}
  \caption{The measured differential fiducial cross sections and normalized differential fiducial cross sections as a function of $E^{\gamma}_{\mathrm{T}}$ for the $\ell \nu \gamma$ and $\ell^{+} \ell^{-} \gamma$ processes using combined electron and muon measurements in the extended fiducial region defined in Table~\ref{tab:fiducialcut}: inclusive with $N_{\mathrm{jet}} \geq 0$ and exclusive with $N_{\mathrm{jet}} = 0$. The uncertainties given here are the combination of the systematic and statistical uncertainties.
Absolute uncertainties are presented for the measured differential fiducial cross sections,
and relative uncertainties are presented for the measured normalized differential fiducial cross sections.
}
  \label{tab:nor_Vgamma}
\end{table*}

\begin{table*}[htbp]
  \begin{tabular}{ccccccc}
       \hline

$N_{\mathrm{jet}}$         & $0$&  $1$  &  $2$ &  $3$    \\
            \hline 
    &  \multicolumn{4}{c}{$pp \rightarrow \ell \nu \gamma$, $E_{\mathrm{T}}^\gamma \geq 15$ \GeV{}} \\
       
1/$\sigma_{W\gamma} \times d \sigma_{W\gamma}/d N_{\mathrm{jet}}$     &   0.58     &    0.27       &  0.12    &  0.037       \\   

Rel. Uncertainty      &   5.2\%     & 11\%          &    11\%       &   22\%       \\   
       \hline
    &  \multicolumn{4}{c}{$pp \rightarrow \ell \nu \gamma$, $E_{\mathrm{T}}^\gamma \geq 60$ \GeV{}} \\
       
1/$\sigma_{W\gamma} \times d \sigma_{W\gamma}/d N_{\mathrm{jet}}$      &   0.33     & 0.40          & 0.16          & 0.11        \\   
Rel. Uncertainty      &   10\%     & 6.4\%          & 11\%          & 22\%         \\   
       \hline
    &  \multicolumn{4}{c}{$pp \rightarrow \ell^+ \ell^- \gamma$, $E_{\mathrm{T}}^\gamma \geq 15$ \GeV{}} \\
       
1/$\sigma_{Z\gamma} \times d \sigma_{Z\gamma}/d N_{\mathrm{jet}}$      &   0.80    & 0.15 &   0.052       & -        \\     
Rel. Uncertainty      &   3.4\%     & 11\%         & 22\%          & -         \\   
       \hline
    &  \multicolumn{4}{c}{$pp \rightarrow \ell^+ \ell^- \gamma$, $E_{\mathrm{T}}^\gamma \geq 60$ \GeV{}} \\
       
1/$\sigma_{Z\gamma} \times d \sigma_{Z\gamma}/d N_{\mathrm{jet}}$      &   0.60     &  0.28         & 0.12          & -        \\   
Rel. Uncertainty    &   6.4\%     &  9.4\%         & 16\%          & -        \\   
       
       \hline
    
\end{tabular}
\caption{The measured differential fiducial cross sections as a function of jet multiplicity for
         $\ell \nu \gamma$ and $\ell^{+} \ell^{-} \gamma$ processes.
         The measurements are performed in the extended fiducial phase spaces defined in Table ~\ref{tab:fiducialcut},
         with $E_{\mathrm{T}}^\gamma > 15$~\GeV{} and with $E_{\mathrm{T}}^\gamma > 60$~\GeV{}.
         The relative uncertainty is computed from the combination of cross sections
         from the electron and muon channels.}
  \label{tab:nor_njet}
\end{table*}

\begin{table*}[htbp]
  \begin{tabular}{ccccccc}
    \hline

$m_{\mathrm{T}}^{W\gamma}$ [GeV]        & [50,100]& [100,150]  & [150,200] & [200,300] & [300,400] & [400,1000]   \\
       \hline
1/$\sigma_{W\gamma} \times d \sigma_{W\gamma}$       &   0.037     & 0.27          & 0.32          & 0.24         & 0.079         & 0.050   \\   
Rel. Uncertainty      &   27\%    &  8.0\%          & 7.4\%          & 6.4\%       & 11\%   & 17\%   \\   
       
       \hline
    
\end{tabular}
\caption{ The measured differential fiducial cross sections as a function of
          $m_{\mathrm{T}}^{W\gamma}$ for inclusive $\ell \nu \gamma$ process.
          The relative uncertainty is computed from the combination of cross sections
          from the electron and muon channels.}
  \label{tab:mass_Wgamma}
\end{table*}

\begin{table*}[htbp]
  \begin{tabular}{ccccccc}
    \hline

$m^{Z\gamma}$ [GeV]        & [50,150]  & [150,200] & [200,300] & [300,1000]   \\
       \hline
1/$\sigma_{Z\gamma} \times d \sigma_{Z\gamma}$       &   0.14     & 0.44          & 0.32  & 0.11      \\     
Rel. Uncertainty      &   13\%     & 5.5\%          & 6.9\%          & 14\%        \\   

       \hline    
\end{tabular}
\caption{The measured differential fiducial cross sections as a function of $m^{Z\gamma}$
         for inclusive $\ell^{+} \ell^{-} \gamma$ process.
         The relative uncertainty is computed from the combination of cross sections
         from the electron and muon channels.}
  \label{tab:mass_zgamma}
\end{table*}

\section{Comparison to Theoretical Predictions}
\label{sec:theory}

To test the predictions of the SM, the cross-section measurements of $pp$ $\to$ $\ell \nu\gamma + X$, $pp$ $\to$ $\ell^{+}$$\ell^{-}$$\gamma + X$  and $pp$ $\to$ $\nu\bar{\nu}$$\gamma + X$ production are compared to NLO and LO calculations using the {\sc mcfm}~\cite{MCFM} program.  
Version 6.3 of {\sc mcfm} includes cross-section predictions for the production of $W\gamma$ + zero partons at NLO and for $W\gamma$ + one parton at LO. For $Z\gamma$ production the predictions are at NLO for both $Z\gamma$ + zero partons and $Z\gamma$ + one parton, and at LO for $Z\gamma$ + two partons. Finally, $\nu\bar{\nu}$$\gamma$ production is calculated at NLO for zero partons and LO for one parton.

Measurements of inclusive $\ell \nu\gamma$ production are compared to the NLO $W\gamma$ prediction with no restriction on the associated quark/gluon. Exclusive $\ell \nu\gamma$ production is compared to the same NLO prediction by requiring no parton with $|\eta|<4.4$ and $p_{\mathrm{T}}>30~\GeV$ in the final state.   
Similarly, measurements of inclusive $\ell^{+}$$\ell^{-}$$\gamma$ production are compared directly to the NLO $Z\gamma$ prediction while the exclusive $\ell \nu\gamma$ measurement is compared to the prediction with no additional parton with $|\eta|<4.4$ and $p_{\mathrm{T}}>30~\GeV$. The exclusive cross section for $\ell^{+}$$\ell^{-}$$\gamma$ production with exactly one jet with $|\eta|<4.4$ and $p_{\mathrm{T}}>30~\GeV$ is compared to the NLO $Z\gamma$ + one-parton prediction with the same kinematic restriction on the single parton. Production of $l^{+}$$l^{-}$$\gamma$ with exactly two jets with $|\eta|<4.4$ and $p_{\mathrm{T}}>30~\GeV$ is compared to the LO $Z\gamma$ + two-parton prediction. The cross sections for $\nu\bar{\nu}$$\gamma$ production are calculated in a similar manner using the {\sc mcfm} NLO prediction for $\nu\bar{\nu}$$\gamma$ + zero partons.

All the {\sc mcfm} predictions include $W$ and $Z$ boson production with photons from direct $W\gamma$ and $Z\gamma$ diboson production, from final-state radiation off the leptons in the $W/Z$ decays and from quark/gluon radiation using the BFGSetII~\cite{BFG} photon fragmentation function. Event generation is done using the default electroweak parameters in the {\sc mcfm} program and the parton distribution functions {\sc ct10}~\cite{CT10}.  The renormalization, factorization and photon fragmentation scales are set equal to $\sqrt{M_V^2 + {E_{\mathrm{T}} ^\gamma}^2}$. Photon isolation is defined using the fractional energy carried by partons in a cone $\Delta R_{\gamma}$ = 0.4 about the photon direction. The fractional parton energy $\epsilon_h$ in the isolation cone (excluding the photon's energy) is required to be less than 0.5. The kinematic requirements for the parton-level generation are the same as those chosen at particle level for the extended fiducial cross-section measurements (see Table~\ref{tab:fiducialcut}).

The parton-level cross-section uncertainties are evaluated by varying the PDFs and the renormalization and factorization scales,
and by changing the definition of photon isolation.
The PDF uncertainty is about 3\%--4\%. It is estimated using the CT10 error eigenvectors at their 68$\%$ C.L. limits, and varying the $\alpha_s$ values in the range 0.116 - 0.120.
The variation of the renormalization and factorization scales from the nominal $\sqrt{M_V^2 + {E_{\mathrm{T}} ^{\gamma}}^2}$
up and down by a common factor of two gives an uncertainty about 3\%--7\%.
For the exclusive channels with no central jets with $p_{\mathrm{T}}$ greater than 30 GeV,
the method suggested in Ref.~\cite{ST} is used to estimate the uncertainty due to the
energy scale of the process.
The uncertainty due to the definition of photon isolation varies in the range 1\%--5\%. It is evaluated by varying the fractional parton energy $\epsilon_{h}^{p}$ from 0.0 to 1.0.

To compare these NLO SM predictions to the measured cross sections,
they must be corrected for the differences between the parton-level and particle-level definitions of the jet and photon isolation, as done for data.
The {\sc alpgen}+{\sc herwig} (for $W\gamma$) and {\sc sherpa} (for $Z\gamma$) MC samples
are used to estimate the parton-to-particle scale factors. 
The scale factor ($S_{W\gamma}$ or $S_{Z\gamma}$) is defined as the number of simulated events passing the
fiducial region selection cuts at the particle level divided by the number of simulated events passing the fiducial region selection cuts at the parton level.
They increase the parton-level cross sections by up to $13\%$ with uncertainties that
vary from 3$\%$ to 7$\%$ depending on the channel.
A typical value of the scale factor predicted for the $W\gamma$ inclusive phase space by the {\sc algpen} ({\sc sherpa}) generator is $1.05$ ($1.00$). The uncertainties for $W\gamma$ events are evaluated by comparing the differences in predictions made using {\sc alpgen} and {\sc sherpa}. The uncertainties for $Z\gamma$ events are evaluated by comparing two signal samples: the nominal sample uses the {\sc sherpa} generator, the alternative sample is obtained from the {\sc madgraph}~\cite{madgraph} generator interfaced to {\sc pythia} for parton shower and fragmentation processes. A typical value of the scale factor predicted for the $Z\gamma$ inclusive phase space by the {\sc sherpa} ({\sc madgraph}) generator is $1.02$ ($1.03$).

\subsection{Integrated cross section predictions}

The inclusive and exclusive production cross sections in the extended fiducial regions defined in Table~\ref{tab:fiducialcut} for the $\ell \nu \gamma$, 
$\ell^+ \ell^- \gamma$ and $\nu\bar{\nu} \gamma$ final states are compared as described above to the NLO predictions made by the {\sc mcfm} generator. The parton-level predictions corrected to the particle level are listed in Table~\ref{tab:cs} together with the measured cross sections for events with $E_{\mathrm{T}}^{\gamma}$ $>$15 GeV.  The {\sc mcfm} NLO predictions agree well with the measured $\ell^+ \ell^-\gamma$ and $\nu\bar{\nu}\gamma$ cross sections. For the $\ell \nu \gamma + X$ channel the measured exclusive ($N_{\mathrm{jet}}$ =0) cross section is slightly higher and the inclusive ($N_{\mathrm{jet}}$ $\geq$0) cross section significantly higher than the {\sc mcfm} predictions. The discrepancy between the NLO prediction and data in the $\ell \nu \gamma + X$ channel is due to significant contributions from multijet production that are not observed in $\ell^+ \ell^-\gamma$ as discussed in more detail below. In $W\gamma$ production there are contributions from processes with direct photon emission from the $W$ boson which are absent in $Z\gamma$ production (see Fig.~\ref{fig:fey_dia}d). These additional $W\gamma$ production processes tend to have a higher jet multiplicity, and these contributions are not included in the current NLO calculations.

\subsection{Differential cross sections for $pp \to \ell^+ \ell^- \gamma$}

The differential cross sections for
$\ell^+ \ell^- \gamma$ production can be compared to the NLO {\sc mcfm} predictions and to
those of the LO {\sc sherpa} generator scaled with an overall normalization factor obtained from data.
The $E_{\mathrm{T}}^{\gamma}$ spectra from the inclusive and exclusive $\ell^+ \ell^- \gamma$ channel are shown in Fig.~\ref{fig:diff_pt_wgamma_xsection_combined}(b).  There is good agreement between the data and the {\sc sherpa} and {\sc mcfm} predictions over the 
full $E_{\mathrm{T}}^{\gamma}$ range. The normalized differential spectrum for the $m^{Z\gamma}$
is compared to {\sc sherpa} and {\sc mcfm} in Fig.~\ref{fig:diff_mass}(b). The NLO {\sc mcfm} prediction
reproduces the measured $m^{Z\gamma}$ somewhat better than the LO {\sc sherpa} MC. The normalized jet multiplicity spectrum from the $\ell^+ \ell^- \gamma + X$ events is shown in Figs.~\ref{fig:diff_njet}(c) and (d).
This can be compared to the LO {\sc sherpa} generator with up to three partons, as well as to the {\sc mcfm} generator with NLO predictions for zero and one parton, and a LO prediction for two partons. As shown in Figs.~\ref{fig:diff_njet}(c) and (d), both the {\sc mcfm} and {\sc sherpa} generators are in good agreement with data.

\subsection{Differential cross sections for $pp \to \ell \nu \gamma$}

The background-subtracted, unfolded differential cross sections for
$\ell \nu \gamma$ production can be compared to the NLO {\sc mcfm} prediction and to both the {\sc alpgen} and {\sc sherpa} MC generators. The predictions from {\sc mcfm} are absolute cross sections, while those from both {\sc alpgen} and {\sc sherpa} are scaled by an overall normalization factor obtained from data. 
The measured $E_{\mathrm{T}}^{\gamma}$ spectrum is shown in Fig.~\ref{fig:diff_pt_wgamma_xsection_combined}(a) for both inclusive and exclusive event selections.
The {\sc mcfm} prediction agrees with the data in the lowest photon $E_{\mathrm{T}}^{\gamma}$ bin but there are significant discrepancies in all higher $E_{\mathrm{T}}^{\gamma}$ bins, the effect being more enhanced for the inclusive event selection. The MC generators ({\sc alpgen} and {\sc sherpa}) reproduce the shape of the $E_{\mathrm{T}}^{\gamma}$ spectrum reasonably well over the full $E_{\mathrm{T}}^{\gamma}$ range. The normalized differential cross section for $\ell \nu \gamma$ as a function of $m_{\mathrm{T}}^{W\gamma}$ is shown in Fig.~\ref{fig:diff_mass}(a). 
The {\sc mcfm}, {\sc alpgen}, and {\sc sherpa} generators all provide a good description of the data.

The better description of {\sc alpgen} and {\sc sherpa} compared to the {\sc mcfm} prediction
for the $E_{\mathrm{T}}^{\gamma}$ spectrum from $\ell \nu \gamma$ production, can be attributed to processes with large parton multiplicities, which correspond to tree-level diagrams of higher order in the strong coupling constant. A comparison of the jet multiplicities in the low-$E_{\mathrm{T}}^{\gamma}$ region (Fig.~\ref{fig:diff_njet}(a)) and in the high-$E_{\mathrm{T}}^{\gamma}$ region (Fig.~\ref{fig:diff_njet}(b)), shows that those processes with more than one parton (jet) contribute more in higher $E_{\mathrm{T}}^{\gamma}$ regions. The {\sc mcfm} NLO cross-section prediction for
$\ell \nu \gamma$ production includes real parton emission processes only up to one radiated quark or gluon. The lack of higher-order QCD contributions results in an underestimate of the predicted cross sections.
For the same reason, the improvement of the description by {\sc alpgen} compared to {\sc sherpa} for the predictions of the jet multiplicity spectrum can be attributed to the fact that there are more additional hard partons included in the matrix element calculation with the {\sc Alpgen} generator.

\section{Limits on  Anomalous Triple-Gauge-Boson Couplings}
\label{sec:ATGC}
The reconstructed $E_{\mathrm{T}}^\gamma$ distributions from $V\gamma$ events with the exclusive zero-jet selection are used to set limits on $WW\gamma$, $ZZ\gamma$ and $Z\gamma\gamma$ anomalous triple-gauge-boson coupling parameters. Assuming C and P conservation separately, the aTGCs are generally chosen as $\lambda_{\gamma}$, $\Delta \kappa_{\gamma}$ ($\Delta \kappa_{\gamma} = \kappa_{\gamma} -1$) for the $WW\gamma$ vertex~\cite{BaurLO,BaurNLO}, $h^{Z}_{3}$, $h^{Z}_{4}$ for the $ZZ\gamma$ vertex~\cite{PhysRevD.47.4889}, and  $h^{\gamma}_{3}$, $h^{\gamma}_{4}$ for the $Z\gamma\gamma$ vertex~\cite{PhysRevD.47.4889}.

Form factors are introduced to avoid unitarity violation at very high energy.
Typical choices of these form factors for the $WW\gamma$ aTGCs are:
$\Delta\kappa_\gamma(s) = \Delta\kappa_{\gamma}/(1+\hat{s}/\Lambda^2)^2$ and
$\lambda_\gamma(s) = \lambda_{\gamma}/(1+\hat{s}/\Lambda^2)^2$~\cite{BaurNLO}.
For the $ZZ\gamma$ aTGCs, conventional choices of form factors are
$h^{Z}_{3}(s) =  h^{Z}_{3}/(1+\hat{s}/\Lambda^2)^3$ and
$h^{Z}_{4}(s) =  h^{Z}_{4}/(1+\hat{s}/\Lambda^2)^4$~\cite{PhysRevD.47.4889}. Similar choices of form factors are used for $Z\gamma\gamma$ aTGCs.
Here $\sqrt{\hat{s}}$ is the $W\gamma$ or $Z\gamma$ invariant mass
and $\Lambda$ is the new-physics energy scale. To conserve unitarity, $\Lambda$ is chosen as 6 \TeV{} in the $W\gamma$ analysis and 3 \TeV{} in the $Z\gamma$ analysis. The results with energy cutoff $\Lambda=\infty$ are also presented as a comparison in the unitarity violation scheme.

\begin{table}[h]
\begin{center}
 \begin{tabular}{cccc}
\hline
  processes               & \multicolumn{2}{c}{$pp \rightarrow \ell \nu \gamma$}    \\ 
  $\Lambda$               &   \multicolumn{2}{c}{$\infty$}   \\
\hline
                 & Measured    & Expected \\ 
\hline
 $\Delta\kappa_{\gamma}$  & $(-0.41,0.46)$       &   $(-0.38,0.43)  $   \\
$\lambda_{\gamma}$       &   $(-0.065,0.061)$     &   $(-0.060,0.056)$  \\ \hline
  $\Lambda$          & \multicolumn{2}{c}{6 TeV}  \\
\hline
                 & Measured    & Expected \\ 
\hline
 $\Delta\kappa_{\gamma}$  & $(-0.41,0.47)$       &   $(-0.38,0.43)$     \\
$\lambda_{\gamma}$       &   $(-0.068,0.063)$     &  $(-0.063,0.059)$   \\ \hline
  processes               & \multicolumn{2}{c}{$pp \rightarrow \nu \nu \gamma$ and $pp \rightarrow \ell^+ \ell^-\gamma$}   \\
  $\Lambda$          &  \multicolumn{2}{c}{$\infty$}  \\
\hline
                 & Measured    & Expected \\ 
\hline
  $h_{3}^{\gamma}$ &   $(-0.015, 0.016)$    &    $(-0.017, 0.018)$   \\
  $h_{3}^{Z}     $   &   $(-0.013, 0.014)$     &  $(-0.015, 0.016)$     \\
  $h_{4}^{\gamma}$  &   $(-9.4 \times 10^{-5}, 9.2 \times 10^{-5})$   &   $(-1.0 \times 10^{-4}, 1.0 \times 10^{-4})$\\
  $h_{4}^{Z}$      &   $(-8.7 \times 10^{-5}, 8.7 \times 10^{-5})$    &  $(-9.7 \times 10^{-5}, 9.7 \times 10^{-5})$\\ \hline
  $\Lambda$          & \multicolumn{2}{c}{3 TeV}  \\
\hline
                 & Measured    & Expected \\ 
\hline
  $h_{3}^{\gamma}$ &   $(-0.023, 0.024) $   &    $(-0.027, 0.028)$   \\
  $h_{3}^{Z}     $   &   $(-0.018, 0.020)$     &  $(-0.022, 0.024)$     \\
  $h_{4}^{\gamma}$  &   $(-3.7 \times 10^{-4}, 3.6 \times 10^{-4})$   &   $(-4.3 \times 10^{-4}, 4.2 \times 10^{-4})$\\
  $h_{4}^{Z}$      &   $(-3.1 \times 10^{-4}, 3.1 \times 10^{-4}) $   &  $(-3.7 \times 10^{-4}, 3.6 \times 10^{-4})$\\ \hline

\end{tabular}
 \caption[]{The measured and expected 95\% C.L. intervals on the charged
($\Delta\kappa_{\gamma}$, $\lambda_{\gamma}$) and
neutral ($h_{3}^{\gamma}$, $h_{3}^{Z}$, $h_{4}^{\gamma}$, $h_{4}^{Z}$) anomalous couplings.
The results obtained using different $\Lambda$ values are shown with all the other couplings are set to the SM values. The two numbers in each
parenthesis denote the 95\% C.L. interval.}
 \label{Tab:wg_atgc}\vspace{12pt}
\end{center}
\end{table}

Deviations of the aTGC parameters from the SM predictions would nearly all lead to an excess of
high-energy photons associated with the $W$ and $Z$ bosons. Thus, measurements of the exclusive extended fiducial cross sections for $W\gamma$ production with $E_{\mathrm{T}}^\gamma>100$ \GeV{} are used to extract aTGC limits. The cross-section predictions with aTGCs ($\sigma^{\mathrm{aTGC}}_{W\gamma}$ and $\sigma^{\mathrm{aTGC}}_{Z\gamma}$) are obtained from the {\sc mcfm} generator.
The number of expected $W\gamma$ events in the exclusive extended fiducial region
($N^{\mathrm{aTGC}}_{W\gamma}(\Delta\kappa_\gamma, \lambda_\gamma)$) for a given aTGC strength is obtained using Eq.~(\ref{equ:natgc})
\begin{equation}
N^{\mathrm{aTGC}}_{W\gamma}(\Delta\kappa_\gamma, \lambda_\gamma) = \sigma^{\mathrm{aTGC}}_{W\gamma} \times C_{W\gamma}  \times A_{W\gamma} \times S_{W\gamma} \times \int {\cal L} dt.
\label{equ:natgc}
\end{equation}
For the $Z\gamma$ case, $N^{\mathrm{aTGC}}_{Z\gamma}(h_3^\gamma,h_4^\gamma)$ or $N^{\mathrm{aTGC}}_{Z\gamma}(h_3^Z,h_4^Z)$ are obtained in a similar way.
The anomalous couplings influence the kinematic properties of $W\gamma$ and $Z\gamma$ events and thus the corrections for event reconstruction ($C_{W\gamma}$ and $C_{Z\gamma}$). The maximum variations of $C_{W\gamma}$ and $C_{Z\gamma}$ within the measured aTGC limits are quoted as additional systematic uncertainties.

The limits on a given aTGC parameter are extracted from a frequentist profile likelihood test, as explained in Sec.~\ref{sec:cs}, given the extended fiducial measurements.
The profile likelihood combines the observed number of exclusive $V\gamma$ candidate events with $E_{\mathrm{T}}^\gamma>100$ \GeV{}, the expected signal as a function of the aTGC [Eq.~(\ref{equ:natgc})] and the estimated number of background events. 
 A point in the aTGC space is accepted (rejected) at the 95\% C.L. if less (more) than 95\% of the randomly generated pseudo-experiments exhibit larger profile likelihood ratio values than those observed in data. The systematic uncertainties are included in the likelihood function as nuisance parameters with correlated Gaussian constraints, and all nuisance parameters are fluctuated in each pseudoexperiment.

\begin{figure*}
  \centering

  \subfigure[]{\includegraphics[width=0.99\columnwidth]{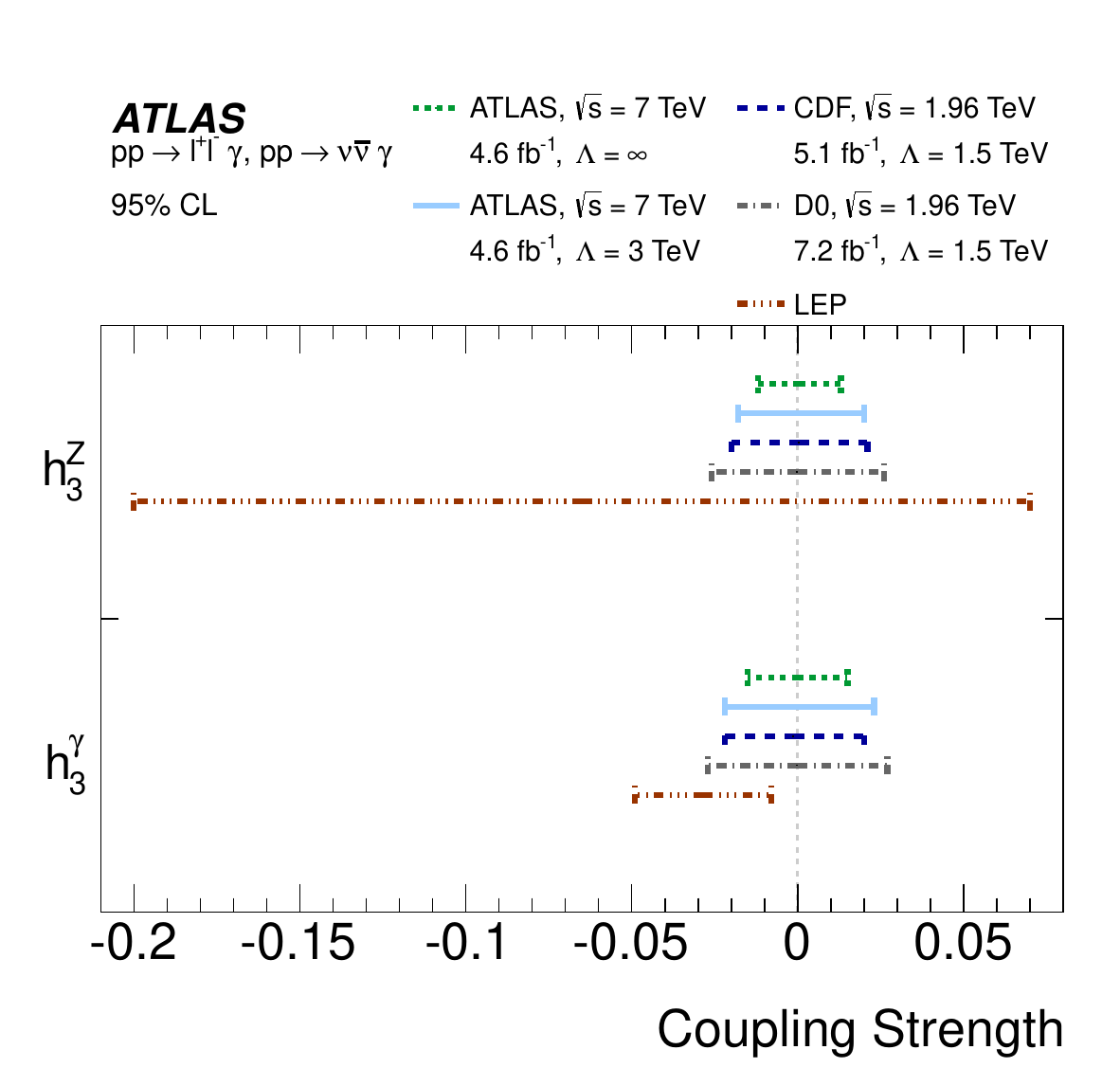}}
  \subfigure[]{\includegraphics[width=0.99\columnwidth]{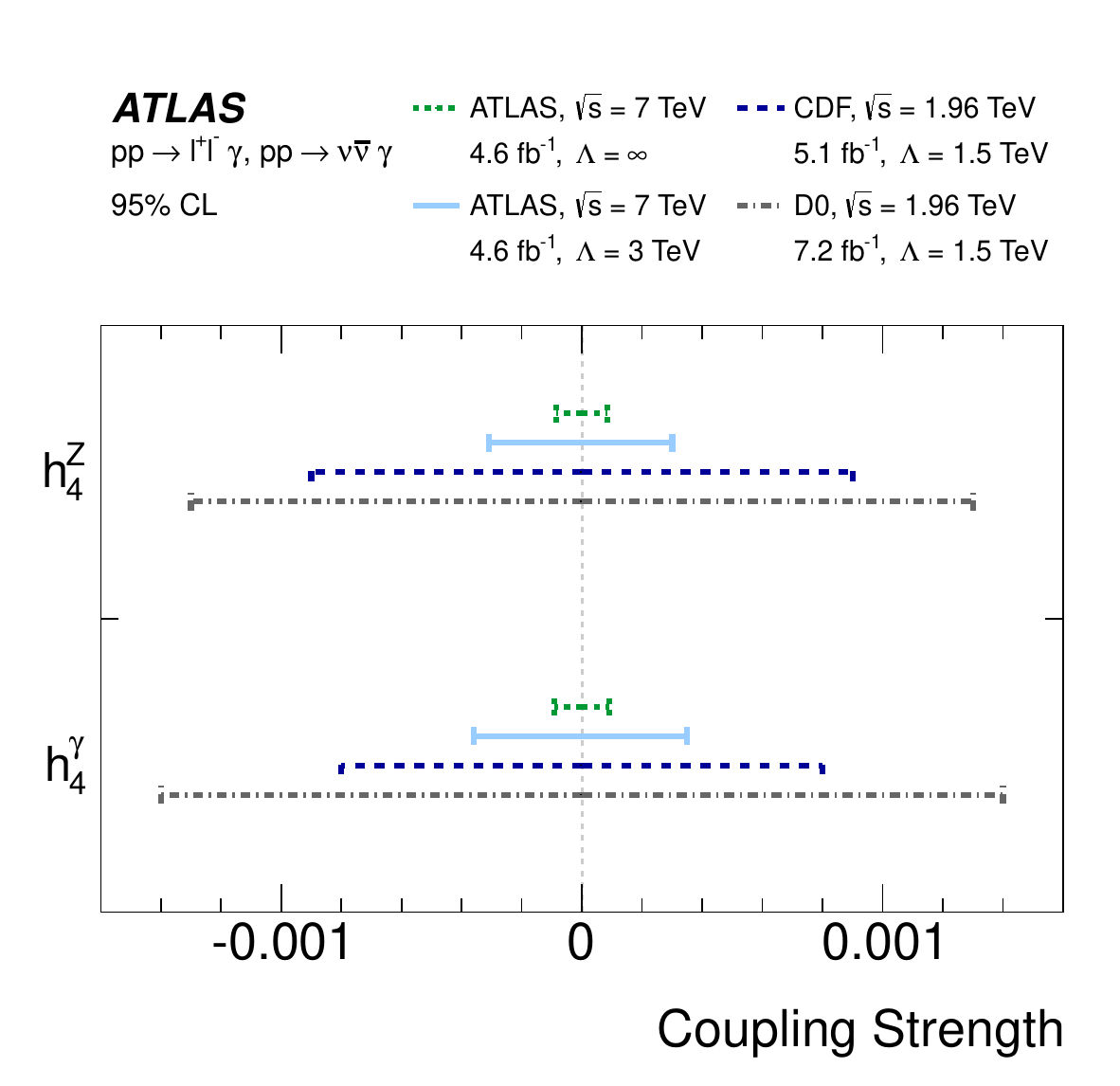}}
  \subfigure[]{\includegraphics[width=0.99\columnwidth]{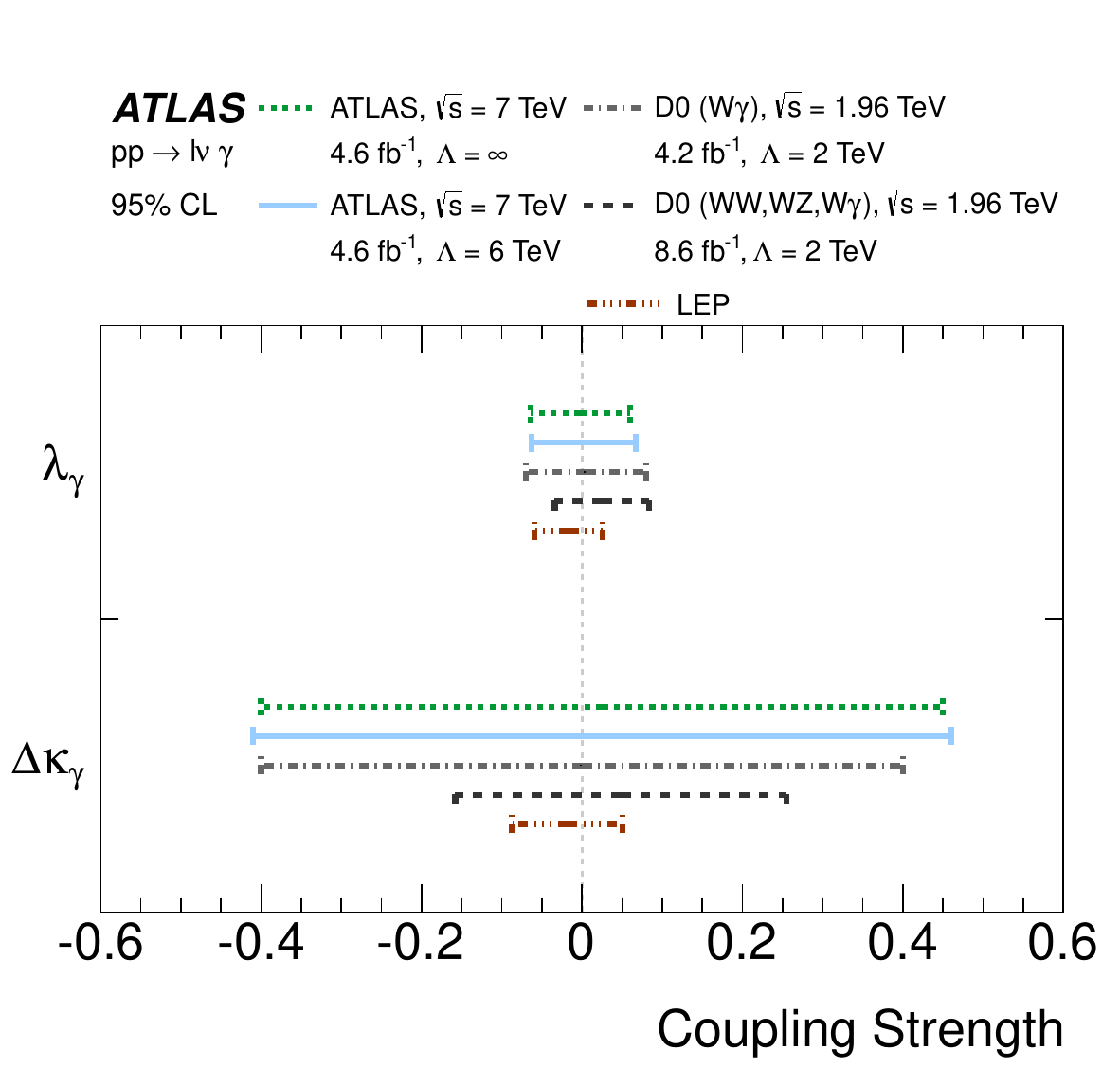}}
  \caption{The 95\% C.L. intervals for anomalous couplings from ATLAS, D0~\cite{D0paper,D0paper2},
CDF~\cite{CDFpaper} and LEP~\cite{LEP} for (a);(b) the neutral aTGCs $h_{3}^{\gamma}$, $h_{3}^{Z}$, $h_{4}^{\gamma}$, $h_{4}^{Z}$ as obtained from $Z\gamma$ events; and (c) the charged aTGCs $\Delta\kappa_{\gamma}$, $\lambda_{\gamma}$. The integrated luminosities and new-physics scale parameter $\Lambda$ are shown. The ATLAS and D0 results for the charged aTGCs measured from $W\gamma$ production are shown. Except for the coupling under study, all other anomalous couplings are set to zero. The LEP charged aTGCs results were obtained from $WW$ production, which is also sensitive to the $WWZ$ couplings and therefore required some assumptions ($\lambda_{Z}=\lambda_{\gamma}$, $\Delta \kappa_{\gamma}= (cos^2\theta_{W}/sin^2\theta_{W})(\Delta g^{Z} - \Delta \kappa_{Z})$) about the relations between the $WW\gamma$ and $WWZ$ aTGCs~\cite{LEP,LEP1,LEP2,LEP3}, but did not require assumptions about the scale $\Lambda$. The combined aTGC results from the D0 experiment are obtained from $WW+WZ \rightarrow \ell\nu jj$, $WW+WZ \rightarrow \ell\nu \ell^+ \ell^-$, $W\gamma \rightarrow \ell\nu\gamma$ , and $WW \rightarrow \ell \nu \ell \nu$ events~\cite{D0comb}. The LEP limits on neutral aTGC's are much larger than those from hadron colliders and are not included in (b).}
  \label{fig:limit}
\end{figure*}

The limits are defined as the values of aTGCs that demarcate the central 95\% of the
integral of the likelihood distribution. The resulting allowed ranges for the anomalous 
couplings are shown in Table~\ref{Tab:wg_atgc} for $WW\gamma$, $ZZ\gamma$ and $Z\gamma\gamma$. These results are also compared in Fig.~\ref{fig:limit} with the results from LEP~\cite{LEP} and the Tevatron~\cite{CDFpaper, D0paper, D0paper2}.

\begin{figure*}[htbp]
  \centering

\subfigure[]{\includegraphics[width=0.99\columnwidth]{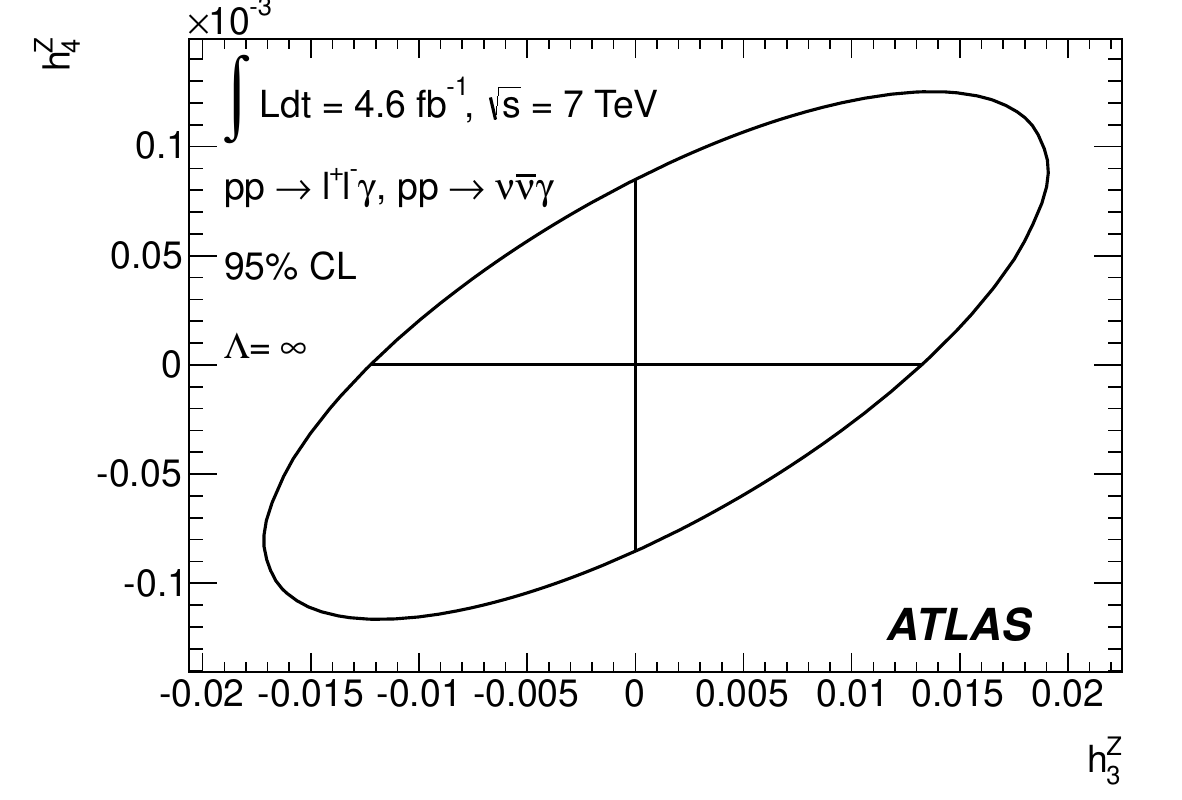}}
\subfigure[]{\includegraphics[width=0.99\columnwidth]{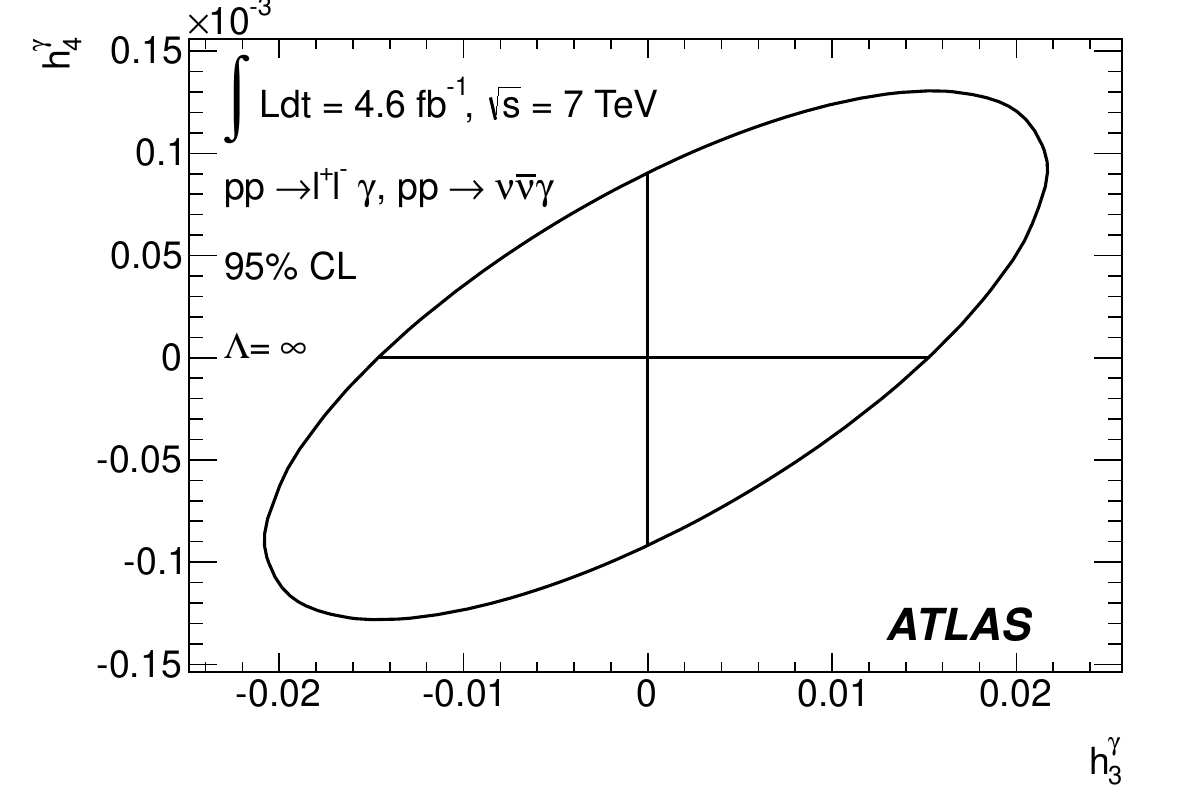}}
\subfigure[]{\includegraphics[width=0.99\columnwidth]{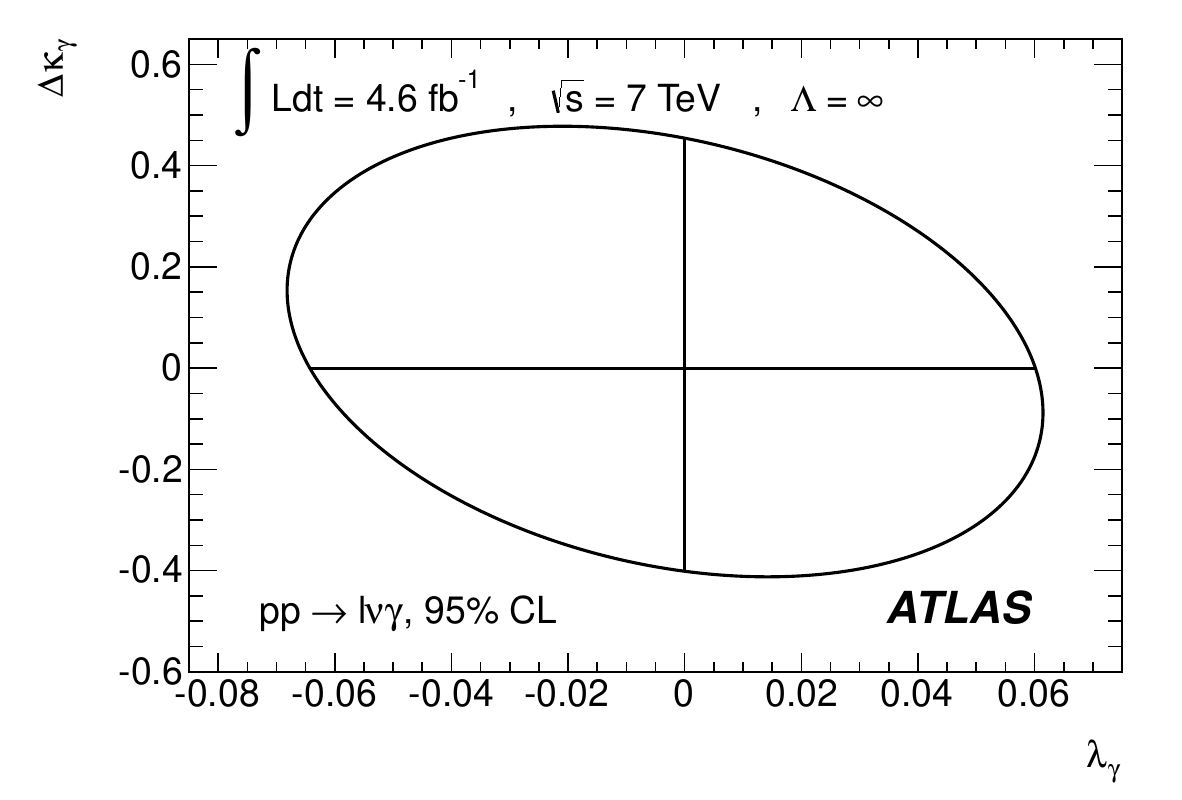}}
  \caption{Observed 95\% C.L. limits shown in the two-parameter planes for pairs of anomalous couplings (a) $h_{3}^{Z}$ and $h_{4}^{Z}$, (b) $h_{3}^{\gamma}$ and $h_{4}^{\gamma}$, and (c) $\Delta k_{\gamma}$ and $\lambda_{\gamma}$, corresponding to an infinite cutoff scale. The horizontal and vertical lines inside each contour correspond to the limits found in the one-parameter fit procedure, and the ellipses indicate the correlations between the one-parameter fits. Since all sensitivity of the measurement is contained in a single measurement of the $V\gamma$ cross sections in the high-$E_{\mathrm{T}}^\gamma$ regions, the likelihood ratio used to obtain the two-dimensional limits has one effective degree of freedom.}
  \label{fig:2Dlimit}
\end{figure*}

The limits on each aTGC parameter are obtained with the other aTGC parameters set to their SM values using a one-dimensional profile likelihood fit. The limits on each pair of aTGC are also evaluated by the same method. The 95\% C.L. regions in two-dimensional aTGC space are shown as contours on the ($\Delta\kappa_\gamma, \lambda_\gamma$), ($h_3^\gamma,h_4^\gamma$) and ($h_3^Z$,$h_4^Z$) planes in Fig.~\ref{fig:2Dlimit}. Since all sensitivity of the measurement is contained in a single measurement of the $V\gamma$ cross sections in the high-$E_{\mathrm{T}}^\gamma$ regions, the likelihood ratio used to obtain the two-dimensional limits has one effective degree of freedom. Therefore the results of the aTGC frequentist limits found in the one-dimensional fit are identical to the corresponding limits obtained from the two-dimensional fits at the points where the other aTGC is zero as shown in Fig.~\ref{fig:2Dlimit}. 
 
\section{Search for narrow resonances}
\label{sec:ExoticSearch}

Models such as technicolor (TC) predict spin-1 mesons that have significant branching ratios to \Wgamma\ and \Zgamma. The discovery of a particle compatible with the SM Higgs~\cite{Higgs:ATLAS,Higgs:CMS} does not exclude the full phase space of the TC models~\cite{Eichten:HiggsImpostor,Sanino:LightHiggs,Sanino:LightHiggs2}. Therefore, they are used here as a benchmark for new physics processes that would appear as new resonant \Wgamma\ and \Zgamma\ states. 

Exotic resonance signals and SM backgrounds are modeled using probability density functions as described below. The model is then fit to the data to test for the presence of new-physics. The electron ($e^{+}e^{-}\gamma$ and $e\nu\gamma$) and muon ($\mu^{+}\mu^{-}\gamma$ and $\mu\nu\gamma$) channels are evaluated independently in the search for \Wgamma\ and \Zgamma\ resonances.

\subsection{Generation and event selection}
The technicolor strawman~\cite{Lane:2002sm} model implemented in {\sc pythia}~\cite{pythia} is used to describe the production and decay of neutral and charged techni-mesons: \tomega{} $\to$ \Zgamma\ and \techa{} $\to$ \Wgamma{}. The following parameters are used in the event generation: number of technicolors $N_{TC}=4$; techni-quark charges $Q_{U}=1$ and $Q_{D}=0$ for the \Zgamma\ final state and $Q_{U}=1/2$ and $Q_{D}=-1/2$ for the \Wgamma\ final state\footnote{This parameterization of the techni-quark charges is used in order to keep only the dominant \techa\ contribution in the \Wgamma\ final state to avoid the model dependence that could result from having two nearby peaks in the signal. In this way the \trho\ contribution is removed.}; mixing angle between the techni-pions and electroweak gauge boson longitudinal component $\sin\chi=1/3$. In addition, the mass splittings between the techni-mesons are set to be as follows: $m_{\rho_{\rm{T}}}=m_{\tomega}$, $m_{\techa}\approx{}1.1\times{}m_{\trho}$ and $m_{\rho_{\rm{T}}}-m_{\tpion}=m_{W}$.

This set of parameters follows those introduced for previous low scale technicolor (LSTC)~\cite{Lane:1999uh,Eichten:2007sx} searches in the $WZ$ and dilepton final states at the Tevatron and at the LHC. Using these parameters, the intrinsic widths of the resonances are of order $1~\GeV$, which is less than the measurement resolution. The results obtained in this study are therefore generic, as long as the resonances studied are narrow. The best limits on techni-meson production have been set 
at the LHC. Studying dilepton final states~\cite{aad:2012kk} in $4.9~\ifb$ of $\sqrt{s}=7~\TeV$ data, the ATLAS experiment excluded at the 95\% C.L. the production of \tomega{} and \trho{} with masses $m_{\trho/\tomega}<855~\GeV$. In the $\Wboson\Zboson$ final state~\cite{Chatrchyan:2012kk}, the CMS collaboration obtained an exclusion $m_{\trho/\tomega}<938~\GeV$ based on $5.0~\ifb$ of $\sqrt{s}=7~\TeV$ data.

The searches for narrow resonances in the \Wgamma\ and \Zgamma\ final states are performed using the event selections defined in Sec.~\ref{sec:eff} but with the photon transverse energy $E^{\gamma}_{\mathrm{T}}$ required to be greater than 40~\GeV . This choice is made to optimize the signal over SM background ratio since the decay products of a heavy resonance would be boosted. In order to keep the results as generic as possible, there is no further optimization of the cuts.

This study uses five mass points for $m_{\tomega}$ ranging from $200$~\GeV{} to $650$~\GeV{} for the \Zgamma\ channel, and seven mass points for $m_{\techa}$ ranging from $275$~\GeV{} to $800$~\GeV{} for the \Wgamma\ channel. The signal samples are produced using the {\sc pythia}~\cite{pythia} generator interfaced to the full ATLAS {\sc geant4}~\cite{geant4} simulation~\cite{ATLASsim} with events reconstructed as for the data. Table~\ref{tab:LSTCcrosssectionsATLAS} summarizes the expected number of events at each mass point after all selection cuts.

\begin{table}[htb]
  \begin{center}
{
    \begin{tabular}{cccc}

      \hline
      
$m_{\tomega}$ $[\GeV]$  & $m_{\techa}$ $[\GeV]$   & \Wgamma\ events &  \Zgamma\ events\\ 

      \hline
      200    & 225 &  -               & $ 47.8 \pm 4.8 $\\ 

      250    & 275 & $ 85.2 \pm 8.4 $ &   -  \\ 
      300    & 330 & $ 58.2 \pm 5.8 $ & $ 16.4 \pm 1.3 $\\ 
      350    & 385 & $ 39.3 \pm 4.1 $ &   - \\ 
      400    & 440 & $ 27.1 \pm 2.2 $ & $ 6.9 \pm 0.4 $\\ 
      450    & 490 & $ 18.9 \pm 1.6 $ &   - \\ 
      500    & 550 & $ 13.6 \pm 1.2 $ & $ 3.4 \pm 0.2 $\\ 

      650    & 720 & -                & $ 1.4 \pm 0.1 $\\ 
      725    & 800 & $ 3.4 \pm 0.3 $  &  - \\ 
      
\hline
    \end{tabular}
 }
 \caption{Expected number of events after all selection cuts for the generated signal mass values. The quoted uncertainties are the combined statistical and systematic uncertainties. The number of expected events is indicated only for points used in the analysis.}
  \label{tab:LSTCcrosssectionsATLAS}
  \end{center}
\end{table}

\subsection{Signal modeling}

For the \tomega{} $\to$ \Zgamma\ channel, the $m^{\Zgamma}$ distribution is fit by the sum of a crystal-ball function (CB)~\cite{CrystalBall1,CrystalBall2,CrystalBall3}, which simulates the core mass resolution plus a non-Gaussian tail for low mass values, and a small wider Gaussian component that takes into account outliers in the mass distribution. The mean values of the CB and Gaussian functions are fixed to be equal. For simulated events, the mean fitted mass is found to be within $0.6~\GeV$ of the generated resonance mass for both the $\mu^{+}\mu^{-}\gamma$ and $e^{+}e^{-}\gamma$ channels. At the reconstruction level, the full width at half maximum of the signal grows linearly from $9~\GeV$ at $m_{\tomega}=200~\GeV$ to $30~\GeV$ at $m_{\tomega}=650~\GeV$. In order to scan for resonance signals in the data, $m^{\Zgamma}$ mass distributions are constructed in $5~\GeV$ steps from 200 to $650~\GeV$ by linearly interpolating the signal line shape fit parameters for the $m^{\Zgamma}$ distributions. 

For the \techa{} $\to$ \Wgamma\ channel, the $m_{\mathrm{T}}^{\Wgamma}$ distribution is fit by a CB function. The mean value of the distribution is measured to be lower than the generated mass of the resonance as expected for the transverse mass. The signal resolution grows linearly from $20~\GeV$ at $m_{\techa}=275~\GeV$ to about $35~\GeV$ at $m_{\techa}=800~\GeV$. The data are scanned for \techa{} $\to$ \Wgamma\ resonance signals using $m_{\mathrm{T}}^{\Wgamma}$ mass distributions constructed in $10~\GeV$ steps from 275 to $800~\GeV$ by linearly interpolating the signal line shape fit parameters for the $m_{\mathrm{T}}^{\Wgamma}$ distributions.

\subsection{Background modeling}

The background estimations for the $V\gamma$ resonance searches use the techniques described in Sec.\ref{sec:background}. The distributions of the SM predictions and the data after the event selection cuts are shown in Figs.~\ref{fig:AllMassDistributionsW}(a) and (b) for $m_{\mathrm{T}}^{e\nu\gamma}$ and $m_{\mathrm{T}}^{\mu\nu\gamma}$, and in Figs.~\ref{fig:AllMassDistributionsZ}(a) and (b) for $m^{e^{+}e^{-}\gamma}$ and $m^{\mu^{+}\mu^{-}\gamma}$. All the mass distributions have a broad maximum at about 150 GeV. Since the search for a resonant structure on top of a peaking background is more complex than on a falling distribution, the search is conducted only on the tails of the $m^{\Zgamma}$ and $m_{\mathrm{T}}^{\Wgamma}$ mass distributions for masses larger than $180~\GeV$.

\begin{figure*}[htbp]
  \centering
\subfigure[]{\includegraphics[trim=35 20 60 5,clip=true,width=1.\columnwidth]{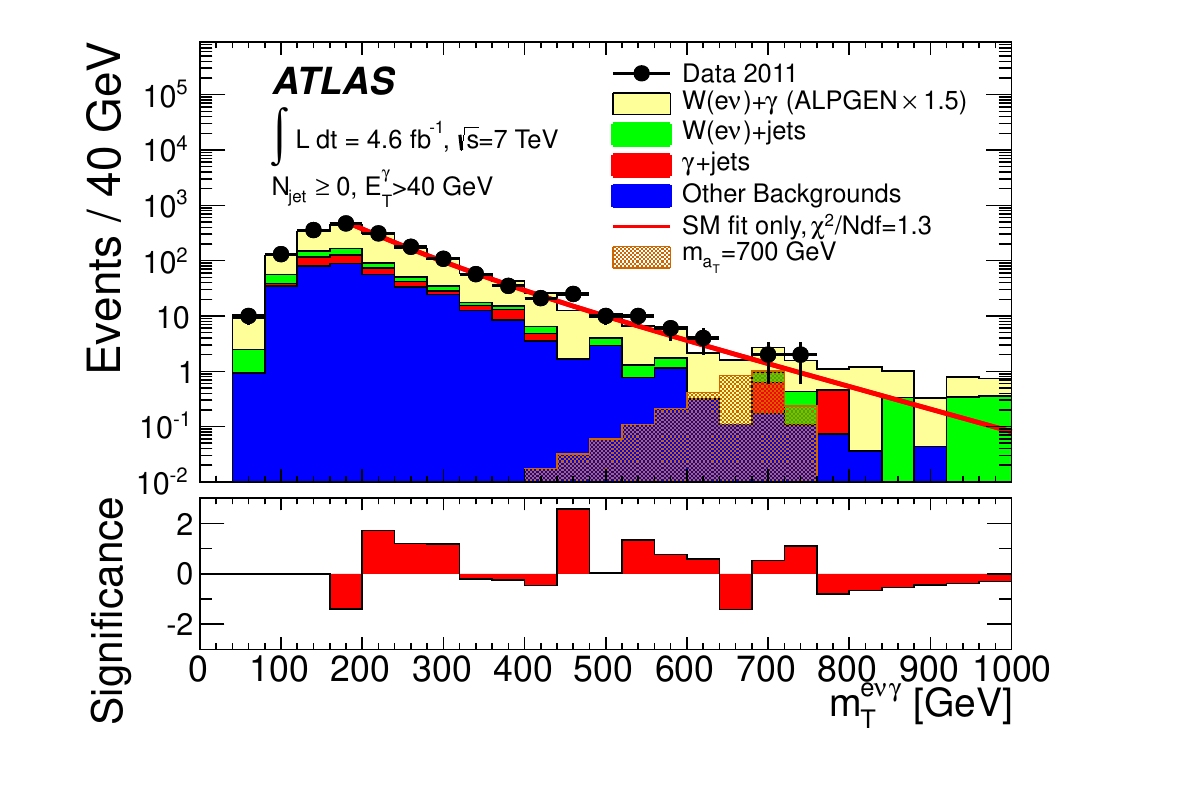}}
\subfigure[]{\includegraphics[trim=35 20 60 5,clip=true,width=1.\columnwidth]{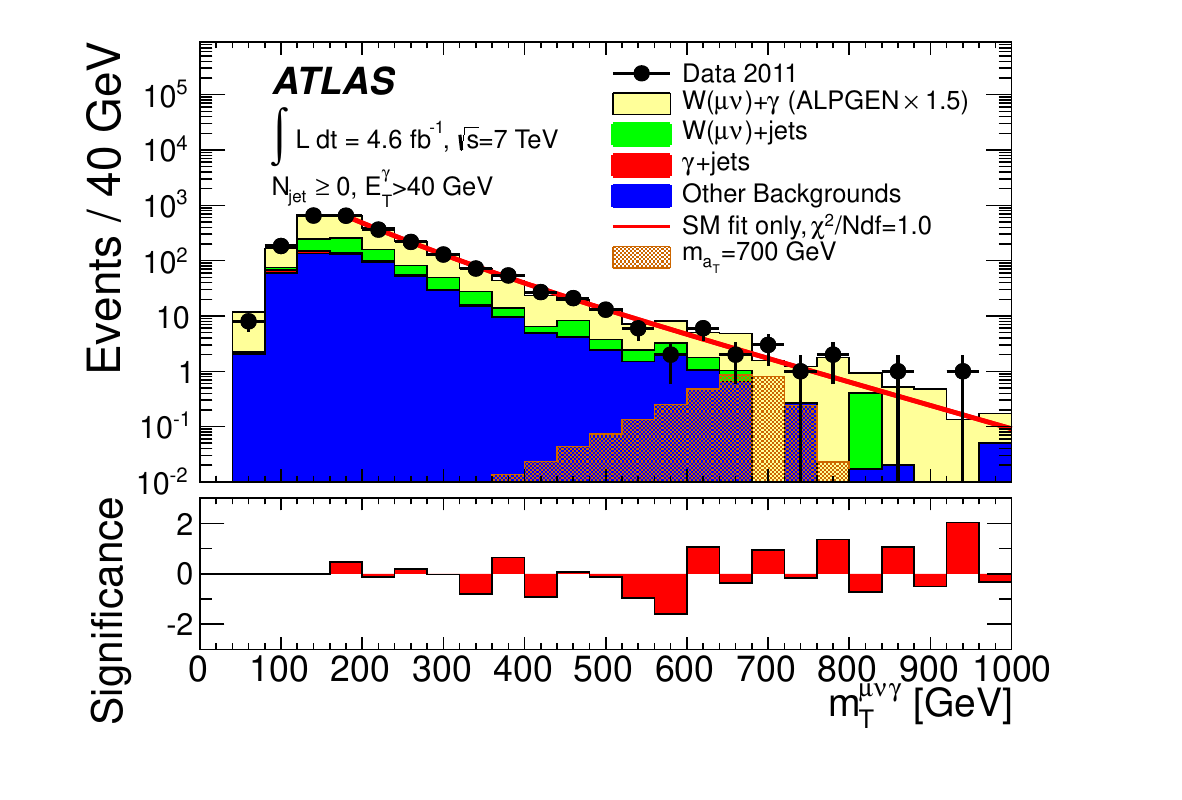}}
  \caption{Three-body transverse mass for (a) the $e\nu\gamma$ and (b) $\mu\nu\gamma$ final states. The background-only fit to the data is shown. The significance quoted is defined as $\frac{(D-B)}{\sqrt{B}}$, where D is the number of data events and B the number of predicted events by the fit in the bin considered. The background distributions for the expected $\ell \nu\gamma$ events are taken from the MC simulation (generated with {\sc alpgen}) and normalized to the extracted number of $\ell \nu\gamma$ events. The signal near the limit at $m_{\techa}=700~\GeV$ is also shown.}
  \label{fig:AllMassDistributionsW}
\end{figure*}

\begin{figure*}[htbp]
  \centering
\subfigure[]{\includegraphics[trim=35 20 60 5,clip=true,width=1.\columnwidth]{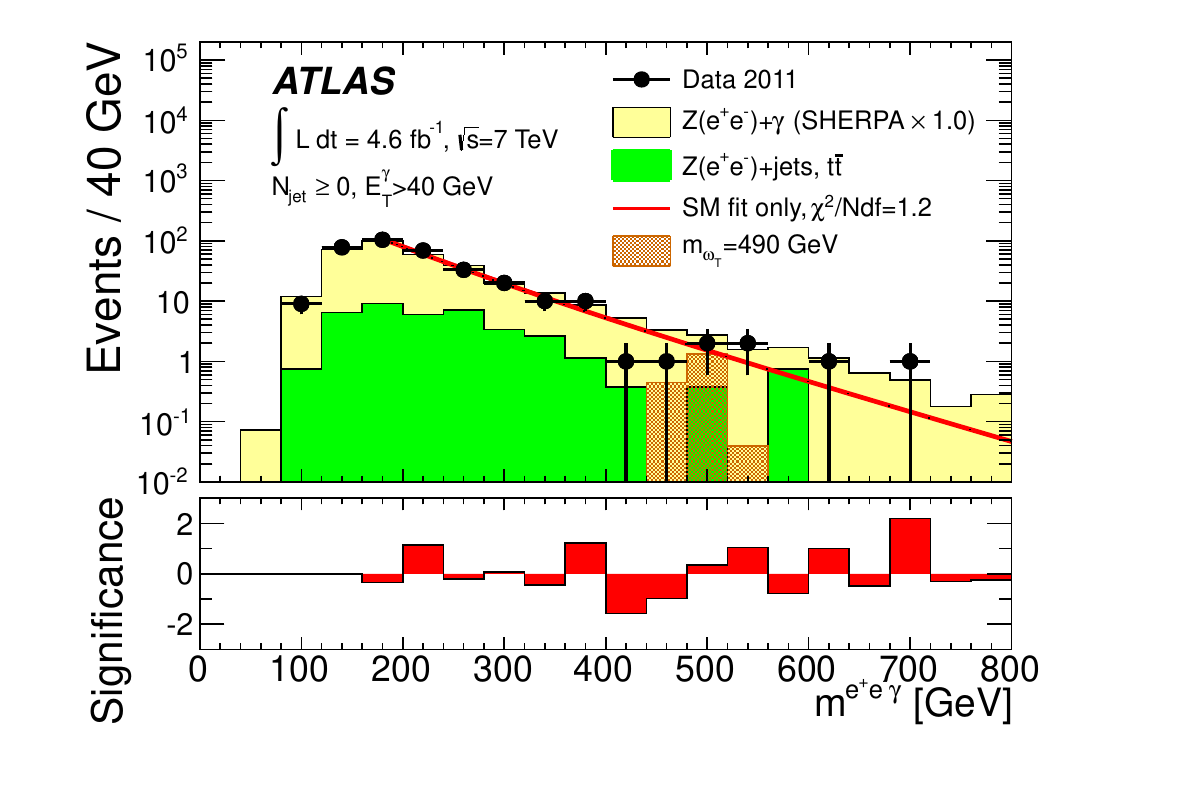}}
\subfigure[]{\includegraphics[trim=35 20 60 5,clip=true,width=1.\columnwidth]{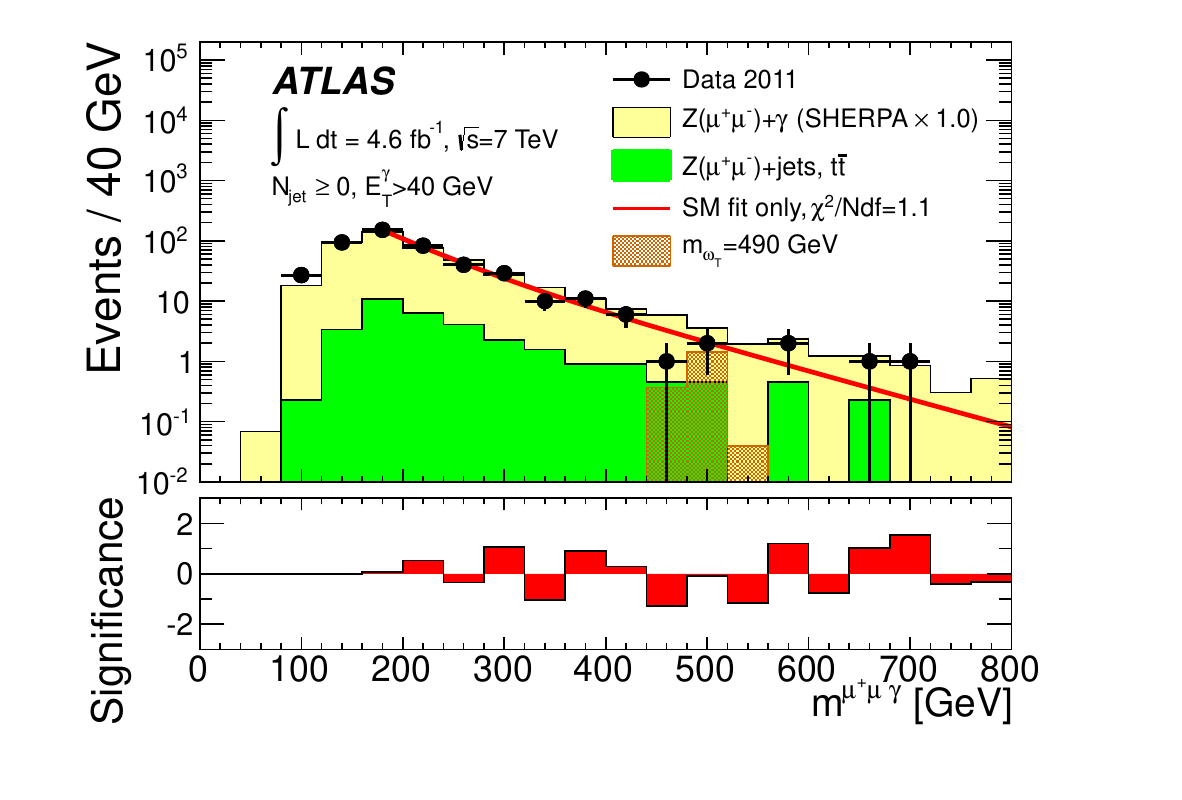}}
  \caption{Three-body invariant mass for (a) the $e^{+}e^{-}\gamma$ and (b) $\mu^{+}\mu^{-}\gamma$ final states. The background-only fit to the data is shown. The significance quoted is defined as $\frac{(D-B)}{\sqrt{B}}$, where D is the number of data events and B the number of predicted events by the fit in the bin considered. The background distributions for the expected $\ell^{+}\ell^{-}\gamma$ events are taken from the MC simulation (generated with {\sc sherpa}) and normalized to the extracted number of $\ell^{+}\ell^{-}\gamma$ events. The signal near the limit at $m_{\tomega}=490~\GeV$ is also shown.}

  \label{fig:AllMassDistributionsZ}
\end{figure*}

A blinded search is conducted in the signal region.
Agreement between the data and the Monte Carlo modeling is checked in two control regions
for each final state. One control region is obtained by reversing the cut on the photon transverse energy ($E_{T}^{\gamma}<40~\GeV$), and the other one by reversing the cut ($<170~\GeV$) on the discriminating variable (i.e.: $m^{\Zgamma}$ for the \Zgamma\ channel
and  $m_{\mathrm{T}}^{\Wgamma}$ for the \Wgamma\ channel). Good agreement is found
between data and Monte Carlo samples in these control regions.

A probability density function is created to describe the SM background in the signal region. This approach has two advantages. The shape of the SM background is taken directly from the sidebands of the fit. The probability density function obtained is also less sensitive to statistical fluctuations in the tail than techniques relying on Monte Carlo templates. For both the \Wgamma\ and \Zgamma\ channels, the overall shape of the SM background in the signal region is due to the sum of components with different shapes. A double-exponential function provides the best model in the signal region:

\begin{equation}\label{eq:dbleexpo}
f_{{bkg}}=N_{\rm bkg}\times(e^{\alpha_{1} \cdot {}m} + e^{\alpha_{2} \cdot {}m}).
\end{equation}

The background model is tested against a $1~\ifb$ data sample that has been previously analyzed~\cite{WZg1fbpaper}. In addition it is tested with the nominal Monte Carlo distribution and the shape of the Monte Carlo distribution obtained by varying the background composition within systematic uncertainties. The double-exponential function is found to reproduce the shapes of all these distributions properly, and is therefore used for the SM background estimation. The results of the unbinned fit to the data can be seen as the solid curve on each of the mass spectra in Figs.~\ref{fig:AllMassDistributionsW} and~\ref{fig:AllMassDistributionsZ}. The $\chi^{2}$ per degree of freedom  obtained for the background-only fit is close to unity for all the distributions.

\subsection{Fit model and statistical methods}

The normalization, $N_{\rm bkg}$, and the two exponential coefficients, $\alpha_{1}$ and $\alpha_{2}$, in the SM background probability density function [Eq.(\ref{eq:dbleexpo})] are all free to vary.  Another term takes into account a systematic uncertainty on the background shape. In order to ensure there are enough events in the sidebands on each side of the distribution, the SM background fit is performed in the range $[180,800]~\GeV$ for the $m^{\Zgamma}$ distribution and the range $[180,1000]~\GeV$ for the  $m_{\mathrm{T}}^{\Wgamma}$ distribution.
For both the data and the pseudodata experiments, a maximum log-likelihood method is used to fit the SM background probability density function to the observed event distribution.

The parameters of the signal probability density functions are all fixed to their nominal values, except the normalization of the signal and two nuisance terms that account for systematic uncertainty on the signal event rate and resolution as explained below. The search is conducted by scanning the $m^{\Zgamma}$ and  $m_{\mathrm{T}}^{\Wgamma}$ distributions every $5~\GeV$ for the \Zgamma\ channel and every $10~\GeV$ for the \Wgamma{} channel using the signal template.
The normalization of the signal is fit according to the equation
\begin{equation}\label{eq:fitproc}
 N_{S}=\sigma_{\mbox{\tiny{Fid}}}\times\epsilon_{\mbox{\tiny{Reco}}}\times{}\int{}{\cal L}dt
\end{equation}
where the factor $\sigma_{\mbox{\tiny{Fid}}}$, the signal fiducial cross section, is the only free parameter in this equation. The factor $\epsilon_{\mbox{\tiny{Reco}}}$ is the signal reconstruction efficiency\footnote{$\epsilon_{\mbox{\tiny{Reco}}}$ contains both the acceptance $A_{W\gamma(Z\gamma)}$ and the correction factor $C_{W \gamma(Z\gamma)}$ in Eq.(\ref{Equ:cs_fid}).}, defined as the number of signal events passing the detector simulation and the full event selection divided
by the number of events generated in the extended fiducial volume defined in Table~\ref{tab:fiducialcut} but applying $E^{\gamma}_{\mathrm{T}}>40~\GeV$. The factor $\epsilon_{\mbox{\tiny{Reco}}}$ accounts for the selection efficiency for signal events generated within the fiducial region. It includes, for example, effects due to the detector resolution on the lepton and photon transverse momentum and energies, and on the missing transverse energy. The normalization of the signal is determined simultaneously in the electron and muon samples for the combination. The results obtained are therefore less sensitive to statistical fluctuations in a given channel.\\

The parameter of interest used in this analysis is the fiducial cross section of an eventual new-physics signal.  $\sigma_{\mbox{\tiny{Fid}}}$ is scanned to check the compatibility of the data with a background-only or a signal-plus-background hypothesis. 

The statistical test used is based on the profile likelihood ratio~\cite{PCLpaper} $L(\sigma_{\mbox{\tiny{Fid}}})$, to test different hypothesized values of $\sigma_{\mbox{\tiny{Fid}}}$. $L(\sigma_{\mbox{\tiny{Fid}}})$ is built from the likelihood function describing the probability density function of $m^{\Zgamma}$ and $m_{T}^{\Wgamma}$ under a signal-plus-background hypothesis and the systematic uncertainties. It combines both electron and muon final states. The statistical tests are then performed on the $m_{T}^{\Wgamma}$ and  $m^{\Zgamma}$ distributions.

The data are interpreted using a modified frequentist approach ($CL_{s}$)~\cite{CLpaper} for setting limits. A fiducial cross section is claimed to be excluded at 95\% C.L. when $CL_{s}$ is less than 0.05. The probability of the background-only hypothesis, or $p_{0}$ value, is obtained using a frequentist approach. The latter gives the probability that the background fluctuates to the observed number of events or above.\\

\subsection{Systematic uncertainties}

Systematic uncertainties on the signal resonances are taken into account as nuisance parameters in the likelihood function used for the signal-plus-background model. Two different effects are evaluated for each source of systematic uncertainty, one for the signal event rate and one for the resolution of the signal. Each systematic effect is investigated by propagating the corresponding uncertainty to the signal sample. These are computed separately for each of the simulated resonance mass points. The four categories of systematic uncertainties and their impacts on the resonant signals are summarized below for $m_{\tomega}=300~\GeV$ in the \Zgamma\ channel and $m_{\techa}=330~\GeV$ in the \Wgamma\ channel.

The systematic effects due to the photon isolation, identification, energy resolution, and energy scale are considered. The impact of the photon geometric position in the detector on the peak resolution is also investigated to account for differences that could arise from changes of the photon pseudorapidity distribution in different theoretical models. The impact of this effect is minor and found to be about $0.2~\GeV$. The systematic uncertainties due to the photon reconstruction and identification contribute most to the systematic uncertainties on the signal. The total effect on the event rate is measured to be $5.7\%$ in all the channels and contributes about $0.5~\GeV$ to the systematic uncertainty on the resolution of the central mass of the resonance.

The systematic effects due to the electron energy resolution and electron energy scale are treated as fully correlated with the photon energy scale and resolution in the final states containing electrons ($e^{+}e^{-}\gamma$ and $e\nu\gamma$). The effects of the muon energy scale and muon energy resolution, lepton identification and trigger efficiency are also investigated. The total effect of the lepton reconstruction and identification on the signal event rate is about $1.8\%$ in the muon channels and about $1.2\%$ in the electron channels. The effect on the peak resolution is only about $0.2~\GeV$.

Systematic effects due to the jet energy scale and resolution and the calibration of the missing transverse energy impact only the \Wgamma\ channel. These are found to cause uncertainties in the event rate of about $1\%$ and on the peak resolution of about $1~\GeV$.

Finally there is a systematic uncertainty on the resonance production rate due to the $3.9\%$ uncertainty on the integrated luminosity~\cite{ATLASLumi1}. 

The effects of all the systematic uncertainties are combined in quadrature. The total systematic uncertainty on the event rate is found to be approximately $7\%$ for all the mass points in the two channels. The systematic uncertainty on the peak resolution is found to be approximately $1 (2)$~\GeV{} for the \Zgamma\ channel at $m_{\tomega}=300 (650)~\GeV$ and $1.5 (3)$~\GeV{} at $m_{\techa}=330 (800)$~\GeV{} in the \Wgamma\ channel.

Since the SM backgrounds are determined using a sideband fit to the data, uncertainties in the detector resolution and physics object reconstruction or identification have a negligible effect on the background in this analysis.
 However, a systematic effect from the background modeling is investigated. The method considered consists of generating background-only pseudoexperiments, and fitting each pseudodataset with the signal-plus-background model to measure a residual signal strength. 
For each final state, 1000 background-only pseudoexperiment samples are generated with the expected number of SM background events. For each pseudo-experiment, the signal-plus-background model is fit in steps of $\Delta{}m_{\tomega}=1~\GeV$ for the \Zgamma\ channel and $\Delta{}m_{\techa}=1~\GeV$ for the \Wgamma\ channel to measure a residual signal strength. For each mass point the mean value of the fitted strength is measured. If there is no bias in the fit model, this distribution should be centered exactly at 0. Since this is not the case, the systematic uncertainty on the background shape is taken to be the difference between 0 and the most discrepant fitted strength obtained anywhere in the mass range, augmented by the 1$\sigma$ uncertainty on that fitted strength. The size of this effect is measured to be 0.2~fb{} for the \Zgamma\ analysis. This represents about $5\%$ on the limit at low masses and up to $20\%$ at high masses. It is measured to be 1.2~fb{} for the \Wgamma\ analysis, which represents about $6\%$ on the limit at low masses and up to $25\%$ at high masses. This dominant systematic effect is taken into account in the fit model, by allowing the backgrounds to fluctuate like the signal, but constrained by these values.

Finally, systematic effects are evaluated on the signal theoretical cross sections due to the limited knowledge of the proton PDFs and the energy scale of the process. These are computed by comparing predictions of the nominal LO PDF set {\sc mrst2007}~\cite{pdfmrst2007} to the $68\%$ C.L. error set of the {\sc mstw2008}~\cite{pdfmstw} PDF sets using the {\sc lhapdf} framework~\cite{lhapdf}. The deviation of the predictions from the central value are added in quadrature and taken to be the size of the uncertainty. The magnitude of the PDF uncertainties on the cross sections is about $3\%$ for the $\Zgamma$ channel and $5\%$ for the $\Wgamma$ channel.

\subsection{Results}

The reconstruction efficiencies, $\epsilon_{\mbox{\tiny{Reco}}}$, and the expected and observed limits on the fiducial cross section times branching ratio for the \tomega{} $\to$ \Zgamma\ and
\techa{} $\to$ \Wgamma\ resonance signals are summarized in Tables~\ref{tab:LSTCAcceptanceZg} and~\ref{tab:LSTCAcceptanceWg}, respectively. The efficiencies are relatively flat versus the mass of the resonances.

The search is used to set $95\%$ C.L. limits on the production of techni-mesons. Figure~\ref{fig:AllLimitsLSTC}~(a) shows the expected and observed limits obtained for \tomega{} $\to$ \Zgamma{}.  The two largest deviations are observed at $m_{\tomega}=465~\GeV$ where a downward fluctuation is seen with a pvalue of $p_{0}\approx{}0.01$ or a local significance of $2.7\sigma$ and at $m_{\tomega}=205~\GeV$ where an upward fluctuation is seen with a pvalue of $p_{0}\approx{}0.02$ or a local significance of $2.4\sigma$. In the \Zgamma\ channel the expected mass limit on the LSTC production of \tomega{} is $m_{\tomega}=483~\GeV$, while the observed limit is $m_{\tomega}=494~\GeV$.

Figure~\ref{fig:AllLimitsLSTC}~(b) shows the expected and observed limits obtained for \techa{} $\to$ \Wgamma\ . The largest deviation is observed at $m_{\techa}=285~\GeV$ where an upward fluctuation is recorded with a pvalue of $p_{0}\approx{}0.05$ or a local significance of $2.0\sigma$. In the \Wgamma\ channel the expected mass limit on the LSTC production of \techa{} is $m_{\techa}=619~\GeV$, while the observed limit is $m_{\techa}=703~\GeV$.

These results are similar to those from previous searches for LSTC~\cite{aad:2012kk,Chatrchyan:2012kk} in other channels. They are more stringent than previous limits from vector resonance searches~\cite{D0paperExoSearch} in the \Zgamma\ final state and they are the first limits to be set from single resonance searches in the $W\gamma$ channel.

\begin{table}[htb]
  \begin{center}
{\footnotesize
    \begin{tabular}{cccc}
      \hline
      
  $m_{\tomega}$ [\GeV]      &  $\epsilon_{\mbox{\tiny{Reco}}}$  &  Expected (fb)       &  Observed (fb)        \\    
      \hline

      200    & $0.52 \pm 0.05$  & $4.3^{+1.9}_{-1.3}$ & 8.3               \\ 
      300    & $0.54 \pm 0.05$  & $2.6^{+1.2}_{-0.8}$ & 2.2                \\ 
      400    & $0.54 \pm 0.05$  & $1.8^{+0.9}_{-0.6}$ & 2.2               \\ 
      500    & $0.55 \pm 0.05$  & $1.4^{+0.7}_{-0.5}$ & 1.5                \\ 
      650    & $0.57 \pm 0.05$  & $1.0^{+0.6}_{-0.3}$ & 0.9                \\ 

      \hline

    \end{tabular}
 }
 \caption{Reconstruction efficiency in the extended fiducial volume as defined in Table~\ref{tab:fiducialcut} but for $E^{\gamma}_{\mathrm{T}}>40~\GeV$. The expected and observed $95\%$ C.L. upper limits for the different signal points in the \Zgamma\ final states are given.}
  \label{tab:LSTCAcceptanceZg}
  \end{center}
\end{table}

\begin{table}[htb]
  \begin{center}
{\footnotesize
    \begin{tabular}{cccc}
      \hline

      $m_{\techa}$ [\GeV]      &  $\epsilon_{\mbox{\tiny{Reco}}}$  &  Expected (fb)                          &  Observed (fb)        \\    \hline

      275    & $0.45 \pm 0.04$  & $16.8^{+7.1}_{-4.9}$  & 31.9                \\ 
      330    & $0.43 \pm 0.04$  & $13.4^{+5.7}_{-3.9}$  & 9.8                 \\ 
      385    & $0.42 \pm 0.04$  & $10.0^{+4.3}_{-2.9}$  & 7.2                 \\ 
      440    & $0.41 \pm 0.04$  & $8.0^{+3.4}_{-2.3}$   & 8.1                 \\ 
      490    & $0.41 \pm 0.04$  & $6.9^{+3.0}_{-2.0}$   & 8.9                 \\ 
      550    & $0.40 \pm 0.04$  & $5.9^{+2.5}_{-1.7}$   & 6.1                 \\ 
      800    & $0.39 \pm 0.04$  & $3.3^{+1.5}_{-0.9}$   & 2.5                 \\

\hline
    \end{tabular}
 }
 \caption{Reconstruction efficiency in the extended fiducial volume as defined in Table~\ref{tab:fiducialcut} but for $E^{\gamma}_{\mathrm{T}}>40~\GeV$. The expected and observed $95\%$ C.L. upper limits are given for the different signal points in the \Wgamma\ final states. }
  \label{tab:LSTCAcceptanceWg}
  \end{center}
\end{table}

\begin{figure*}[htbp]
  \centering
\subfigure[]{\includegraphics[width=1.\columnwidth]{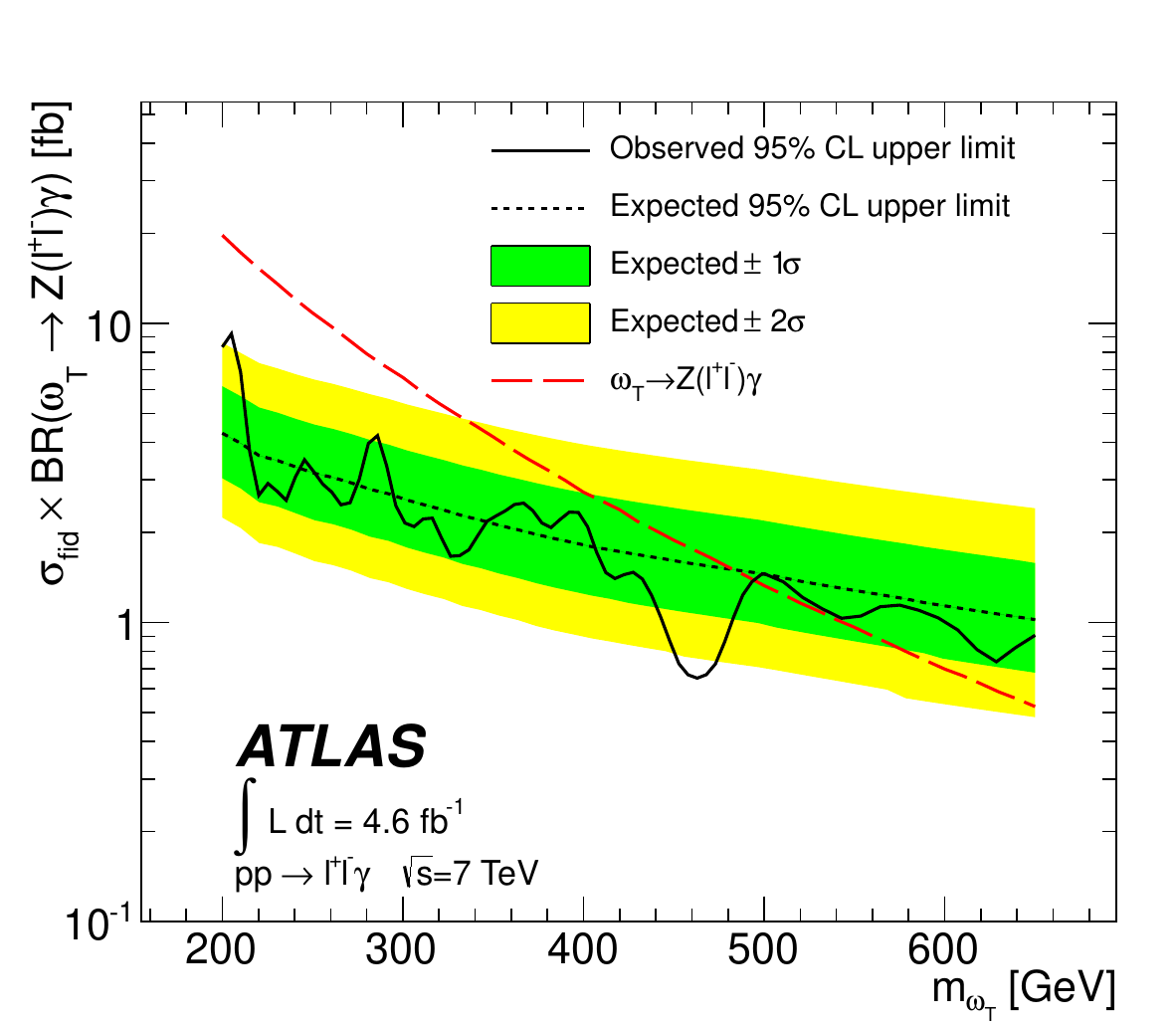}}
\subfigure[]{\includegraphics[width=1.\columnwidth]{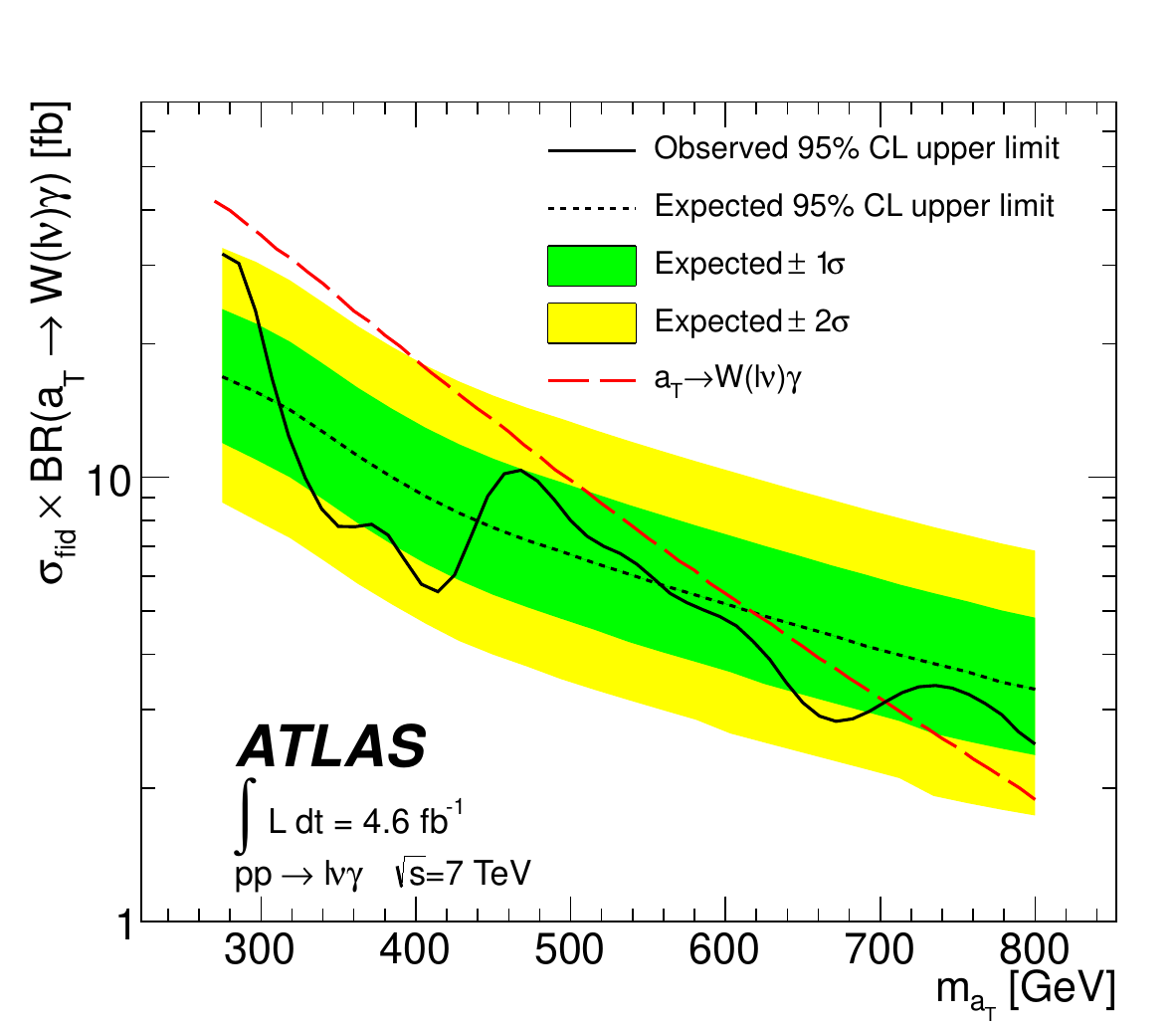}}
  \caption{$95\%$ C.L. limits on narrow vector resonance production obtained using $\mathcal{L}=4.6~\ifb{}$ of data for (a) the $pp \to \ell^{+}\ell^{-}\gamma$ final state and for (b) the $pp \to \ell\nu\gamma$ final state. The parameterization of LSTC~\cite{Lane:1999uh,Eichten:2007sx} used to benchmark the observed limit is obtained using $m_{\trho}=m_{W}-m_{\tpion}$.}
  \label{fig:AllLimitsLSTC}
\end{figure*}

\section{Summary}
\label{sec:summary}

The production of $W\gamma$ and $Z\gamma$ boson pairs in 7 TeV{}
$pp$ collisions is studied using 4.6 fb$^{-1}$ of data
collected with the ATLAS detector. The measurements are made using the leptonic decays of the $W$ and $Z$ bosons ($W(e\nu,\mu\nu)$ and $Z(e^{+}e^{-},\mu^{+}\mu^{-},\nu\bar{\nu})$) with associated high-energy isolated photons.

The results are compared to SM predictions using the NLO parton-level generator {\sc mcfm}. 
In general, the NLO SM predictions for the exclusive $W\gamma$ and $Z\gamma$ production cross sections agree with measurements. However, as the photon $E_{\mathrm{T}}^\gamma$ threshold is raised for inclusive $pp \to \ell\nu\gamma$ production, the associated jet multiplicity increases and there are disagreements with the NLO predictions, which do not include multiple quark/gluon emission.
The measurements are also compared to LO MC generators ({\sc algpen} or {\sc sherpa}) with multiple quark/gluon emission in the matrix element calculations. These LO MC predictions reproduce the shape of the photon $E_{\mathrm{T}}^\gamma$ spectrum and the kinematic properties of the leptons and jets in the $W\gamma$ and $Z\gamma$ measurements.

The measurements of exclusive $W\gamma$ and $Z\gamma$ production with $E_{\mathrm{T}}^\gamma >$ 100 \GeV{} are used to constrain anomalous triple-gauge-boson couplings ($\lambda_{\gamma}$, $\Delta$$\kappa_{\gamma}$, $h^{V}_{3}$ and $h^{V}_{4}$). They are also used to search for narrow resonances in the $V+\gamma$ final state with $E_{\mathrm{T}}^\gamma >$ 40 \GeV{} and compared to low scale technicolor models. No evidence for physics beyond the SM is observed. The limits obtained from this study of anomalous triple-gauge-boson couplings improve on previous LHC and Tevatron results. The results of the vector resonance search are the first ones reported for the study of the $W\gamma$ final state and the most stringent in the $Z\gamma$ final state. Using the LSTC benchmark model, the production of $\techa{}$ is excluded up to $m_{\techa}=703~\GeV$ in the $W\gamma$ mode and the production of $\tomega{}$ is excluded up to $m_{\tomega}=494~\GeV$ in the $\Zgamma$ channel.

\section{Acknowledgements}

We thank CERN for the very successful operation of the LHC, as well as the
support staff from our institutions without whom ATLAS could not be
operated efficiently.

We acknowledge the support of ANPCyT, Argentina; YerPhI, Armenia; ARC,
Australia; BMWF and FWF, Austria; ANAS, Azerbaijan; SSTC, Belarus; CNPq and FAPESP,
Brazil; NSERC, NRC and CFI, Canada; CERN; CONICYT, Chile; CAS, MOST and NSFC,
China; COLCIENCIAS, Colombia; MSMT CR, MPO CR and VSC CR, Czech Republic;
DNRF, DNSRC and Lundbeck Foundation, Denmark; EPLANET, ERC and NSRF, European Union;
IN2P3-CNRS, CEA-DSM/IRFU, France; GNSF, Georgia; BMBF, DFG, HGF, MPG and AvH
Foundation, Germany; GSRT and NSRF, Greece; ISF, MINERVA, GIF, DIP and Benoziyo Center,
Israel; INFN, Italy; MEXT and JSPS, Japan; CNRST, Morocco; FOM and NWO,
Netherlands; BRF and RCN, Norway; MNiSW, Poland; GRICES and FCT, Portugal; MERYS
(MECTS), Romania; MES of Russia and ROSATOM, Russian Federation; JINR; MSTD,
Serbia; MSSR, Slovakia; ARRS and MVZT, Slovenia; DST/NRF, South Africa;
MICINN, Spain; SRC and Wallenberg Foundation, Sweden; SER, SNSF and Cantons of
Bern and Geneva, Switzerland; NSC, Taiwan; TAEK, Turkey; STFC, the Royal
Society and Leverhulme Trust, United Kingdom; DOE and NSF, United States of
America.

The crucial computing support from all WLCG partners is acknowledged
gratefully, in particular from CERN and the ATLAS Tier-1 facilities at
TRIUMF (Canada), NDGF (Denmark, Norway, Sweden), CC-IN2P3 (France),
KIT/GridKA (Germany), INFN-CNAF (Italy), NL-T1 (Netherlands), PIC (Spain),
ASGC (Taiwan), RAL (UK) and BNL (USA) and in the Tier-2 facilities
worldwide.

\bibliographystyle{atlasnote}
\bibliography{prd_wzgamma2011}

\providecommand{\href}[2]{#2}\begingroup\raggedright\begin{thebibliography}{10}


\bibitem{Weinberg:1979bn}
S.~Weinberg, Phys. Rev. D {\bf 19}, 1277 (1979).

\bibitem{Susskind:1978ms}
L.~Susskind, Phys. Rev. D {\bf 20}, 2619 (1979).


\bibitem{DetectorPaper:2008}
{ATLAS} Collaboration, JINST {\bf 3}, S08003 (2008).


\bibitem{CDFpaper}
F.~Abe{~\em et al.}, {CDF} Collaboration, Phys. Rev. D {\bf 107}, 051802 (2011).  
\bibitem{D0paper}
V.~Abazov{~\em et al.}, {D0} Collaboration, Phys. Rev. Lett. {\bf 100}, 241805 (2008)  .

\bibitem{D0paper2}
V.~Abazov{~\em et al.}, {D0} Collaboration, Phys. Rev. D {\bf 85}, 052001 (2012).

\bibitem{ATLASpaper}
{ATLAS} Collaboration, JHEP {\bf 09}, 072 (2011).


\bibitem{WZg1fbpaper}
{ATLAS} Collaboration, Phys. Lett. B {\bf 717}, 49 (2012).

\bibitem{CMSpaper}
{CMS} Collaboration, Phys. Lett. B {\bf 701}, 535 (2011).

\bibitem{D0paperExoSearch}
V.~Abazov{~\em et al.}, {D0} Collaboration, Phys. Lett. B {\bf 671}, 349 (2009).




\bibitem{ATLASLumi1}
{ATLAS} Collaboration, Eur. Phys. J. C {\bf 71}, 1630 (2011).

\bibitem{ATLASLumi2}
{ATLAS} Collaboration, ATLAS-CONF-2011-116, (2011).
\href{http://cdsweb.cern.ch/record/1376384}{{http://cdsweb.cern.ch/record/1376384}}.

\bibitem{ATLASsim}
{ATLAS} Collaboration, Eur. Phys. J. C {\bf 70}, 823 (2010).

\bibitem{geant4}
S.~Agostinelli{~\em et al.}, Nucl. Instrum. Meth. A {\bf 506}, 250 (2003).

\bibitem{alpgen}
M.~L. Mangano{~\em et al.}, JHEP {\bf 0307}, 001 (2003).

\bibitem{herwig}
G.~Corcella{~\em et al.}, JHEP {\bf 0101}, 010 (2001).

\bibitem{jimmy}
J.~M. Butterworth, J.~R. Forshaw, and M.~H. Seymour, Z. Phys. C {\bf 72}, 637 (1996).

\bibitem{sherpa}
T.~Gleisberg{~\em et al.}, JHEP {\bf 0402}, 056 (2004).

\bibitem{cteq6l1}
J.~Pumplin{~\em et al.}, JHEP {\bf 0207}, 012 (2002).

\bibitem{cteq66m}
P.~M. Nadolsky{~\em et al.}, Phys. Rev. D {\bf 78}, 013004 (2008).

\bibitem{photos}
P.~Golonka and Z.~Was, Eur. Phys. J. C {\bf 45}, 97 (2006).

\bibitem{sherpafsr}
M.~Schonherr and F.~Krauss, JHEP {\bf 0812}, 018 (2008).

P.~Golonka and Z.~Was, Eur. Phys. J. C {\bf 45}, 97 (2006).

\bibitem{pythia}
T.~Sj\"ostrand, S.~Mrenna, and P.~Z. Skands, JHEP {\bf 05}, 026 (2006).

\bibitem{pdfmrst2007}
A.~Sherstnev and R.S. Thorne, Eur. Phys. J. C {\bf 55}, 553 (2008).


\bibitem{tau}
N.~Davidson, G.~Nanava, T.~Przedzinski, E.~Richter-Was, and Z.~Was, (2010)
\newblock \href{http://arxiv.org/abs/1002.0543}{{\tt arXiv:1002.0543}}.

\bibitem{powheg}
S.~Frixione, P.~Nason, and C.~Oleari, JHEP {\bf 11}, 070 (2007).

\bibitem{MCatNLO1}
S.~Frixione and B.~R. Webber, JHEP {\bf 06}, 029 (2002).

\bibitem{MCatNLO2}
S.~Frixione, F.~Stoeckli, P.~Torrielli, and B.~R. Webber, JHEP {\bf 01}, 053 (2011).

\bibitem{CT10}
H.~-L.~Lai, M.~Guzzi, J.~Huston, Z.~Li, P.~M.~Nadolsky, J.~Pumplin and C.~-P.~Yuan, Phys. Rev. D {\bf 82}, 074024 (2010).

\bibitem{Hamberg:1990np}
R.~Hamberg, W.~L. van Neerven, and T.~Matsuura,  Nucl. Phys. B {\bf 359}, 343 (1991); Erratum-ibid. B{\bf 644}, 403 (2002).

\bibitem{Anastasiou:2003ds}
C.~Anastasiou, L.~J. Dixon, K.~Melnikov, and F.~Petriello, Phys. Rev. D {\bf 69}, 094008 (2004).

\bibitem{Bonciani:1998vc}
R.~Bonciani, S.~Catani, M.~L. Mangano, and P.~Nason, Nucl. Phys. B{\bf 529}, 424 (1998).

\bibitem{Moch:2008qy}
S.~Moch and P.~Uwer, Phys. Rev. D {\bf 78}, 034003 (2008).

\bibitem{atlas_electron}
{ATLAS} Collaboration, Eur. Phys. J. C. {\bf 72}, 1909 (2012).

\bibitem{photonpaper}
{ATLAS} Collaboration, Phys. Rev. D {\bf 83}, 052005 (2011).

\bibitem{ATLASPhotonIDConf}
{ATLAS} Collaboration, ATLAS-CONF-2012-123, (2012).
\href{http://cdsweb.cern.ch/record/1473426}{http://cdsweb.cern.ch/record/1473426}


\bibitem{WZpaper}
{ATLAS} Collaboration, JHEP {\bf 1012}, 060 (2010).

\bibitem{antikt}
M.~Cacciari, G.~P. Salam, and G.~Soyez, JHEP {\bf 04}, 063 (2008).

\bibitem{METpaper}
{ATLAS} Collaboration, Eur. Phys. J. C. {\bf 72}, 1844 (2012).

\bibitem{PhotonJetPaper} {ATLAS} Collaboration, \prd {\bf 85}, 092014 (2012).

\bibitem{35pbPhotonPaper} {ATLAS} Collaboration, Phys. Lett. B, {\bf 706}, 150 (2011).

\bibitem{BaurLO}
U.~Baur and E.~L. Berger, Phys. Rev. D {\bf 41}, 1476 (1990).

\bibitem{BaurNLO}
U.~Baur, T.~Han, and J.~Ohnemus, Phys. Rev. D {\bf 48}, 5140 (1993).


\bibitem{ATLAS_Wjet}
{ATLAS} Collaboration, Phys. Lett. B {\bf 698}, 325 (2011).

\bibitem{ATLAS_leptrigger}
{ATLAS} Collaboration, Eur. Phys. J C {\bf 72}, 1849 (2012).

\bibitem{ATLASMuonTrigger}
{ATLAS} Collaboration, ATLAS-CONF-2012-099, (2012).
\href{http://cdsweb.cern.ch/record/1462601}{http://cdsweb.cern.ch/record/1462601}

\bibitem{ATLASElectronTrigger}
{ATLAS} Collaboration, 	ATLAS-CONF-2012-048, (2012).
\href{http://cds.cern.ch/record/1450089/}{http://cds.cern.ch/record/1450089/}


\bibitem{unfolding}
G.~D'Agostini, Nucl. Instrum. Methods Phys. Res., Sect. A {\bf 362}, 487 (1995).

\bibitem{MCFM}
J.~M. Campbell, R.~Ellis and C.~Williams, JHEP {\bf 1107}, 018 (2011).


\bibitem{BFG}
L.~Bourhis, M.~Fontannaz. and J.~P. Guillet, Eur. Phys. J. C {\bf 2}, 529 (1998).


\bibitem{ST}
I.~W. Stewart and F.~J. Tackmann, Phys. Rev. D {\bf 85}, 034011 (2012).

\bibitem{madgraph}
J.~Alwall, M.~Herquet, F.~Maltoni, O.~Mattalaer and T.~Stelzer, JHEP {\bf 06}, 128 (2011).



\bibitem{PhysRevD.47.4889}
U.~Baur and E.~L. Berger, Phys. Rev. D {\bf 47}, 4889 (1993).

\bibitem{LEP}
The LEP Collaborations: ALEPH, DELPHI, L3, OPAL, and the LEP Electroweak Working Group.
\newblock \href{http://arxiv.org/abs/hep-ex/0612034}{{\tt arXiv:0612034}}.


\bibitem{LEP1}
K.~Hagiwara, S.~Ishihara, R.~Szalapski, and D.~Zeppenfeld, Phys. Lett. B {\bf 283}, 353 (1992).

\bibitem{LEP2}
K.~Hagiwara, S.~Ishihara, R.~Szalapski, and D.~Zeppenfeld, Phys. Rev. D {\bf 48}, 2182 (1993).

\bibitem{LEP3}
G.~Gounaris{~\em et al.}, Physics at LEP2, vol. 1, 525-576, CERN-96-01-V-1 \href{http://arxiv.org/abs/hep-ph/9601233}{{\tt
  arXiv:hep-ph/9601233}}.
\bibitem{D0comb}
V.~Abazov{~\em et al.}, {D0} Collaboration, Phys. Lett. B {\bf 718}, 451 (2012).




\bibitem{Higgs:ATLAS}
ATLAS Collaboration, Phys. Lett. B {\bf 716}, 1 (2012).

\bibitem{Higgs:CMS}
CMS Collaboration, Phys. Lett. B {\bf 716}, 30 (2012).

\bibitem{Eichten:HiggsImpostor}
E.~Eichten, K.~Lane, and A.~Martin, \href{http://arxiv.org/abs/1210.5462}{{\tt arXiv:hep-ph/1210.5462}}.

\bibitem{Sanino:LightHiggs}
D.~D.~Dietrich, F.~Sannino, and K.~Tuominen, Phys. Rev. D {\bf 72}, 055001 (2005).

\bibitem{Sanino:LightHiggs2}
R.~Foadi, M.~T.~Frandsen, and F.~Sannino, \href{http://arxiv.org/abs/1211.1083}{{\tt arXiv:hep-ph/1211.1083}}.


\bibitem{Lane:2002sm}
K.~Lane and S.~Mrenna, Phys. Rev. D {\bf 67}, 115011 (2003).

\bibitem{Lane:1999uh}
K.~D.~Lane, Phys. Rev. D {\bf 60}, 075007 (1999).

\bibitem{Eichten:2007sx}
E.~Eichten and K.~Lane, Phys. Lett. B {\bf 669}, 235 (2008).



\bibitem{aad:2012kk}
ATLAS Collaboration, (2012) \href{http://arxiv.org/abs/1209.2535v1}{{\tt arXiv:1209.2535}}.

\bibitem{Chatrchyan:2012kk}
CMS Collaboration, (2012) \href{http://arxiv.org/abs/1206.0433v1}{{\tt arXiv:1206.0433}}.


\bibitem{CrystalBall1}
M.~J. Oreglia, Ph.D. Thesis, SLAC-R-236 (1980), Appendix D., SLAC.

\bibitem{CrystalBall2}
J.~E. Gaiser, Ph.D. Thesis, SLAC-R-255 (1982), Appendix F., SLAC.

\bibitem{CrystalBall3}
T.~Skwarnickia, Ph.D. Thesis, F31-86-02 (1986), Appendix E., DESY.

\bibitem{PCLpaper}
G.~Cowan, K.~Cranmer, E.~Gross, and O.~Vitells, Eur. Phys. J. C {\bf 71}, 1554 (2011).

\bibitem{CLpaper}
A.~L.~Read, J. Phys. G {\bf 28}, 2693–2704 (2002).


\bibitem{pdfmstw}
A.~D. Martin, W.~J. Stirling, R.~S. Thorne, and G.~Watt, Eur. Phys. J. C {\bf 63}, 189 (2009).


\bibitem{lhapdf}
M.~R.~Whalley, D.~Bourilkov and R.~C.~Group, \href{http://arxiv.org/abs/hep-ph/0508110}{{\tt arXiv:hep-ph/0508110}}.
\end{thebibliography}\endgroup
\clearpage 
\onecolumngrid
\clearpage 
\begin{flushleft}
{\Large The ATLAS Collaboration}

\bigskip

G.~Aad$^{\rm 48}$,
T.~Abajyan$^{\rm 21}$,
B.~Abbott$^{\rm 111}$,
J.~Abdallah$^{\rm 12}$,
S.~Abdel~Khalek$^{\rm 115}$,
A.A.~Abdelalim$^{\rm 49}$,
O.~Abdinov$^{\rm 11}$,
R.~Aben$^{\rm 105}$,
B.~Abi$^{\rm 112}$,
M.~Abolins$^{\rm 88}$
% ...
\bigskip
\\
% ...
$^{11}$ Institute of Physics, Azerbaijan Academy of Sciences, Baku, Azerbaijan\\
$^{12}$ Institut de F{\'\i}sica d'Altes Energies and Departament de F{\'\i}sica de la Universitat Aut{\`o}noma de Barcelona and ICREA, Barcelona, Spain\\
% ...
$^{21}$ Physikalisches Institut, University of Bonn, Bonn, Germany\\
% ...
$^{48}$ Fakult{\"a}t f{\"u}r Mathematik und Physik, Albert-Ludwigs-Universit{\"a}t, Freiburg, Germany\\
$^{49}$ Section de Physique, Universit{\'e} de Gen{\`e}ve, Geneva, Switzerland\\
% ...
$^{88}$ Department of Physics and Astronomy, Michigan State University, East Lansing MI, United States of America\\
% ...
$^{105}$ Nikhef National Institute for Subatomic Physics and University of Amsterdam, Amsterdam, Netherlands\\
% ...
$^{111}$ Homer L. Dodge Department of Physics and Astronomy, University of Oklahoma, Norman OK, United States of America\\
$^{112}$ Department of Physics, Oklahoma State University, Stillwater OK, United States of America\\
% ...
$^{115}$ LAL, Universit{\'e} Paris-Sud and CNRS/IN2P3, Orsay, France\\
% ...
\end{flushleft}


\end{document}
%